\documentclass[aps,preprint,superscriptaddress]{revtex4}
\usepackage{amsfonts}
\usepackage{amsmath}
\usepackage{amssymb}
\usepackage{mathrsfs}
\usepackage{subfigure}
\usepackage{graphicx}
\usepackage{epstopdf}
\usepackage{float}
\usepackage{color}
\usepackage{bm}
\usepackage{ulem}
\usepackage{xcolor}
\usepackage{appendix}
\usepackage{booktabs}
\usepackage{hyperref}
\usepackage{array}
\usepackage{multirow}

\hypersetup{colorlinks=true, linkcolor=blue, filecolor=blue, urlcolor=blue, citecolor=blue}
\begin{document}
\title{Probing the thermodynamics of charged Gauss Bonnet AdS black holes with the Lyapunov exponent}
\preprint{CTP-SCU/2023039}
\author{Xin Lyu}
\email{lyuxin@stu.scu.edu.cn}
\affiliation{Center for Theoretical Physics, College of Physics, Sichuan University, Chengdu, 610065, China}
\author{Jun Tao}
\email{taojun@scu.edu.cn}
\author{Peng Wang}
\email{pengw@scu.edu.cn}
\affiliation{Center for Theoretical Physics, College of Physics, Sichuan University, Chengdu, 610065, China}
\begin{abstract}
   In this paper, we investigate the thermodynamic properties of charged AdS Gauss-Bonnet black holes and the associations with the Lyapunov exponent. The chaotic features of the black holes and the isobaric heat capacity characterized by Lyapunov exponent are studied to reveal the stability of black hole phases. With the consideration of both timelike and null geodesic, we find the relationship between Lyapunov exponent and Hawking temperature can fully embody the feature of the Small/Large phase transition and the triple point even further. Then we briefly reveal the properties of Lyapunov exponent as an order parameter and explore the black hole shadow with it. 
\end{abstract}
\maketitle
\tableofcontents
\section{Introduction}
The black hole thermodynamics can be utilized to probe the profound relationship between general gravity, quantum field theory, thermodynamics, and statistical mechanics, revealing the nature and properties of black holes, which would provide guidelines for harmonizing the quantum theory and gravitational theory. The study of black hole thermodynamics begins when Hawking proved that the area of the event horizon of a black hole does not decrease, which is known as the area theorem in 1971 \cite{Hawking:1971tu}. After 1972, similar to the thermodynamic entropy increase principle, Bekenstein put forward an entropy describing black holes which is proportional to the area at black hole's horizon, named after Bekenstein entropy \cite{Bekenstein:1973ur,Bekenstein:1974ax}. In 1974, Hawking proposed Hawking radiation from curved spacetime quantum field theory, establishing a correlation between the temperature of a black hole and its surface gravity \cite{Hawking:1974rv}. Then Bardeen, Carter and Hawking established the four laws of black hole thermodynamics \cite{Bardeen:1973gs}, similar to the four laws of ordinary thermodynamics. 
 
The anti-de Sitter spacetime gains a great deal of attention, because of the convincing arguments about the AdS/CFT correspondence inspired by considering the equivalence of a high-dimensional theory of quantum gravity in an anti-de Sitter spacetime and a low-dimensional conformal field theory \cite{Maldacena:1997re,Cai:1998vy}. 
In the anti-de Site space with negative cosmology constant, one of the important research is the Hawking-Page phase transition accompanied with the evolution from pure anti-de Sitter(AdS) space to black holes in asymptotically AdS space \cite{Hawking:1982dh}. Subsequent researches show that the phase behavior of Reissner-Nordstöm-AdS black hole is similar to van der Waals gas-liquid phase transition, and it is a first-order phase transition, where large black holes correspond to the gas phase and small black holes correspond to the liquid phase \cite{Chamblin:1999tk,Chamblin:1999hg}. For these kinds of charged AdS black holes, free energy and heat capacity are frequently used to study phase transition processes in numerous papers \cite{He:2010zb,Guo:2021enm,He:2016fiz,Yao:2020ftk,Huang:2021iyf,Bai:2022hti,Wang:2019urm,Li:2019dai,Wang:2019kxp,Wang:2018xdz, Bai:2023woh}. 

Research on Gauss-Bonnet AdS black holes can be traced before 1971 when Lovelock first proposed that the Gauss-Bonnet term, naturally consistent with the first-order $\alpha'$ correction of the closed-string low-energy effective action \cite{Boulware:1985wk,Zwiebach:1985uq}, in which the gravitational field equation can hold second-order derivatives in higher dimensions($D>4$) \cite{Lovelock:1971yv,Lovelock:1972vz, Wang:2020eln,Bai:2022klw}. 
The generalized Einstein tensor by Lovelock is dimension related, qualifying the symmetric, divergence-free and ghost-free conditions. The Gauss-Bonnet term is associated with the Euler characteristic number through the formula $\mathcal{L}_{GB}=4 \pi \chi$. The Gauss-Bonnet black hole solutions and their thermodynamic properties have been investigated extensively \cite{Oikonomou:2020sij,Liu:2022aqt,Cai:2001dz,Oikonomou:2021kql,Odintsov:2020sqy,Banados:1993ur,Wiltshire:1985us,Hendi:2017lgb,Cvetic:2001bk,Cai:2013qga,Wei:2014hba,Hendi:2016yof,Haroon:2020vpr,Qu:2022nrt,Wang:2019urm,Wei:2012ui}. In four dimensional spacetime, the Gauss-Bonnet term is always zero and is regarded as topological invariant, there's no black hole. Nevertheless, a new method was applied in the study of the four dimensional Gauss-Bonnet gravity theory by rewriting the coefficient of Gauss-Bonnet term $\alpha =\frac{\alpha }{D-4}$ and taking the limit $D\to 4$, which is proposed first by  D. Glavan and C. Lin \cite{Glavan:2019inb}. This work arouses the investigation of 4D Einstein-Gauss-Bonnet gravity including the black hole solutions and its properties \cite{Fernandes:2020rpa,Konoplya:2020qqh,Kumar:2020owy,Yang:2020jno,Wang:2020pmb,Marks:2021fpe}

The chaos theory is constructed for the study of nonlinear systems, aiming to discovers ordered structures and laws hidden in some seemingly disordered phenomena, such as the fractals and the sensitivity of the initial value of a dynamical system, also known as the butterfly effect.  There have been ample studies conducted on black hole's spacetime perturbations and particle orbits around \cite{Bombelli:1991eg,Suzuki:1996gm,Hartl:2002ig,Lu:2018mpr}, which is recognized to be nonlinear and non-integrable in the chaos theory.  Utilized to explore the properties of these systems, the Lyapunov exponent as an indicator of the separation rate between neighboring trajectories, can reflect the sensibility of the system to the initial condition. The positive Lyapunov exponent indicate a system of chaotic property, which means an initially slight difference will leads to exponential separation of trajectories. When Lyapunov exponent $\lambda =0$, the system is kind of stable with the neighboring trajectories keeping at a distance and will neither diverge or converge. For $\lambda <0$, the particle orbit will be asymptotic stability, which means the trajectories nearby will tend to overlap. The Lyapunov exponent can be applied to probe the orbits stability and rate of orbits divergence of the massive and massless particles in the outer space of black holes, which has been already investigated in Schwarzschild-Melvin spacetime \cite{Wang:2016wcj}, accelerating and rotating black holes \cite{Chen:2016tmr} and Born-Infeld AdS spacetime \cite{Yang:2023hci} .etc. 

The first public display of a black hole photo was the M87* black hole in the central giant elliptical galaxy, taken by the Event Horizon Telescope on April 10, 2019\cite{EventHorizonTelescope:2019dse,EventHorizonTelescope:2019uob,EventHorizonTelescope:2019jan,EventHorizonTelescope:2019ths,EventHorizonTelescope:2019pgp,EventHorizonTelescope:2019ggy}, and on May 12, 2022, the first photo of Sgr A*, the supermassive black hole at the center of the Milky Way, was released\cite{EventHorizonTelescope:2022wkp,EventHorizonTelescope:2022apq,EventHorizonTelescope:2022wok,EventHorizonTelescope:2022exc,EventHorizonTelescope:2022urf,EventHorizonTelescope:2022xqj}. 
The study of Schwarzschild black hole whose shadow is a black circular disk attributed to the spherically symmetric condition can be acquired \cite{Synge:1966okc}. It is accompanied with many other studies of black hole shadow \cite{deVries:1999tiy,Bardeen:1972fi,Grenzebach:2014fha,Guo:2018kis,Hennigar:2018hza,Amir:2017slq,Jusufi:2020cpn}. The black hole shadows can reveal many characteristics of black holes, among which we care most is the chaotic properties of black holes. Furthermore, by taking into account the observational effect of the Lyapunov exponents, we are looking forward to explore the relationship between radius of the black hole shadow and the Lyapunov exponents, hoping to observe the chaotic properties of the orbits and then the thermodynamic of black holes, as the thermodynamic properties will be investigated with its Lyapunov exponent. 

The organization of this paper is as follows: In Sect. \ref{properties_GB}, we investigate the thermodynamic of the charged Gauss-Bonnet AdS black hole in 4, 5 and 6D spacetime. In Sect. \ref{Lyapunov_exponents_GB}, the expression of the Lyapunov exponent will be derived.  After obtaining the chaotic properties of both timelike and null geodesic around the d-dimensional Gauss-Bonnet black holes, we take a consideration of isobaric heat capacity with Lyapunov exponents to probing the stability of the black hole phase. Then we obtain the relationship of the Lyapunov exponents versus the Hawking temperature aiming to reveal the phase behavior. The feasibility of the difference of Lyapunov exponent at the phase transition point as an order parameter will also be investigated in this section.  

\section{Charged Gauss-Bonnet AdS black holes}
\label{properties_GB}
The Lagrangian form of the Einstein tensor found by Lovelock can be shown as \cite{Lovelock:1971yv},
\begin{align}
	\label{hl}
	\mathcal{L}=\sum _{i=0}^n \alpha _i \mathcal{L}_i,
\end{align}
in which
\begin{align}
	\label{chl}
	\mathcal{L}_i= \frac{1}{2^i} a_i \ {\delta _{\lambda _1 \rho _1 \cdots  \lambda _i \rho _i}}^{\mu _1 \nu _1 \cdots  \mu _i \nu _i} {R_{\mu _1 \nu _1}}^{\lambda _1 \rho _1} \cdots {R_{\mu _i \nu _i}}^{ \lambda _i \rho _i},
\end{align}
it is symmetric, divergence-free and ghost-free, $\alpha_i$ is the arbitrary constant coefficient and the $\alpha_0$ is the cosmological constant. $\mathcal{L}_i$ is Euler density, in which $\mathcal{L}_0=1$, $\mathcal{L}_1$ and $\mathcal{L}_2$ are the Einstein Hilbert and Gauss Bonnet terms respectively. According to Gauss Bonnet Theory, the Euler characteristic number $\chi$ is equal to the integral of its Gaussian curvature over the whole surface, divided by $4 \pi$, and with the substituting of Riemann tensor, the Euler characteristic number $\chi$ is
\begin{align}
	\label{ecn}
	\chi =\frac{1}{4 \pi }\int _M  R_{\mu  \nu  \rho  \sigma } n^{\mu } n^{\nu } n^{\rho } n^{\sigma }  d A ,
\end{align}
where $R_{\mu  \nu  \rho  \sigma }$ is the Riemann tensor. Then the explicit expression of Gauss Bonnet term with the formula $\mathcal{L}_{\text{GB}}=4 \pi \chi$ take the form as 
\begin{align}
	\label{col}
	\mathcal{L}_{G B}= R_{\mu  \nu  \rho  \sigma } R^{\mu  \nu  \rho  \sigma }-4 R_{\mu  \nu } R^{\mu  \nu }+R^2.
\end{align}
Then, the action of Gauss Bonnet AdS spacetime for $D$-dimensional($D$ \textgreater 4) spacetime can be derived that 
\begin{align}
	\label{ehl}
	S=\int  d^Dx \ \sqrt{-g} \ \left[\frac{1}{16 \pi  G_D}(R-2 \Lambda +\alpha \mathcal{L}_{GB})-\mathcal{L}(F) \right],
\end{align}
here $R$ is the Rician scalar,  and the electromagnetic Lagrangian term is $\mathcal{L}(F)$, which is equal to $ 4 \pi  F^{\mu \nu } F_{\mu \nu }$. The $\Lambda$ is cosmological constant expressed by the AdS radius $l$, which is $\Lambda =-(D-1) (D-2)/2 l^2$. There is a constant $\alpha$ in the action of Gauss Bonnet AdS spacetime, which indicates the coupling strength with Gauss Bonnet invariant term.   

\subsection{Phase transition of 4D charged Gauss-Bonnet AdS black holes}
The Gauss-Bonnet term is well defined when the dimension is greater than four, but it will tend to zero due to its topology properties if $D=4$, which blocks the study of the nature of black holes in four-dimensional Gauss Bonnet black hole until the discovery from D. Glavan and C. Lin, who introduce $\alpha/(D-4)$ to replace the ordinary $\alpha$ with the limit $D\rightarrow 4$, in order to make the four dimensional Gauss Bonnet black hole solution non-trivial. Thus, one can obtain the action of charged 4D Gauss-Bonnet AdS gravity presented as \cite{Glavan:2019inb}, 
\begin{align}
	\label{e4l}
	S=\int  d^4x \ \sqrt{-g}\ \left[ \frac{R-2 \Lambda +\alpha \mathcal{L}_{GB}}{16 \pi  G}- \mathcal{L}(F)\right]. 
\end{align}
Under the static spherical symmetry condition, the metric of 4-dimensional Gauss-Bonnet AdS black hole is \cite{Qu:2022nrt,Cai:2013qga,Bousder:2021aek},
\begin{align}
	\label{static_spherical_metric}
	ds^2=-f(r) dt^2+\frac{1}{f(r)} d r^2 + r^2 d \Omega^{2}_{2}, 
\end{align}
here,
\begin{align}
	\label{metric_function_of 4D}
	f(r)=1+\frac{r^2}{2 \alpha } \left[1-\sqrt{1 + 4 \alpha \left(\frac{2 M}{r^3}-\frac{q^2}{r^4}-\frac{1}{l^2}\right)}\right],
\end{align}
where $q$ , $M$ is the charge and the mass of the  black hole respectively. The form of the metric function indicates that these parameters $M$, $q$, $l$ and $\alpha$ need to fit the real number condition of 
$
	4 \alpha (\frac{2 M}{r^3}-\frac{q^2}{r^4}-\frac{1}{l^2}) +1 > 0,
$
developing the upper bound $\alpha_M$. 

From the condition $f(r)=0$, we can get the horizon radius of the black hole $r_+$, and the explicit expression of black hole mass $M$ can be obtained as,  
\begin{align}
	\label{mass_of_4D}
	M=\frac{r_+}{2}+\frac{q^2}{2 r_+}+\frac{r_+^3}{2 l^2}+\frac{\alpha }{2 r_+},
\end{align}
\begin{table}[t]
	\centering
	\setlength{\tabcolsep}{12mm}
	\begin{tabular}{c|ccc}
		\midrule[2pt]
		\toprule
		$q$ & $\alpha_c$ & $r_{+c}$ & $T_{c}$ \\
		\toprule[1pt]
		0 & 0.0106 & 0.3700 & 0.2484 \\
		0.02 & 0.0104 & 0.3706 & 0.2485 \\
		0.05 & 0.0096 & 0.3740 & 0.2495 \\
		0.10 & 0.0066 & 0.3853 & 0.2528 \\
		0.1667 & 0 & 0.4082 & 0.2599 \\
		\midrule[2pt]
	\end{tabular}
	\caption{Data for critical values of different charge $q$ in 4D spacetime. The table shows the critical values of different fixed $q$, including the Gauss-Bonnet constant $\alpha_c$, the horizon $r_{+c}$ and the Hawking temperature $T_c$. With the increase of charge $q$ from $0$ to $q_c=0.1667$, the critical values of Gauss-Bonnet constant $\alpha$ will drop to zero from $0.0106$, with the event horizon $r_{+c}$ and Hawking temperature $T_c$ both trending up. When $q$ is greater than $0.1667$, $\alpha_c$ will turn to negative, which is out of our league. }\label{critical_value_4D}
\end{table}
The Hawking temperature for charged Gauss-Bonnet AdS black holes can be derived  \cite{Hawking:1975vcx,Bekenstein:1972tm},
\begin{align}
	\label{HT_of_4D}
	T=\frac{r_+^2-q^2-\alpha +3 r_+^4 l^{-2}}{4 \pi r_+ (r_+^2+2\alpha)},
\end{align}
and the Bekenstein-Hawking entropy derived as follows,
\begin{align}
	\label{BH_of_4D}
	S=\pi r_+^2+4\pi \alpha \ln \left(\frac{r_+}{\alpha}\right).
\end{align}
Moreover, the first law of black hole thermodynamics reads,
\begin{align}
	\label{the _first_law_of_GB_bh}
	dM=\Phi dq+T dS.
\end{align}
By computing the Euclidean action in the semiclassical approximation, the free energy of the black hole can be obtained by $F=M-T S$, 
\begin{align}
	\label{free_energy_of_4D}
	F=&\frac{1}{4 l^2 (2 \alpha  r_+ +r_+^3)}\{ [4 \alpha ^2+q^2 (4 \alpha +3 r_+^2)+7 
	\alpha  r_+^2+r_+^4]l^2 \nonumber\\ 
	&+4 \alpha   \ln \left(r_+/\sqrt{\alpha }\right) \left[ 
	\left(\alpha +q^2-r_+^2\right)l^2-3 r_+^4\right]+
	4 \alpha  r_+^4-r_+^6\}.
\end{align}
For the sake of dimension, the powers of $l$ can be used as the physical quantities scale \cite{Guo:2022kio}
\begin{align}
	\label{dimensionless_op}
	\tilde{q}=\frac{q}{l},\ \tilde{r_+}=\frac{r_+}{l},\ \tilde{T}=l T,\ \tilde{F}=\frac{F}{l},\ \tilde{M}=\frac{M}{l},\ \tilde{r}=\frac{r}{l}. 
\end{align}
Therefore, all the quantities are dimensionless under these operations, and we will directly display these quantities without tilde above for convenience in the following content. 

\begin{figure}[t]
	\begin{center}
		\subfigure[$\ \text{4D} $]{\includegraphics[height=5.15cm]{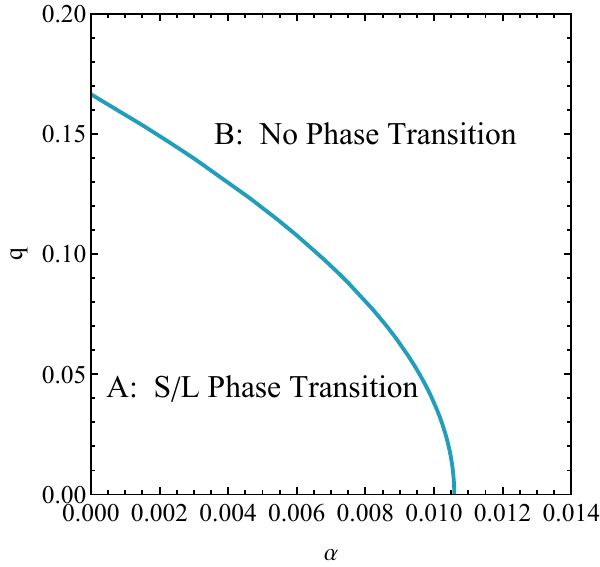}\label{4par}}
		\subfigure[$\ \text{5D} $]{\includegraphics[height=5.0cm]{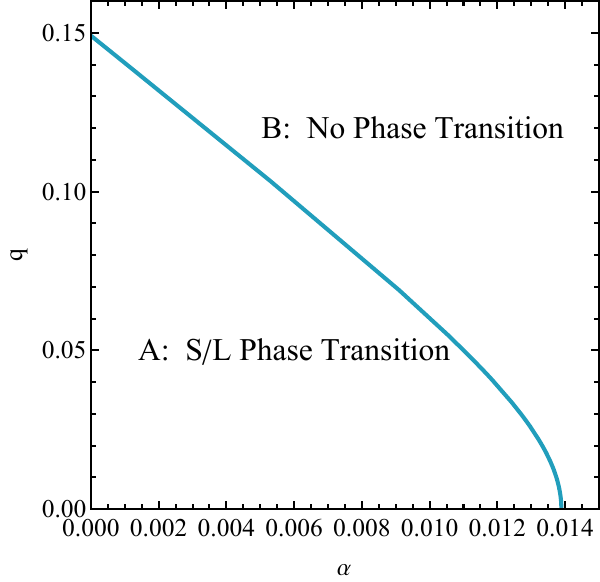}\label{5par}}
		\subfigure[$\ \text{6D} $]{\includegraphics[height=5.1cm]{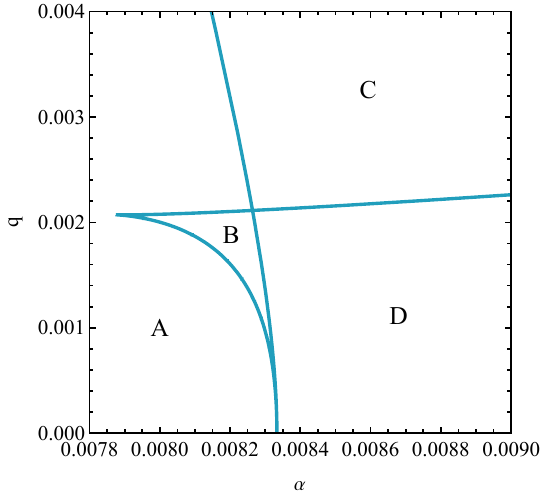}\label{6par}}
	\end{center}
	\caption{The critical curves in $\alpha-q$ parameter space for 4D, 5D and 6D charged Gauss-Bonnet AdS spacetime. In 4D and 5D, the critical curve divide the parameter space into two regions A and B, corresponding to Small/Large phase transition space and no phase transition space respectively, and the curve is the position where the second order phase transition occur. However, the parameter space is divided into four parts marked by A, B, C and D in 6D spacetime. Region A, D represents the Van der Waals-like phase transition, while region C stand for the no phase transition area. Region B has two Small/Large phase transitions, and triple points for the proper value of $q$ and $\alpha$. }\label{par}
\end{figure}

Afterwards, we are going to investigate the phase transition of 4D charged Gauss-Bonnet AdS black hole. The Hawing temperature expressed in Fig. \eqref{HT_of_4D} is of great importance in the investigation of the black hole thermodynamics. With the following critical condition of temperature,
\begin{align}
	\label{critical_condition}
	\frac{\partial T}{\partial r_+}=0,\ \frac{\partial ^2T}{\partial r_+^2}=0,
\end{align}
we can get influence of parameter $\alpha$ and $q$ on the phase transition by the obtained parameter diagram in Fig. \ref{4par}. 

\begin{figure}[t]
	\begin{center}
		\subfigure[$\ T-r_+$]{\includegraphics[height=6.5cm]{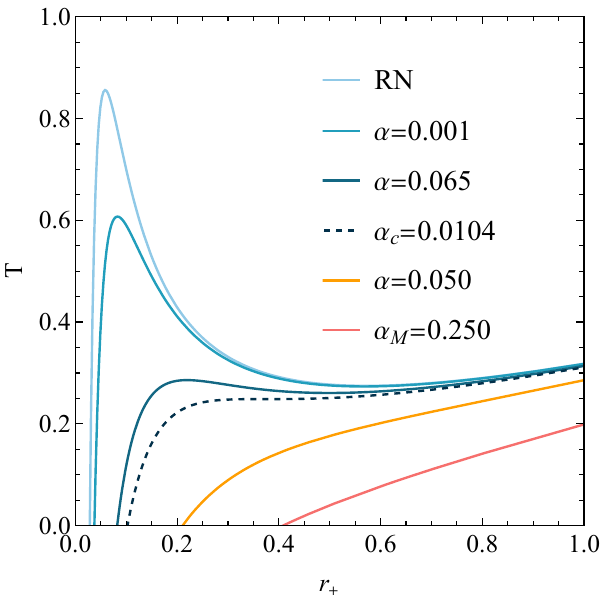}\label{T_r_figure_4D_a}}
		\subfigure[$\ T(r_+,\alpha) $]{\includegraphics[height=6.6cm]{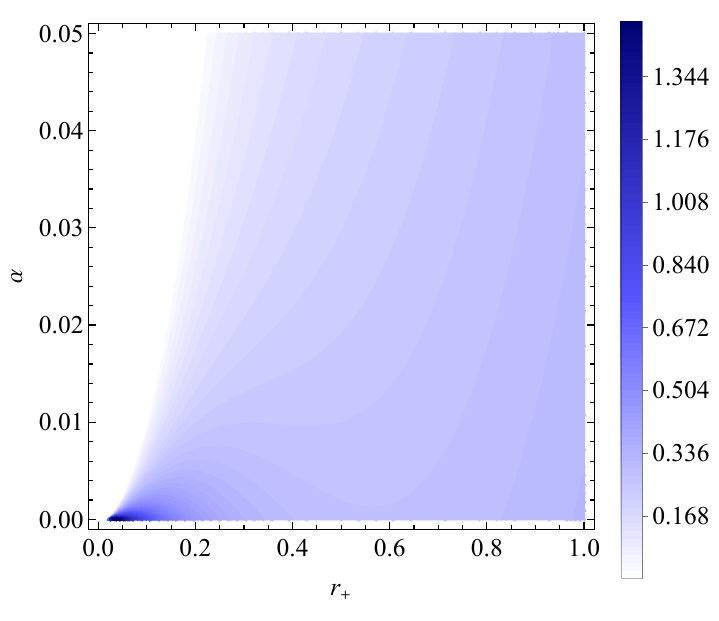}\label{T_r_figure_4D_b}}
	\end{center}
	\caption{The $T-r_+$ diagram for different Gauss-Bonnet constant $\alpha$ and the contour map in $\alpha-r_+$ space for the fixed $q=0.02$ in 4D spacetime. (a) As $\alpha=\alpha_c=0.010427$, the dashed curve in the diagram, divides the curves into two branches. The blue curves have two extreme points and another branch are a monotonically increasing functions of horizon $r_+$. (b) The contour map is depicted with $T(r_+,\alpha)>0$, and the color changing from white to dark blue represents the rising of Hawking temperature $T$. The cutoff line in the right column shows that with the increase of $\alpha$, the minimum value of horizon $r_+$ will become greater. }\label{T_r_figure_4D}
\end{figure}
\begin{figure}[t]
	\begin{center}
		\subfigure[$ \ \alpha \rightarrow 0$]{\includegraphics[height=4.5cm]{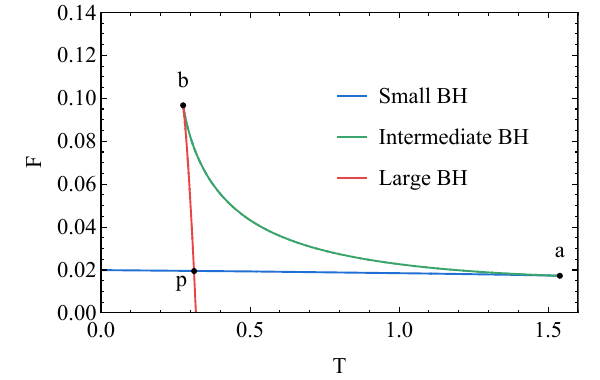}\label{F_T_4_0}}
		\subfigure[$ \ \alpha<\alpha_c$]{\includegraphics[height=4.5cm]{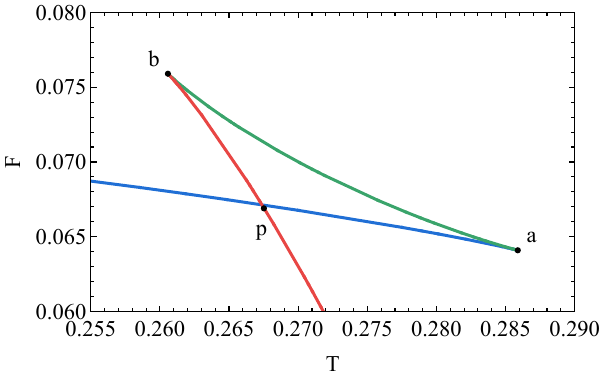}\label{F_T_4_1}}
		\subfigure[$ \ \alpha=\alpha_c$]{\includegraphics[height=4.5cm]{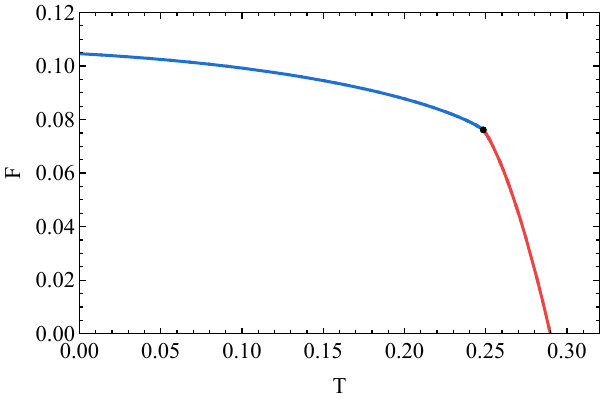}\label{F_T_4_2}}
		\subfigure[$ \ \alpha>\alpha_c$]{\includegraphics[height=4.5cm]{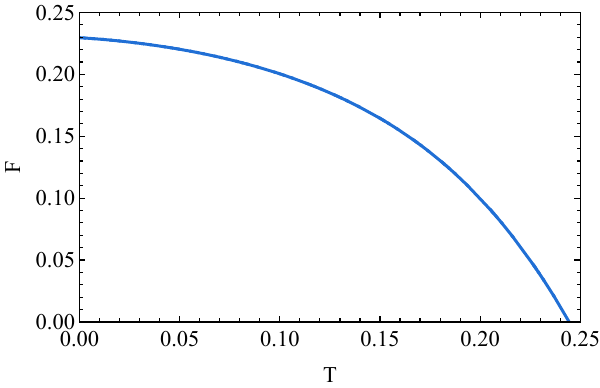}\label{F_T_4_3}}
	\end{center}
	\caption{The diagram of free energy versus Hawking temperature for fixed $q=0.02$ in 4D spacetime. (a)(b) For $\alpha \rightarrow 0$ and $\alpha=0.0065<\alpha_c$, there are three types of black hole which is marked by blue, green and red corresponding to Small BH, Intermediate BH and Large BH, respectively. There is a Small/Large phase transition on $p$. Point $a$ and $b$ represent the evolution from Small BH to Intermediate BH and from Intermediate BH to Large BH respectively. (c) When $\alpha=\alpha_c$, the points $a$ and $b$ will tern to one point, where Small BH will change to Large BH smoothly. (d) For $\alpha>\alpha_c$, the behavior of $F-T$ curve is a monotone decreasing function with no phase transition. }\label{F_T_figure_4D}
\end{figure}

There will be a Van de Waals-like phase transition in Region A, while there is only one phase in Region B. From the detailed information in Table \ref{critical_value_4D}, as $q<0.1667$, there's always a critical value of the Gauss-Bonnet constant $\alpha_c$. For $\alpha<\alpha_c$ as demonstrated in Region A of Fig. \ref{4par}, there's a Small/Large phase transition. Conversely, if $\alpha>\alpha_c$, there's no Small/Large Phase transition, while it will have a second order phase transition when $\alpha=\alpha_c$. In the case of $q>0.1667$, there's no critical value $\alpha_c$ and will be no phase transition generated. Aiming to have a  more subtle observation, we choose the black hole charge $q$ to be $0.02$ for the investigation of thermodynamic properties with varying $\alpha$. Thus, the critical values can be read from Table \ref{critical_value_4D},  
$
	\tilde{\alpha }_c\approx 0.0104,\ \tilde{r}_{+c}\approx 0.3702 \ \text{and}\ \tilde{T}_c\approx 0.2492.
$

To demonstrate the properties of black hole thermodynamics through Hawking temperature when $q=0.02$, we obtain the diagram of temperature $T$ against the horizon $r_+$ for the various $\alpha$ as presented in Fig. \ref{T_r_figure_4D}. 
Note that $\alpha_c$ is the critical value for the fixed $q=0.02$ in Fig. \ref{T_r_figure_4D_a}. For $\alpha< \alpha_c$, there are two extreme points and one inflection point, whose thermodynamic properties of the 4D charged Gauss-Bonnet AdS black holes are similar to the Reissner-Nordström-AdS black holes. There's a phase transition temperature between extremely large and extremely small values according to the Maxwell's law of equal area. As $\alpha=\alpha_c$, it is in a state of criticality with the two extreme points converging. For $\alpha>\alpha_c$, no extreme point exists, which means no phase transition will occur. The relationship of Hawking temperature versus $r_+$ and continuous $\alpha$ is shown in Fig. \ref{T_r_figure_4D_b}, we can see that there is a cutoff line with the condition $T(r_+,\alpha)>0$, which shows that in the four-dimensional case, a larger $\alpha$ will make the cutoff value of $r_+$ increase. 

To reflect the stability and phase transition characteristics of black holes, one can illustrate the free energy $F$ against Hawking temperature $T$ according to Eq. \eqref{HT_of_4D} and Eq. \eqref{free_energy_of_4D} with different $\alpha$ in Fig. \ref{F_T_figure_4D}. For $\alpha<\alpha_c$, there is a characteristic swallow tail composed of three phases i.e., Small BH, Intermediate BH and Large BH in Figs. \ref{F_T_4_0} and \ref{F_T_4_1}. According to the stable equilibrium condition, the section p-a-b-p in the diagram is unstable. There is a first order phase transition called Small/Large transition poised to happen on $p$. 
The Small/Large phase transition is considered to be a gas-liquid phase transition in van der Waals fluids, and it is equivalent to Reissner-Nordström type for $\alpha \rightarrow 0$ shown in Fig. \ref{F_T_4_1}. As $\alpha$ approaches $\alpha_c=0.010427$, the states of black hole will change from Small BH into Large BH at the black dot with the growing of $r_+$ in Fig. \ref{F_T_4_2}, which is accompanied by a second order phase transition. For $\alpha =0.05 > \alpha_c$,  the free energy decrease monotonically with Hawking temperature opposite to the situations discussed above, which shows no Small/Large phase transition exists presented in Fig. \ref{F_T_4_3}. 

\subsection{The $D$-dimensional($D>4$) Gauss-Bonnet AdS black hole} 
In this section, we will discuss the thermodynamics of $D$-dimensional($D>4$) charged Gauss-Bonnet AdS black holes. The metric of the black holes can be derived as \cite{Cai:2013qga},
\begin{align}
	\label{line_element_dD}
	ds^2=-f(r) dt^2 + \frac{1}{f(r)} dr^2 + r^2 h_{ij} dx^i dx^j,
\end{align}
where $h_{ij}$ is the line element of a symmetric space with the curvature $(D-2)(D-3)k$ and volume of the $(D-2)$ dimensional subspace $\Sigma _k$, here $k$ can choose $-1$, $0$ and $1$ corresponding to hyperbolic, Ricci-flat and spherical case within generality, respectively. One can choose spherical topology of Gauss-Bonnet AdS black holes with $k=1$, and the metric function reads, 
\begin{align}
	\label{metric_function_dD}
	f(r)=1+\frac{r^2}{2 \tilde{\alpha }} \left(1-\sqrt{1+\frac{64 \pi  M \tilde{\alpha }}{(D-2) r^{D-1} \Sigma _k}-\frac{2 q^2 \tilde{\alpha }}{(D-2) (D-3) r^{2 D-4}}+\frac{8 \tilde{\alpha }  \Lambda}{(D-1) (D-2)}}\right),
\end{align}
here $\tilde{\alpha }=(D-3) (D-4) \alpha$, $q$ is the total charge of the black hole, $M$ is the black hole mass and one can set $\Sigma _k=1$ in the following content. The square root operation in the metric function requires the values $M$, $q$, $l$, and $\alpha$ that 
\begin{align}
	1+\frac{64 \pi  M \tilde{\alpha }}{(D-2) r^{D-1}}+\frac{8 \tilde{\alpha }  \Lambda  }{(D-1) (D-2)}-\frac{2 q^2 \tilde{\alpha }}{(D-2) (D-3) r^{2 D-4}} > 0,
\end{align}
indicating the upper bound $\alpha_M$.  

The horizon radius of $D$-dimensional Gauss-Bonnet AdS black holes $r_+$ is given by $f(r_+)=0$, then we obtain the mass of $D$-dimensional Gauss-Bonnet AdS black holes, 

\begin{align}
	\label{mass_of_dD}
	M=\frac{r_+^{-D-5}}{16 \pi  (D-2) l^2} \left\{(D-3) (D-2) r_+^{2 D} \left[ \left(\alpha  (D-4) (D-3)+r_+^2\right)l^2+r_+^4\right]+2 l^2 q^2 r_+^8 \right\}. 
\end{align} 
The Hawking temperature for these black holes can be obtained as,
\begin{align}
	\label{HT_dD}
	T=\frac{\alpha  (D-5) (D-4) (D-3)-(2 q^2 r_+^{8-2 D}+2 \Lambda  r_+^4)(D-2)^{-1}+(D-3) r_+^2}{4 \pi  r_+ \left[2 \alpha  (D-4) (D-3)+r_+^2\right]},
\end{align}  
then we have the Bekenstein-Hawking entropy,
\begin{align}
	\label{bh_dD}
	S=\frac{r_+^{D-2}}{4}  \left[1+\frac{2\alpha}{ r_+^2}  (D-2)   (D-3) \right].
\end{align}
The first law of $D$-dimensional Gauss-Bonnet AdS black hole thermodynamics can be expressed as \cite{Bekenstein:1973ur}
\begin{align}
	\label{the_first_law_of_dD}
	dM=\Phi  dq+T dS.
\end{align}
The free energy of the $D$-dimensional GB AdS black hole satisfying $F=M-T S$, yields
\begin{multline}
	\label{free_energy_dD}
	F=A \{2 \left(2 D^2-7 D+5\right) q^2 r_+^{5-D}+2 \alpha ^2 (D-3)^3 (D-2)^2 \left(D^2-5 D+4\right) r_+^{D-5}+ \\
	 \left(D^2-5 D+6\right)  (-36 \alpha  \Lambda +12 \alpha  D \Lambda +D-1) r_+^{D-1} +\alpha  \left(D^3-11 D^2+26 D-16\right) (D-3)^2 r_+^{D-3} \\
	+4 \alpha  \left(2 D^4-19 D^3+64 D^2-89 D+42\right) q^2 r_+^{3-D}+2 (D-3) \Lambda  r_+^{ D+1} \},
\end{multline}
where $$A^{-1}=16 \pi  (D-3) (D-2) (D-1) [2 \alpha  \left(D^2-7 D+12\right)+r_+^2].$$
From dimensional analysis, the powers of $l$ can be used as the physical quantities scale similar to Eq. \eqref{dimensionless_op} in four dimension for the simplicity. 
  
\subsubsection{Phase transition of 5D charged Gauss-Bonnet AdS black hole}
In this section, we will probe the thermodynamic properties of 5D charged Gauss-Bonnet AdS black hole. 
The Hawking temperature can reveal abundant information of the black holes, which is influenced by horizon $r_+$, charge of black hole $q$ and Gauss-Bonnet constant $\alpha$ as prepared in Eq. \eqref{HT_dD}. We obtain the diagram of $\alpha-q$ parameter space through the critical condition shown in Eq. \eqref{critical_condition}, which can be presented in Fig. \ref{5par}. It is topological identical to 4D case with two Regions A and B, corresponding to Small/Large phase transition and no phase transition situation respectively. It can be seen in more detail in the Table \ref{critical_value_5D} that, the thermodynamic properties of black holes are more subtle as $q\in (0, 0.1491)$, accompanied with the coexistence of a Small/Large phase transition case and a no-phase transition case. 

For ampler thermodynamic properties of the black holes, we choose $q=0.02$ in order to assure the existence of $\alpha_c$. Thus, other critical values of Hawking temperature, Gauss-Bonnet coupling constant and horizon radius will be determined as demonstrated in Table \ref{critical_value_5D}, 
\begin{table}[t]
	\centering
	\setlength{\tabcolsep}{12mm}
	\begin{tabular}{c|ccc}
		\toprule
		\midrule[2pt]
		$q$ & $\alpha_c$ & $r_{+c}$ & $T_c$ \\
		\toprule[1pt]
		0 & 0.0139 & 0.4082 & 0.3898 \\
		0.02 & 0.0133 & 0.4210 & 0.3924 \\
		0.10 & 0.0057 & 0.5254 & 0.4222 \\
		0.1491 & 0 & 0.5774 & 0.4302 \\
		\bottomrule
		\midrule[2pt]
	\end{tabular}
	\caption{Critical values of 5D charged Gauss Bonnet AdS black hole. The critical value of Gauss-Bonnet constant $\alpha_c$ will decrease with the increasing charge $q$, and it will drop to zero when $q=0.1491$. The critical values of horizon $r_{+c}$ and the Hawking temperature $T_c$ will maintaining the same trend as charge $q$. The phase transition will occur when $q\in (0,q_c)$, where $q_c=0.1491$. }\label{critical_value_5D}
\end{table}
\begin{figure}[t]
	\begin{center}
		\subfigure[$\ T-r_+$]{\includegraphics[height=6.5cm]{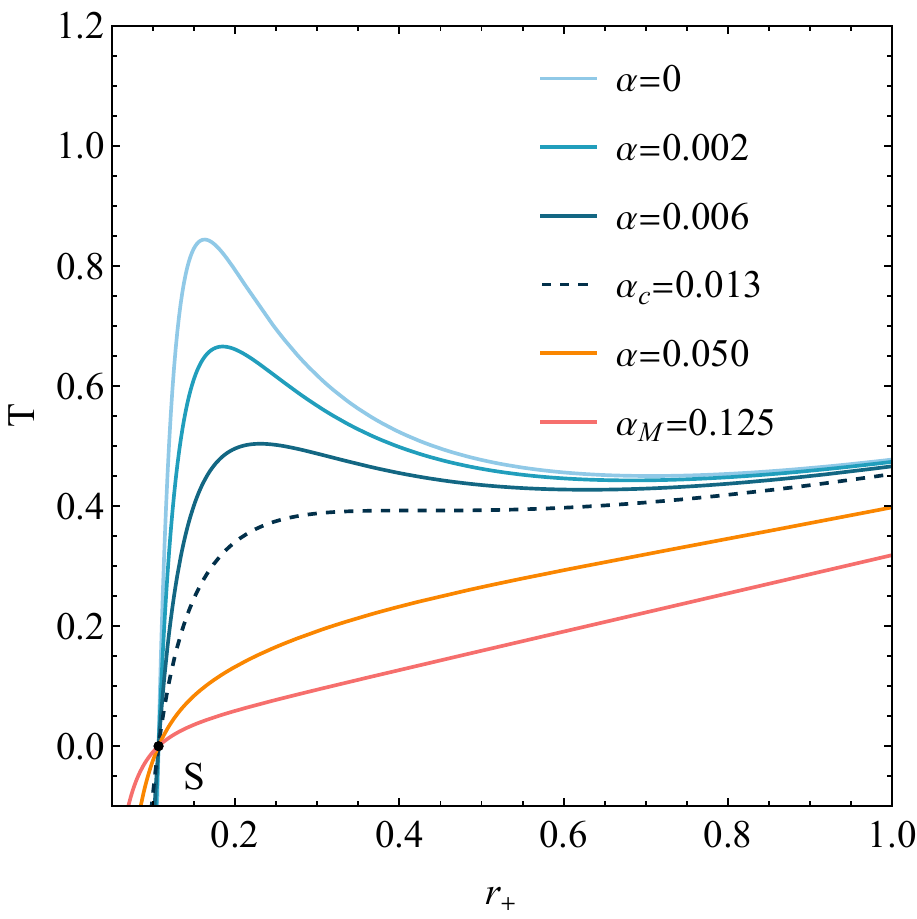}\label{5_tr}}
		\subfigure[$\ T(r_+,\alpha)$]{\includegraphics[height=6.5cm]{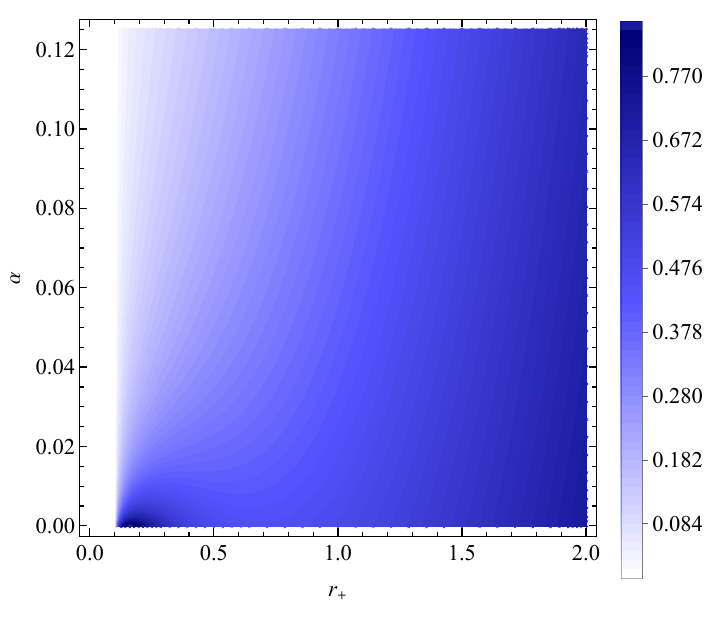}\label{5_tr_c}}
	\end{center}
	\caption{The $T-r_+$ diagram of 5D charged Gauss-Bonnet AdS black hole when $q=0.02$. (a) The set of the $T-r_+$ blue curves whose $\alpha$ are lower than $\alpha_c=0.013323$ have two extreme points, while the dashed curve with extreme points and inflexion point coinciding is the critical situation. The other curves are simple cases with Hawking temperature monotonically increasing. (b) From the color distribution of the contour map, we can see that the possible phase transition will occur when $\alpha$ is less than $\alpha_c$, and characteristic of the color that keeps getting darker when $\alpha$ continue to rise shows the monotonically increasing relationship between $T$ and $r_+$. From the diagram we can see that there's a vertical cutoff line $r_+=0.106852$. }
	\label{T_r_figure_5D}
\end{figure}
$
	\tilde{\alpha }_c\approx 0.0133, \ \tilde{r}_{+c}\approx 0.421, \ \text{and} \ \tilde{T}_c\approx 0.3924.
$

\begin{figure}[t]
	\begin{center}
		\subfigure[$\ \alpha<\alpha_c$]{\includegraphics[height=4.5cm]{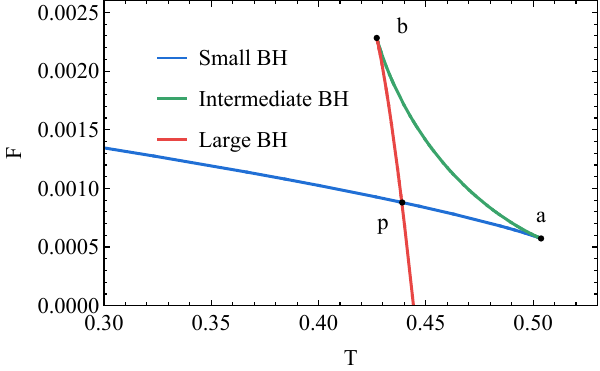}\label{5f_t_1}}
		\subfigure[$\ \alpha>\alpha_c$]{\includegraphics[height=4.5cm]{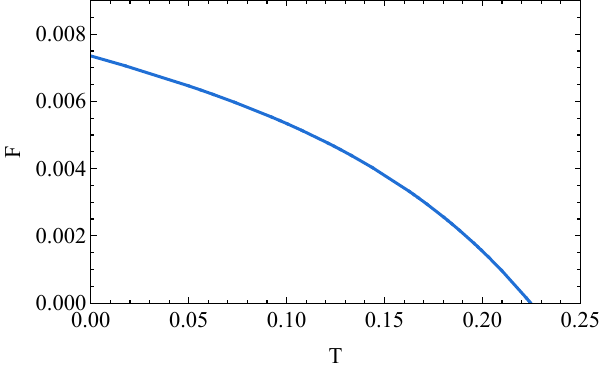}\label{5f_t_2}}
	\end{center}
	\caption{The diagram of free energy versus Hawking temperature in 5D charged Gauss-Bonnet AdS black hole when $q=0.02$. (a) For $\alpha=0.006<\alpha_c$, there are three phase Small BH, Intermediate BH and Large BH in the diagram painted by blue, green and red respectively. Point $a$ and $b$ reflect the change of the black hole phase which is the extreme point in $T-r_+$ diagram. The Small/Large phase transition with a leap of horizon $r_+$ will happen at the point $p$. (b) In right column with $\alpha=0.05>\alpha_c$, the free energy decreases with increasing Hawking temperature, and there is no phase transition. }
	\label{free_energy__HT_5D}
\end{figure}

For the presentation of thermodynamics with Hawking temperature, we have the diagram of Hawking temperature versus the event horizon shown in Fig. \ref{5_tr}, whose values $\alpha$ varies from $0$ to $\alpha_M$. The phase transition of 5D charged Gauss-Bonnet AdS black holes with the fixed charge $q=0.02$ can be split into two branches by $\alpha$ similar to 4D, and we will set $\alpha_c=0.013324$ to be the boundary. For $\alpha<\alpha_c$, these $T-r_+$ curves have two extreme points on two sides of the inflection point. For $\alpha >\alpha_c$, no extreme point exists, which means no Small/Large phase transition will occur. The Hawking temperature will tend to positive infinity regardless of $\alpha$ as the growing of the horizon $r_+$. Here's a difference we find from the 4D case that, for various $\alpha$, the corresponding $T-r_+$ curves will intersect at the black dot  $S(0.106852,0)$ in Fig. \ref{5_tr}. It shows that the Hawking temperature will always be equal to zero when $r_+ \rightarrow 0.106852$ for $\alpha \in (0,\alpha_{M})$, which can be inferred from Eq. \eqref{HT_dD} that, 
\begin{align}
	\label{zpr_+}
	r_{+0}=\frac{\sqrt{3}}{4\pi}\sqrt{B+\frac{1}{B}-1}, 
\end{align}
where
$
B=\sqrt[3]{18 q^2+6 \sqrt{q^2 \left(9 q^2-1\right)}-1}, 
$
independent of $\alpha$ and consisting with the $T-r_+$ diagram. 
Actually, the intersection point is in the fourth quadrant of the $T-r_+$ diagram in 4D spacetime. For a displaying of continuing $\alpha$ in the interval ($0, \alpha_M$), we depict the contour map as shown in Fig. \ref{5_tr_c}. We can see that there is a cutoff line assuring the positive Hawking temperature. In terms of colour depth variation, we find $T$ is a monotonically increasing function of $r_+$ when $\alpha>\alpha_c$, corresponding to the no phase transition case.  

A better demonstration of the phase transition of 5D charged Gauss Bonnet AdS black hole can be obtained by combining the free energy Eq. \eqref{free_energy_dD} and the Hawking temperature Eq. \eqref{HT_dD}. Then we can draw the parameter curves of free energy $F$ and Hawking temperature $T$ in Fig. \ref{free_energy__HT_5D} by taking $r_+$ as the parameter. From Fig. \ref{5f_t_1}, we can find that for $\alpha=0.006<\alpha_c$, there is a characteristic swallow tail which can be regarded as a salient feature of the Small/Large phase transition process. The Small/Large phase transition point is the intersection point $p(0.439, 0.00088)$ of the blue and red curves, corresponding to the Small BH and Large BH respectively. The Small/Large phase transition accompanies with a discontinuous variation of horizon $r_+$ at $p$. When $\alpha=0.05>\alpha_c$ as shown in Fig. \ref{5f_t_2}, no phase transition exists, and there is only one phase in the process. 

\subsubsection{Phase transition of 6D charged Gauss-Bonnet AdS black hole}

As the four and five dimensional cases investigated, we will keep studying the thermodynamics of 6D charged Gauss-Bonnet AdS black hole. Likewise, from the critical condition Eq. \eqref{critical_condition}, the $q-\alpha$ parameter diagram can be obtained in Fig. \ref{6par}. Different from four and five dimension, there are four regions marked by A, B, C and D. Regions A and D are the familiar Van der Waals-like phase transition area, separated by regions B and C. It's worth noting that there are two phase transitions occurring in the region B, and suitable values of $\alpha$ and $q$ will lead to the triple point case, like \cite{Wei:2014hba}. Region D is relatively barren by comparison accompanied with no Small/Large phase transition.  

Combining Fig. \ref{6par} and Table \ref{critical_value_6D}, one can analyze the phase transition in the $q-\alpha$ parameter space in more detail. When $q>0.139521$, there's no critical value of $\alpha$ with only no phase transition occurring, for $\alpha>\alpha_M$ is not allowed. If $0.012666<q<0.139521$, it is comparable to the 4 and 5D case accompanied by one critical value $\alpha_c$. 

For $0.00211365<q<0.012666$, $\alpha$ have two critical values $\alpha_{cs}$ and $\alpha_{cl}$, whose notation $\alpha_{cs}$ and $\alpha_{cl}$ represent the small value and the large value of Gauss-Bonnet constant respectively. The Small/Large phase transition will only happen as $\alpha \in (0,\alpha_{cs})$ and $\alpha \in (\alpha_{cl}, \alpha_M)$. There isn't Small/Large phase transition when $\alpha \in (\alpha_{cs},\alpha_{cl})$, which is distinctive. For $q$ belongs to the interval ($0, 0.00211365$), $\alpha$ have two critical values $\alpha_{cs}$, $\alpha_{cl}$ and a special value $\alpha_{T}$, and $\alpha_T$ is corresponding to the triple point case. Especially, the Small/Large phase transition will always exist once $\alpha \in (0,\alpha_M)$. Moreover, the two values, $\alpha_{cs}$ and $\alpha_{cl}$, are almost the same at conventional accuracy when $q=0$. 

To reveal the subtler features in the 6D charged AdS black hole, we require $q=0.00205$, which will have two critical values of $\alpha_{cs}$, $\alpha_{cl}$ and $\alpha_T$ with the coexistence of triple point and Small/Large phase transition. Therefore, we have the critical values presented from Table. \ref{critical_value_6D}.

\begin{table}[t]
	\centering
	\setlength{\tabcolsep}{3.6mm}
	\begin{tabular}{c|cccccc}
		\midrule[2pt]
		$q$ & $\alpha_{cs}$ & $\alpha_{cl}$ & $r_{+cs}$ & $r_{+cl}$ & $T_{cs}$ & $T_{cl}$ \\
		\toprule[1pt]
		0 & 0.0083& 0.0083& 0.31622&0.31622 & 0.5033 & 0.5033 \\
		
		0.00205 & 0.00826729 & 0.0079356 & 0.347028 & 0.229139 & 0.504548 & 0.512826 \\
		
		0.00211365 & 0.0082639 & 0.0082639 & 0.347799 & 0.189987 & 0.50461 & 0.503725 \\
		
		0.126666 & 0.00747791 & 0.0416667 & 0.426374 & 0.259851 & 0.517628 & 0.247016 \\
		
		0.139521 &0 & ~ &0.670574 & ~ & 0.609965 & ~ \\
			
		\midrule[2pt]
	\end{tabular}
	\caption{Critical values of 6D charged Gauss Bonnet AdS black hole. For the black hole charge $q \in (0,0.00211365)$, there are two critical values $\alpha_{cs}$ and $\alpha_{cl}$. In this case, there will always phase transitions whatever the $\alpha \in (0,\alpha_M)$ is. Moreover, there's a special triple point case $\alpha_T$ in the interval $(\alpha_{cs}, \alpha_{cl})$. As $q$ belongs to the interval $(0.00211265, 0.012666)$, there are two critical values as well. However, there's no Small/Large phase transition when $\alpha\in (\alpha_{cs},\alpha_{cl})$. As $q$ in ($0.012666,0.139521$), there is only one critical value $\alpha_c$ with the restriction $\alpha<\alpha_M$. When $q>0.139521$, there's no Small/Large phase transition in the interval ($0,\alpha_M$).   }\label{critical_value_6D}
\end{table}

\begin{figure}[t]
	\begin{center}
		\subfigure[$\ T-r_+$]{\includegraphics[height=6.5cm]{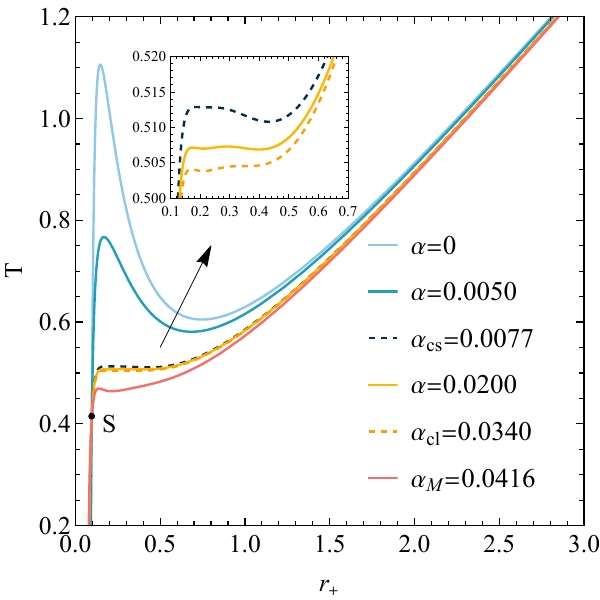}\label{6_tr}}
		\subfigure[$\ T(r_+, \alpha)$]{\includegraphics[height=6.5cm]{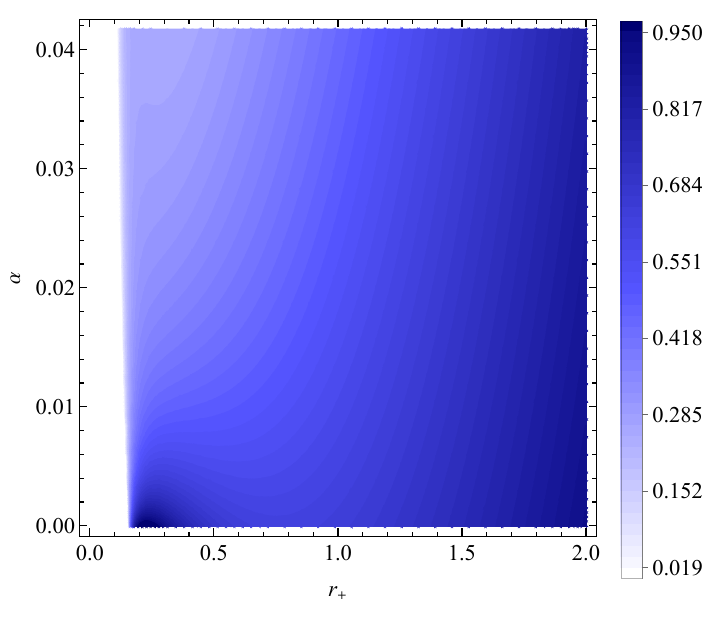}\label{6_tr_c}}
	\end{center}
	\caption{The $T-r_+$ diagram of 6D charged Gauss-Bonnet AdS black hole for the fixed $q=0.00205$. (a) This diagram can be divided into three branches by the critical values $\alpha_{cs}=0.0077$ and $\alpha_{cl}=0.0340$. For $\alpha<\alpha_{cs}$, there are two extreme points and one inflexion point, which is the comparable to the case $\alpha>\alpha_{cl}$. When $\alpha_{cs}<\alpha<\alpha_{cl}$, it will be more subtle with four extreme points coexisting, which indicates an abundant phase transition properties. (b) The color depth variation can used to mark the magnitude of Hawking temperature in $\alpha-r_+$ parameter space. One can easily infer that the greater $\alpha$ causing a stronger restriction of $r_+$ due to condition $T(r_+,\alpha)>0$. The whole area of the diagram can be roughly divided into three parts according to $\alpha$ as seen in the left column. }
	\label{T_r_figure_6D}
\end{figure}

\begin{figure}[t]
	\begin{center}
		\subfigure[$\ \alpha<\alpha_{cs} \ \text{and } \ \alpha>\alpha_{cl}$]{\includegraphics[height=4.5cm]{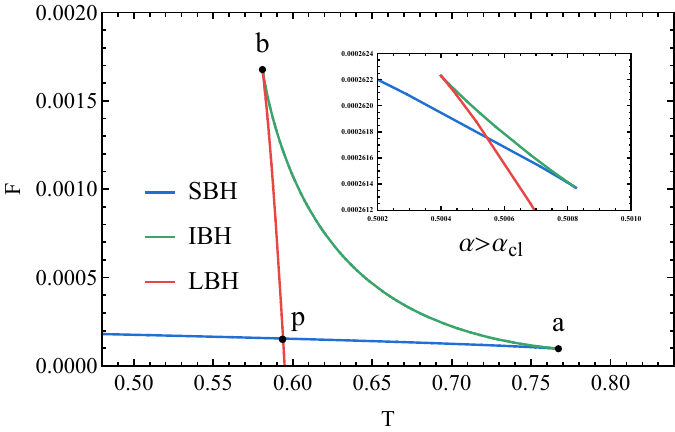}\label{6_ft_1}}
		\subfigure[$\ \alpha_{cs}<\alpha<\alpha_{T}$]{\includegraphics[height=4.6cm]{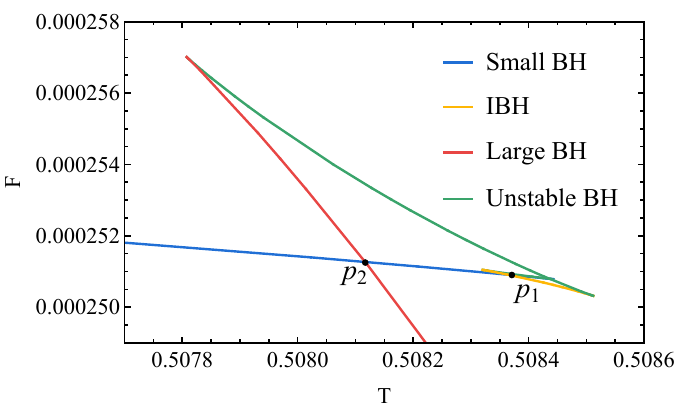}\label{6_ft_2}}
		\subfigure[$\ \alpha=\alpha_{T}$]{\includegraphics[height=4.5cm]{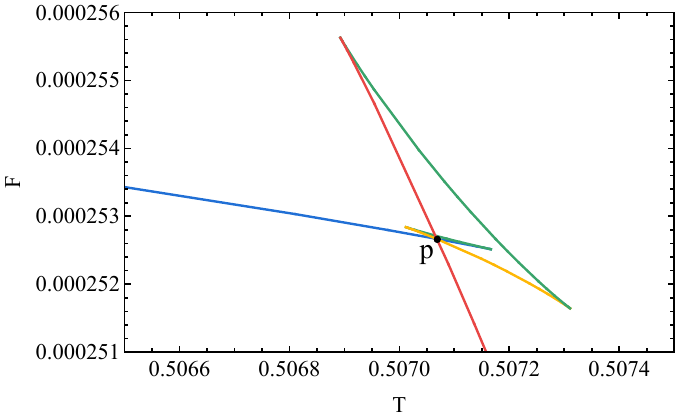}\label{6_ft_3}}
		\subfigure[$\ \alpha_T<\alpha<\alpha_{cl}$]{\includegraphics[height=4.5cm]{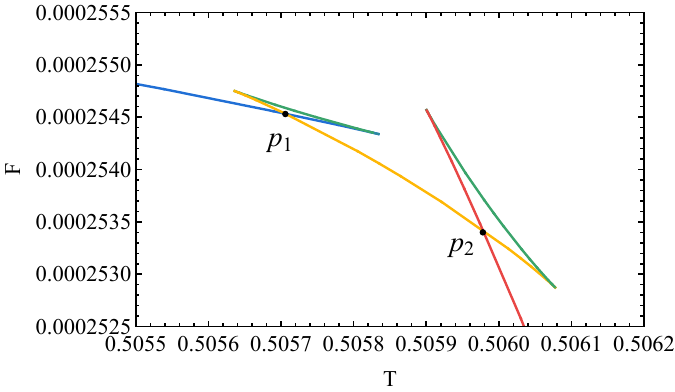}\label{6_ft_4}}
	\end{center}
	\caption{The diagram of free energy versus Hawking temperature for different $\alpha$ in six dimension when $q=0.00205$, and $\alpha_{cs}$, $\alpha_{cl}$ are the frontiers of the parameter space while $\alpha_T$ is the condition of triple point. (a) For $\alpha=0.003$ and $\alpha=0.0084$, there tree phases shown in the swallow tail, and the Small/Large phase transition point is noted by $p$. (b) When $\alpha=0.0081$, there are two swallow tails in the $F-T$ diagram, and the Small/Large phase transition points is marked by $p_1$ and $p_2$, respectively. (c) In the case of $\alpha=0.008149$, the Small/Large phase transition points $p_{p1}$ and $p_{p2}$ coincide with three stable phases, i.e., Small BH, Large BH and Intermediate BH, developing a triple point $p$. (d) For $\alpha = 0.0082$, the behavior of curve is analog to the case $\alpha_{cs}<\alpha<\alpha_{T}$. }
	\label{free_energy_6D}
\end{figure}

\begin{table}[t]
	\centering
	\setlength{\tabcolsep}{9mm}

	\begin{tabular}{c|cccc}
		\midrule[2pt]
		$D$ & $\alpha_M$ & $q_{cs}$ & $q_{cl}$ & $q_{cm}$  \\
		\toprule[1pt]
		7 &0.020833 & 0.004535 & 0.0101589 & 0.134088  \\
		
		8 &0.012500 & 0.004364 & 0.0074119 & 0.130317   \\
		
		9 &0.008333 & 0.003564 & 0.0052932 & 0.127624   \\
		
		10 &0.005952 & 0.002739 & 0.0037539 & 0.125606  \\
		
		\midrule[2pt]
	\end{tabular}
	\caption{Critical values of $D$-dimensional$(D>6)$ charged Gauss Bonnet AdS black hole. It's worth noting that $\alpha_M$ is the upper bound of $D$-dimensional black holes. There will be always Small/Large phase transition as $\alpha \in (0,\alpha_M)$ for $q<q_{cs}$. For $q_{cs}< q < q _{cl}$, there are two critical values $\alpha_{cs}$ and $\alpha_{cl}$. The Small/Large phase transition will occur when $\alpha \in (0,\alpha_{cs}) \cup (\alpha_{cl},\alpha_M)$. For $q_{cl}<q<q_{cm}$, there's only one critical value of $\alpha$, for the restriction of $\alpha_M$. As $q>q_{cm}$, there's no critical value, which means no phase transition as well. }
	\label{critical_value_nD}
\end{table} 

As for now, we are going to probe the thermodynamic properties of black holes by the diagram of Hawking temperature in the presence of a change in $\alpha$, which is presented in Fig. \ref{T_r_figure_6D}. The values of $\alpha$ varying from $0$ to $\alpha_M=0.041667$ can be divided into three branches by critical values $\alpha_{cs}=0.0079356$ and $\alpha_{cl}=0.00826729$ in Fig. \ref{6_tr}. 
Unsurprisingly, for $\alpha<\alpha_{cs}$, the behavior of the temperature is qualitatively similar to the Reissner-Nordström-AdS black holes. These $T-r_+$ curves have a common that they all have two extreme points on both sides of the tipping point, which indicates the existence of Small/Large phase transition from Maxwell's law of area. For $\alpha>\alpha_{cl}$ branch, it is pretty much the same thing as the one just discussed when $\alpha<\alpha_{cs}$,  which also have two extreme points on both sides of the tipping point, implying the Small/Large phase transition. For another branch $\alpha_{cs}<\alpha<\alpha_{cl}$ as shown from the illustration in Fig. \ref{6_tr}, the $T-r_+$ curve is more complicated, which has four extreme points while two inflection points inserting in the gaps. The equation of Hawking temperature in Eq. \eqref{HT_dD} also indicates that, there is a cutoff line of the radius of the event horizon due to the positive Hawking temperature $T>0$ condition as presented in Fig. \ref{6_tr_c}, which will be a restriction of Lyapunov exponent. As what we have discussed in five-dimensional Gauss-Bonnet AdS black holes, there is a black dot $S(0.096, 0.415)$ that all functions will pass though regardless of the value of $\alpha$, and it is different from the 5D case which locates at $(0.106852,0)$.

With the consideration of free energy in Eq. \eqref{free_energy_6D}, one can illustrate the diagram of free energy $F$ against Hawking temperature $T$ with different $\alpha$ to probe the phase transition in Fig. \ref{free_energy_6D}.  
For $\alpha=0.003<\alpha_{cs}$ in Fig. \ref{6_ft_1}, there is a characteristic swallow tail with the existence of three phase structures i.e., Small BH, Intermediate BH and Large BH marked by blue, green and yellow respectively. The behavior of $F-T$ diagram at point $p(0.5936, 0.0001512)$ represent a first order phase transition. For Another branch $\alpha_{cs}<\alpha<\alpha_{cl}$ in Figs. \ref{6_ft_2} and \ref{6_ft_4}, there are three stable phases, i.e., Large, Small and Intermediate BH noted by red, blue and yellow respectively and also two Unstable BHs marked by green. The two swallow tail structures form two Van der Waals-like phase transition processes at $p_1$ and $p_2$. It is worth noting that, the phase transition point $p_2(0.508117, 0.00025125)$ occurs before $p_1(0.508371, 0.0002509)$ when $\alpha=0.0081<\alpha_{T}$ presented in Fig. \ref{6_ft_2}, while $p_1$ is followed by $p_2$ when $\alpha=0.0082>\alpha_{T}$ as shown in Fig. \ref{6_ft_4}. 

For the triple point case when $\alpha=\alpha_T=0.00814854$ as demonstrated in Fig. \ref{6_ft_3}, there are two characteristic swallow tails accompanied with their phase transition points coinciding. The three stable black hole phase will coincide simultaneously at the point $p(0.507069, 0.00025266)$ corresponding to the triple point, which have two discontinuous changes of horizon $r_+$. The phase transitions of triple point will be investigated with the Lyapunov exponents in both timelike and null geodesic.  

Higher dimensions($D>6$) are not as abundant in thermodynamic properties as we might expect \cite{Wei:2014hba}, which are analogous to the 6D black holes except for the absence of two Van der Waals-like phase transitions occurring consecutively, and the triple points. For the partial thermodynamics of higher dimensional charged Gauss-Bonnet AdS balck hole studied as shown in Table \ref{critical_value_nD}, we can see that, the situations of the higher dimension spacetime are similar to each other in the $\alpha-q$ parameter space. For $0<q<q_{cs}$, there will always be a Small/large phase transition when $\alpha\in (0,\alpha_{M})$, and it is different from 6D with another two critical values $\alpha_{cs}$ and $\alpha_{cl}$ and a triple point case happening when $\alpha_{cs}<\alpha<\alpha_{cl}$. For $q \in (q_{cs},q_{cl})$, there are two critical values $\alpha_{cs}$ and $\alpha_{cl}$. The phase transition will only happen when $\alpha<\alpha_{cs} $ and $\alpha>\alpha_{cl}$. There's only one phase in the range $(\alpha_{cs},\alpha_{cl})$. For $q_{cl}<q<q_{cm}$, there's one critical value and the phase transition will happen when $\alpha<\alpha_{c}$, because the other critical value is greater than the upper bound $\alpha_M$. There's no critical value of $\alpha$ and will be no phase transition for $\alpha \in (0,\alpha_M)$ when $q>q_{cm}$. 
The phase structures of higher dimension$(D=7$-$10)$ are no more complex than 6D spacetime, with Van der Waals-like phase transitions. 
Due to space constraints and the similarity of thermodynamics in higher dimensions to the 6D case, we will not discuss it in detail here. Subsequent study focus on 4, 5 and 6D. 

\section{The Lyapunov exponents of $D$-dimensional Gauss-Bonnet AdS black hole}
\label{Lyapunov_exponents_GB}
The Lyapunov exponent is of great importance in researching the sensibility and complexity of dynamic system, which reflect whether the orbitals of the system are separated or converged in phase space. It also can be applied in studying the divergence and convergence rate of trajectories near the black hole. A positive Lyapunov exponent indicates a divergence of nearby trajectories. Lyapunov exponents in background metric have been thoroughly explored, which could describe the stability of geodesics around black holes. 
Especially, for the Lyapunov exponents has been applied to study the Small/Large phase transition of RN AdS black holes and Born-Infeld AdS black hole \cite{Guo:2022kio,Yang:2023hci}, we will extend this method to the application of $D$-dimensional charged Gauss-Bonnet AdS black holes accompanied with other applications such as the stability of the black holes.

We will introduce the effective potential which combines the sum of multiple forms of energy except kinetic energy for particles around the black hole. From the definition, we can write the effective potential as \cite{Cardoso:2008bp,Wei:2023fkn}, 
\begin{align}
	\label{def_effictive _potential}
	V_r=-\dot{r}^2,
\end{align}
where dot represent the derivative of proper time $\tau$ and $E$ is the energy for a unit mass of massive or massless particles. 
The Lyapunov exponents on the stability analysis of geodesic reads,
\begin{align}
	\label{expression_lambda}
	\lambda =\pm \sqrt{-\frac{V_r''}{2 \dot{t}^2}},
\end{align}
for the motion in the equatorial plane and the prime represent the derivative of radius $r$. The stability of circular geodesics is represented by the second derivative on $V_r$ under the condition of $V_r=0$ and $V_r'=0$. Since the signs of Lyapunov exponents doesn't matter in the study of circular stability, we will choose $+$ sign for simplicity on the right-hand side of Eq. \eqref{expression_lambda}. The following content will be divided into three sections: Sect. \ref{timelike_geodesic} is the study of Lyapunov exponents in timelike geodesic with the thermodynamics, and Sect. \ref{null_geodesic} is the investigation of Lyapunov exponents in null geodesic with the thermodynamics and the black hole shadow as a cursory exploration. In Sect. \ref{critical_exponent}, we will consider the difference of Lyaounov exponents as an order parameter, and study its behavior near the critical situation by taking 5D charged Gauss-Bonnet AdS black hole as an example. 

\subsection{Lyapunov exponents and thermodynamics for timelike geodesic}
\label{timelike_geodesic}

For timelike geodesic, we have the Lagrangian of geodesics for 4D charged Gauss Bonnet AdS black holes, 
\begin{align}
	\label{l_timelike_4D}
	2 \mathcal{L}= f(r)\dot{t}^2-\frac{\dot{r}^2}{f(r)}+r^2 \dot{\varphi }^2, 
\end{align}
and the Lagrangian for the 5D case can be expressed as 
\begin{align}
	\label{l_timelike_5D}
	2 \mathcal{L}= f(r)\dot{t}^2-\frac{\dot{r}^2}{f(r)}+r^2\dot{\varphi }^2+r^2\dot{\theta _1}^2+ r^2 \sin ^2\theta _1 \cos ^2\varphi \ \dot{\theta _2}^2,
\end{align}
since the axes in spherical coordinates is $r, \varphi, \theta _1 \ \text{and} \  \theta _2$, respectively. The Lagrangian of 6D black holes employs a similar formula, which can be shown as follows,
\begin{align}
	\label{l_timelike_6D}
	2 \mathcal{L}= f(r)\dot{t}^2-\frac{\dot{r}^2}{f(r)}+r^2\dot{\varphi }^2+r^2\dot{\theta _2}^2 +r^2\dot{\theta _1}^2+ r^2 \cos ^2\theta _1 \sin ^2\theta _2 \ \dot{\theta _3}^2.
\end{align}
We will set a restriction that we fix the timelike geodesic at the equatorial plane ($\theta=\pi/2$ for 4D, $\theta_1=\theta_2=\pi/2$ for 5D case, and $\theta_1=\theta_2=\theta_3=\pi/2$ for the six dimension). Therefore, the Lagrangian for all the dimension will take the same form, 
\begin{align}
	\label{l_timelike_all}
	2 \mathcal{L}= f(r)\dot{t}^2-\frac{\dot{r}^2}{f(r)}+r^2 \dot{\varphi }^2.
\end{align}
From the equation above, we can get the generalized momentum are
\begin{align}
	\label{genealized_momentum_timelike}
	&P_t=\dot{r} f(r)=E,\\ &P_{\varphi }=-r^2 \dot{\varphi }=-L,\\ &P_r=-\frac{\dot{r}}{f(r)},
\end{align}
where $E$ and $L$ is the energy and the angular momentum respectively. From these equations, we can derive the explicit expression of the first order derivative of the time and the angle
\begin{align}
	\label{anglur_time_velosity_timelike}
	\dot{\varphi }=\frac{L}{r^2},\ \dot{t}=\frac{E}{f(r)}.
\end{align}
The Hamiltonian in terms of conserved quantities $L$ and $E$ are
\begin{align}
	\label{h_timelike}
	2 \mathcal{H}=-\frac{\dot{r}^2}{f (r)}-L \dot{\varphi }+E \dot{t}=1.
\end{align}
With the definition of effective potential, we obtain the effective potential 
\begin{align}
	\label{effective_potential}
	V_r=f(r) \left[1-\frac{E}{f(r)}+\frac{L}{r^2}\right].
\end{align}
With Eq. \eqref{expression_lambda} and Eq. \eqref{effective_potential} obtained, we can derive the explicit formula for the following thermodynamic investigation. 

\subsubsection{4D timelike geodesic with Lyapunov exponents}

In this subsection, we will probe the thermodynamics of 4D Gauss-Bonnet AdS black holes with Lyapunov exponents. 
By submitting Eq. \eqref{metric_function_of 4D} and Eq. \eqref{mass_of_4D} the Eq. \eqref{effective_potential}, we have the effective potential shown as, 
\begin{align}
	\label{effective_4D}
	V_r=&1+E^2-\frac{L^2}{r^2} + \frac{\left(L^2+r^2\right)}{2 \alpha} \\ \nonumber
	&-\frac{\left(L^2+r^2\right)r^2}{2 \alpha}\sqrt{ \left[r^4-4 \alpha  q^2-4 \alpha  r^4+ \frac{4 \alpha  r}{ r_+}  \left(\alpha +q^2\right)+  4 \alpha  r r_+^3 + 4 \alpha  r r_+\right]},
\end{align}
where $r$ is the radius of the particle's orbit. 

\begin{figure}[t]
	\begin{center}
		\centering
		\includegraphics[height=7cm]{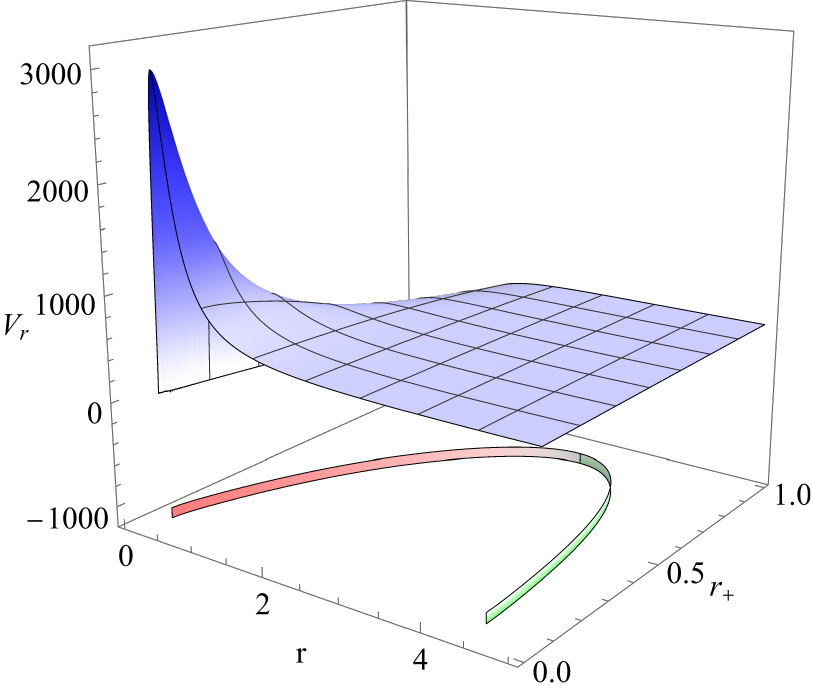}
	\end{center}
	\caption{The diagram of the effective potential and the radius of timelike orbit in four dimensional case with the fixed $q=0.02$ and $\alpha=0.0065$. It is drawn under the condition $r>r_+$ and $T(r_+)>0$. The red and green curves projected onto $V_r = -1000$ plane are drawn to stand for the unstable equilibria and stable equilibria respectively. When $r_+$ is approaching the intersection of the two curves, the unstable equilibrium will coincide with the stable equilibrium, which means the disappearance of unstable equilibrium.  
	}
	\label{4_vr}
\end{figure}

The effective potential is a complicated function affected by the parameters horizon $r_+$, Gauss-Bonnet constant $\alpha$, and charge $q$. 
Therefore, we will observe its property first by numerically obtaining a diagram of the effective potential $V_r$ with respect to $r_+$ and $r$ for the suitable $q$ and $\alpha$ in Fig. \ref{4_vr}. The red curve projected onto $V_r = -1000$ plane, represents the extreme points of effective potential with $V_r'(r_c)=0$ and $V_r''(r_c)<0$, while the green curve projected onto the same plane represent when $V_r'(r_c)=0$ and $V_r''(r_c)>0$, corresponding to unstable equilibria and the stable equilibria respectively. The unstable equilibria are indispensable in the calculation of Lyapunov exponents $\lambda$. With fixed parameter $q$ and $\alpha$, one can find that the unstable equilibrium point will eventually meet the stable equilibrium point in the process of increasing event horizon $r_+$, and after this intersection, there will be no equilibrium position for the massive particles. 

\begin{figure}[t]
	\begin{center}
		\subfigure[$\ \tilde{\lambda}-r_+$]{\includegraphics[height=6.5cm]{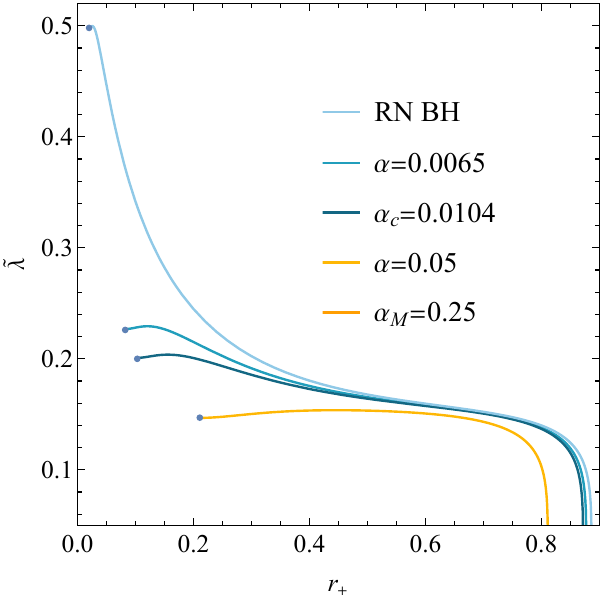}\label{4_lr_1}}
		\subfigure[$\ \tilde{\lambda}(r_+,\alpha)$]{\includegraphics[height=6.6cm]{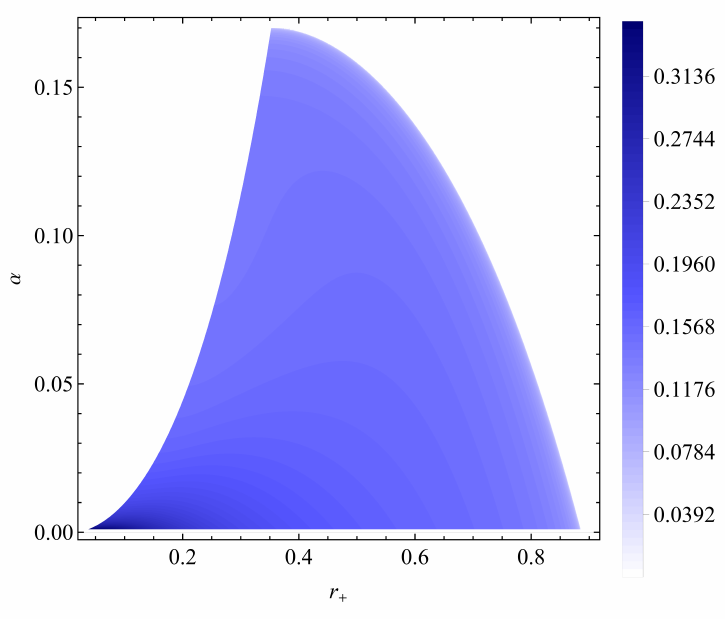}\label{4_lr_2}}
	\end{center}
	\caption{The diagram of logarithm of Lyapunov exponent $\tilde{\lambda}=\log_{100} (\lambda+1)$ versus horizon radius $r_+$ and the contour map in $\alpha-r_+$ space when $q=0.00205$ and $T(r_+,\alpha)>0$ in 4D space. (a) In the diagram of the $\lambda-r_+$ curves for different $\alpha$, the Lyapunov exponent is overall decrease as the increase of $\alpha$. For $\alpha$ fixed, $\lambda$ will increase a little first but reduce continuously to zero with the growing of $r_+$. (b) We can find that when $\alpha>0.17$, Lyapunov exponent will disappear before the upper bound of $\alpha_M=0.25$. }
	\label{Lyapunov_r_4D}
\end{figure}

\begin{figure}[t]
	\begin{center}
		\subfigure[$\  \ C_p-\lambda$]{\includegraphics[height=4.65cm]{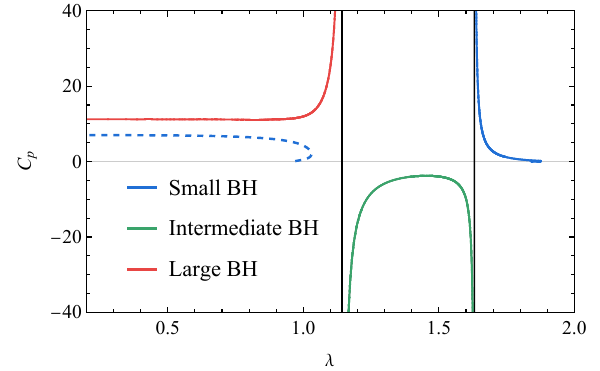}\label{4lc}}
		\subfigure[$\ \ C_p-r_+$]{\includegraphics[height=4.65cm]{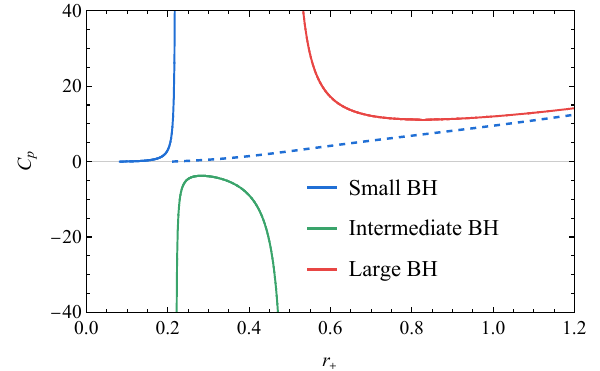}\label{4cr}}
	\end{center}
	\caption{The diagram of isobaric heat capacity when $q=0.02$ in 4D spacetime. The solid line represents the type of Small/Large phase transition and the dashed line is the type when no phase transition occur. We can find that with the Lyapunov exponent decreasing, the phase of black hole will starts with Small BH, going through Intermediate BH and eventually evolves to Large BH. The Small BH and Large BH are stable with $C_p>0$ while the Intermediate BH is unstable with $C_p<0$. The $C_p-\lambda$ diagram of a four-dimensional charged Gauss-Bonnet black hole can be a good alternative to the $C_p-r_+$ diagram to distinguish the stability of different black hole phases. }
	\label{4Dlc}
\end{figure}

Then we are going to derive the expression of Lyapunov exponent at the unstable equilibrium as shown in the $V_r$ diagram Fig. \ref{4_vr}. Characterized by the conditions $V_r=0$ and $V_r'=0$, for the unstable circular orbit, the energy and angular momentum can be expressed,
\begin{align}
	\label{energy_and_anglur_momentum_4D}
	&E^2=\frac{2 f(r_c)^2}{2 f(r_c)-r f'(r_c)},\\ &L^2=\frac{r^3  f'(r_c)}{2 f(r_c)-r f(r_c) },
\end{align}
here 
$
	2 f\left(r_c\right)-r f'(r_c)>0.
$
According to Eq. \eqref{expression_lambda}, the explicit formula of Lyapunov exponent can be derived,
\begin{align}
	\label{lambda_timelike_4D}
	\lambda = \frac{1}{2}  \sqrt{- V_r''(r_c) \left[2 f(r_c)-r_c f'(r_c)\right]}.
\end{align}
One can plug the condition of timelike geodesics in Eq. \eqref{energy_and_anglur_momentum_4D} into Eq. \eqref{lambda_timelike_4D}, which reads,
\begin{align}
	\label{lambda_timelike_4D_2}
	\lambda =\frac{1}{2} \sqrt{-\frac{r_c^3 f'(r_c)  V_r''(r_c)}{L^2}},
\end{align}
where $r_c$ is determined by the circular condition and $L$ is the angular momentum. 

It is clear to see that Lyapunov exponent $\lambda$ is a function of black hole charge $q$ and horizon radius $r_+$ for various Gauss-Bonnet constant $\alpha$. For the absence of valid analytical solution, the behavior of $\lambda$ with respect to $r_+$ is  presented in Fig. \ref{Lyapunov_r_4D} with $q=0.02$ and $\tilde{L}=L/l^2=20$, which is consisting with the dimensionless convection in Eq. \eqref{dimensionless_op}. We can see that these functions have cutoff on the left for different $\alpha$ due to the positive Hawking temperature requirement discussed before  in Fig. \ref{4_lr_1}. 
The Gauss-Bonnet constant between the interval $(0, \alpha_M)$ can be divided into two parts, ($0, 0.17$) and $(0.17, \alpha_M)$ respectively in Fig. \ref{4_lr_2}. For $\alpha\in (0, 0.17)$, the Lyapunov exponent exists, whereas between the interval ($0.17, \alpha_M$), $\lambda$ will always be zero, indicating no chaos in the massive orbits. 
One can find that $\lambda$ will first increase a little then decrease with the rising of $r_+$ and finally be equal to $0$ as $\alpha \in (0,0.17)$, which is caused by the disappearance of the unstable equilibria. The overall trend is monotonically decreasing with the rising $\alpha$, which means the Gauss-Bonnet constant can reduce the level of chaos of the black holes. 

Furthermore, to investigate the relationship between the Lyapunov exponents and the stability of different black hole phase, one can first derive the isobaric heat capacity $C_p$ with the formula,
\begin{align}
	\label{ihc4}
	C_p=-T^2\frac{\partial^2F}{\partial T^2}.
\end{align}
With the submitting of the formulas Eq. \eqref{HT_of_4D} and Eq. \eqref{free_energy_of_4D}, 
we can numerically obtain the diagrams of isobaric heat capacity versus Lyapunov exponents for the two $\alpha$-branches, i.e., $\alpha<\alpha_c$ and $\alpha>\alpha_c$ marked by solid and dashed lines respectively, as shown in Fig. \ref{4lc}. 
We can see that the stability of black holes characterized by the Lyapunov exponent is equivalent to the traditional $C_p-r_+$ diagram in Fig. \ref{4cr}. The red, green and blue curves in the diagram refer to Large BH, Intermediate BH and Small BH, respectively. In the progress of decreasing Lyapunov exponents as $\alpha=0.0065 <\alpha_c$, the black hole phase will evolve from the stable Small BH with positive isobaric heat capacity, going through the unstable black hole with $C_p<0$, to the stable Large BH characterized by the positive $C_p>0$. For $\alpha=0.05>\alpha_c$, there's only one stable phase accompanied with $C_p>0$.  

\begin{figure}[t]
	\begin{center}
		\subfigure[$\ \alpha<\alpha_c$]{\includegraphics[height=4.78cm]{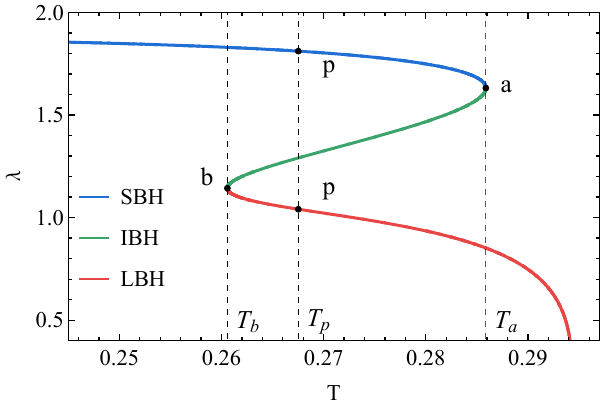}\label{4lt2}}
		\subfigure[$\ \alpha>\alpha_c$]{\includegraphics[height=4.65cm]{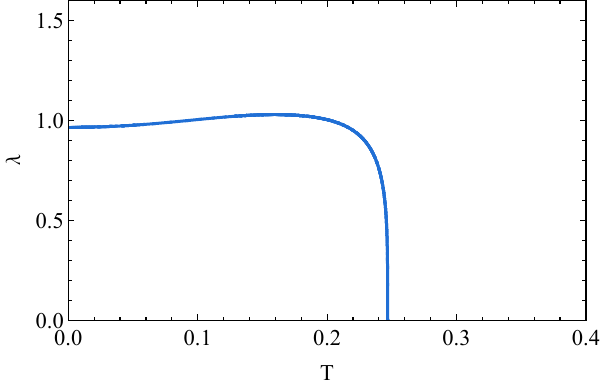}\label{4lt4}}
	\end{center}
	\caption{The diagram of Lyapunov exponent $\lambda$ versus Hawking temperature for different $\alpha$ when $q=0.02$ in 4D spacetime. (a) For $\alpha=0.0065<\alpha_c$, the different colors blue, green and red represent the Small BH, Intermediate BH and Large BH respectively. The multivalued feature of $\lambda$ corresponds to the swallow tail in the $F-T$ diagram, while $T_p=0.26752$ is the Small/Large phase transition temperature. (b) For $\alpha=0.05>\alpha_c$, $\lambda$ is a monotonically decreasing function with respect to $T$, which has no Small/Large phase transition existed. }
	\label{Lyapunov_T_4D}
\end{figure}

As the Lyapunov exponents $\lambda$ with respect to the horizon $r_+$ displayed in Eq. \eqref{lambda_timelike_4D_2}, we can derive the relationship between Lyapunov exponents $\lambda$ and Hawking temperature $T$ analogizing to the $F-T$ diagram. Thus, we can obtained the $\lambda-T$ diagram for various $\alpha$ by taking the horizon $r_+$ as the parameter in Fig. \ref{Lyapunov_T_4D}, where the Large, Intermediate and Small BH corresponding to red, green and blue respectively. Note that we depict the dashed line of Hawking temperature corresponding to the points $a$, and $b$ in Fig. \ref{F_T_figure_4D}, and the Small/Large phase transition points $p$ have also been marked. 

For $\alpha=0.0065<\alpha_c$ as presented in Fig. \ref{4lt2}, we could find that $\lambda$ is a multivalued function of $T$. Moreover, as  the Hawking temperature grows in the interval ($0, T_a$), the Lyapunov exponent corresponding to the Small BH phase decreases a little, which is a stable phase as predicted by the blue solid curve shown in the $C_p-\lambda$ diagram Fig. \ref{4lc}. It will evolve from the stable Small BH into unstable Intermediate BH at $T_a=0.285884$, and the unstable Intermediate BH is represented by the green solid curve in the $C_p-\lambda$ diagram. When Hawking temperature approaches $T_b=0.260578$, the smooth evolution from the unstable Intermediate BH to stable Large BH will occur. The Small/Large phase transition will occur when $T=T_p$, with the discontinuous variation of event horizon $r_+$ and the Lyapunov exponent $\Delta \lambda$. As the growth of $T$ in the interval ($T_p, +\infty$), it is accompanied with a continuing decline of $\lambda$ until zero.  
For $\alpha=0.05>\alpha_c$ as shown in Fig. \ref{4lt4}, the Lyapunov exponent will first rise a little and then continuously decrease to zero, which implies that, there is no Small/Large phase transition, with one stable phase demonstrated in the $C_p$ diagram Fig. \ref{4lc}. The reason that Lyapunov exponent will tend to zero is the disappearance of unstable equilibrium as implied in the $\lambda-r_+$ diagram Fig. \ref{4_lr_2}. 

\subsubsection{5D timelike geodesic with Lyapunov exponents}

\begin{figure}[t]
	\begin{center}
		\centering
		\includegraphics[height=7cm]{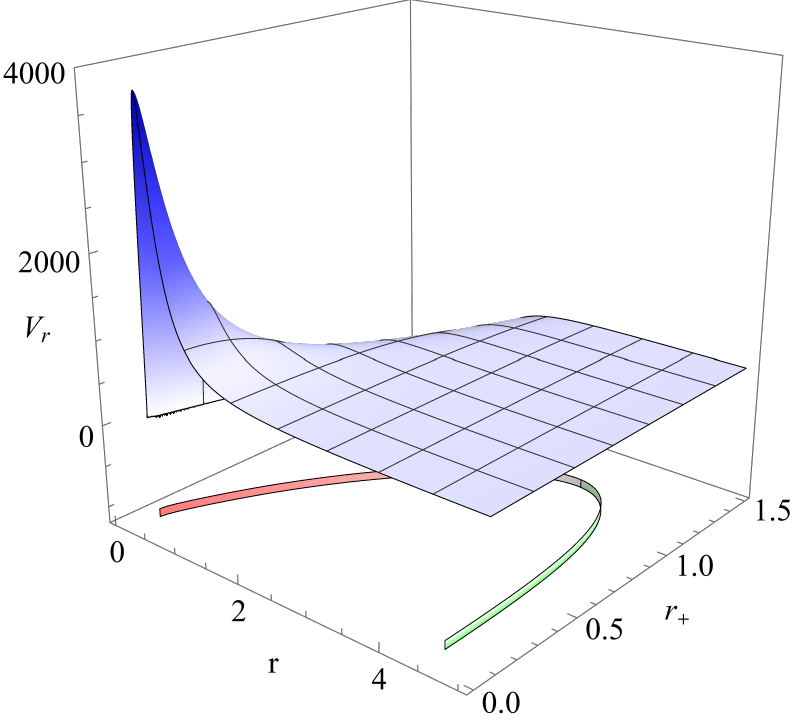}
	\end{center}
	\caption{The diagram of the effective potential versus horizon $r_+$ and radius of the massive orbit $r$ when $q=0.02$ and $\alpha=0.006$ in 5D spacetime. The red and green curves projected onto $V_r = -1000$ plane are corresponding to the unstable equilibria and stable equilibria respectively. When $r_+$ is greater than the $r_+$-coordinate corresponding to the intersection point of the two curves, there's no stable equilibrium and unstable equilibrium, which can be validated in the calculation of the Lyapunov exponents. }
	\label{effective_potential_5D}
\end{figure}

The thermodynamics properties with Lyapunov exponent will be investigated in 5D charged Gauss-Bonnet AdS black hole in this subsection. One can find that there are some parameters, i.e., $q$, $\alpha$ and $r_+$, in the effective potential by submitting Eqs. \eqref{metric_function_dD} and \eqref{mass_of_dD} into Eq. \eqref{effective_potential}. To demonstrate its trend, we make a diagram of $V_r$ against horizon $r_+$ and radius of massive orbits $r$ in Fig. \ref{effective_potential_5D}. We find the extreme points for the stable and unstable equilibria are represented by the green curve and red curves projected onto $V_r = -1000$ plane. The unstable equilibria of our interest can be utilized to calculate the Lyapunov exponents. As in the 4D spacetime, with fixed $q=0.02$ and $L=20$, whose symbol $L$ is a replacement of $\tilde{L}$ for simplification, we can find that the unstable equilibrium point will eventually meet the stable equilibrium point in the process of rising horizon $r_+$, and after this intersection, there will be no unstable equilibrium and stable equilibrium for the massive particles.

We could derive the expression of Lyapunov exponent at the unstable equilibrium of the effective potential through combining Eq. \eqref{expression_lambda}, Eq. \eqref{effective_potential} and simplifying equation Eq. \eqref{energy_and_anglur_momentum_4D}, The simplified expression of Lyapunov exponent similar to Eq. \eqref{lambda_timelike_4D_2},   
where $r_c$ is determined by the conditions $V_r=0$ and $V_r'=0$. 

\begin{figure}[t]
	\begin{center}
		\subfigure[$\ \tilde{\lambda}-r_+$]{\includegraphics[height=6.5cm]{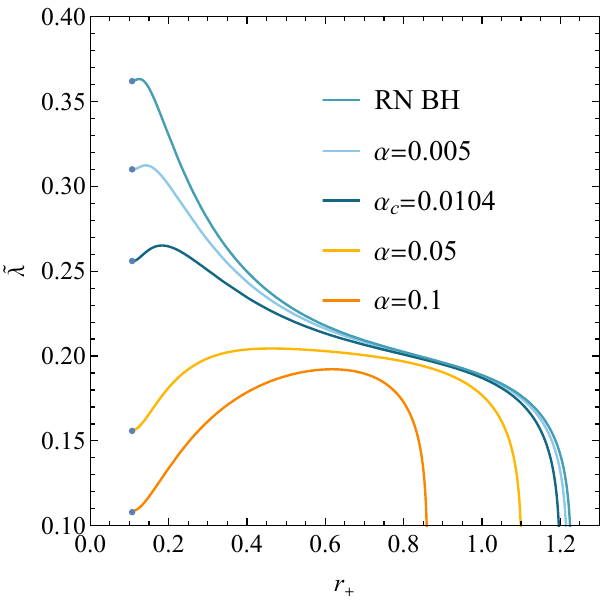}\label{5lr}}
		\subfigure[$\ \tilde{\lambda}(r_+,\alpha)$]{\includegraphics[height=6.5cm]{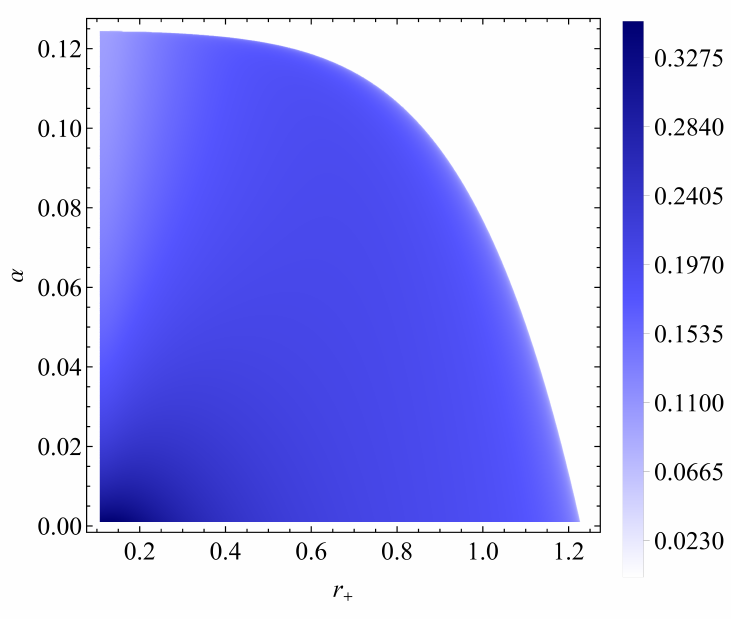}\label{5lrc}}
	\end{center}
	\caption{The diagram of the $\tilde{\lambda}-r_+$ curves and the contour map in $\alpha-r_+$ space for the fixed $q=0.02$ in 5D spacetime. The Lyapunov exponent will have a tendency to increase and then decrease with $r_+$, and $\lambda$ will decrease to zero for the larger $r_+$, implying the disappearance of chaotic orbits of massive particles. As $\alpha$ tends to its upper bound of $\alpha_M=0.125$, $\lambda$ goes to zero overall by the combined effect of $T(r_+)> 0$ and the disappearance of the unstable equilibrium of the effective potential $V_r$. The cutoff values of $r_+$ are mutually equivalent to $0.106852$. }
	\label{Lyapunov_r_5D}
\end{figure}

For the absence of valid analytical solution, the numerically calculated diagram $\lambda-r_+$ can be presented in Fig. \ref{Lyapunov_r_5D}. We can see that these curves all start at $r_+=0.106852$ consisting with the formula Eq. \eqref{zpr_+} and the $T-r_+$ diagram shown in Fig. \ref{T_r_figure_5D}. The Lyapunov exponents will increase a little first and then decline with the rising of $r_+$, and eventually be equal to zero, which implies the non-existence of chaos for the disappearance of the unstable equilibrium. The overall trend is monotonically decreasing when $\alpha$ is increasing in the interval ($0, \alpha_M$) as demonstrated in Fig. \ref{5lrc}, which indicates the stronger Gauss-Bonnet coupling constant make the black holes less chaotic. 
 
\begin{figure}[t]
	\begin{center}
		\subfigure[$\  \ C_p-\lambda$]{\includegraphics[height=4.65cm]{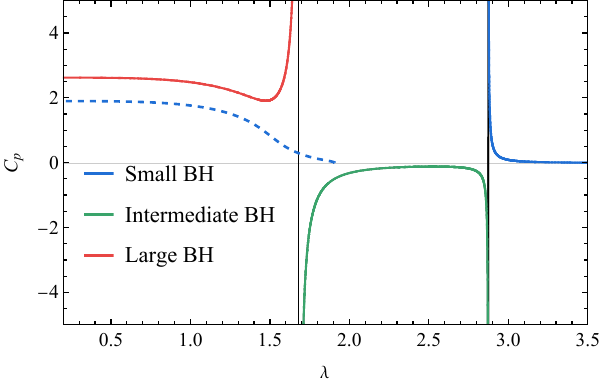}\label{5lc}}
		\subfigure[$\ \ C_p-r_+$]{\includegraphics[height=4.65cm]{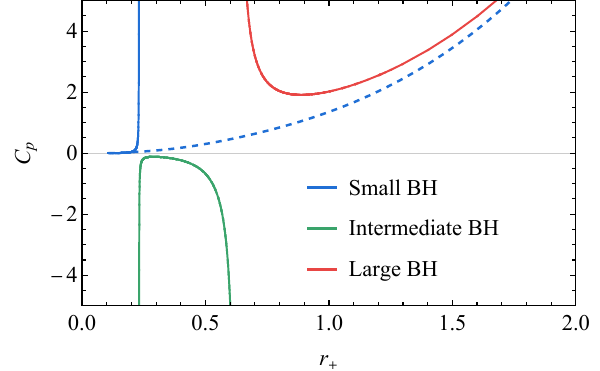}\label{5cr}}
	\end{center}
	\caption{The diagrams of isobaric heat capacity in 5D spacetime. The solid curve represents the type of Small/Large phase transition and the dashed curve is the type that no Small/Large phase transition will occur. We can find that with the Lyapunov exponent decreasing, the black hole phase will starts with Small BH, going through Intermediate BH and eventually evolves to Large BH. The Small BH and Large BH are stable with $C_p>0$ while the Intermediate BH is unstable with $C_p<0$. The five-dimensional $C_p-\lambda$ diagram can be a good alternative to the $C_p-r_+$ diagram to distinguish the stability of different black hole phases. }
	\label{5Dlc}
\end{figure}

To probe the stability of different phases of black holes, we derive two diagrams of $C_p$ with the formula Eq. \eqref{ihc4} in Fig. \ref{5Dlc}, which are $C_p-\lambda$ diagram and $C_p-r_+$ diagram respectively. The solid curve representing the case $\alpha=0.006<\alpha_c$ can be divided into three portions, where the blue portion and red portion, corresponding to Small BH and Large BH respectively, are stable with $C_p>0$. The green portion of the solid curve is the unstable Intermediate BH with the negative isobaric heat capacity. The dashed curve represents the case when $\alpha=0.05>\alpha_c$, and there is only blue portion with positive $C_p>0$. We can deduce that the $C_p-\lambda$ diagram in Fig. \ref{5lc} can fulfill the same role as the traditional $C_p-r_+$ diagram in Fig. \ref{5cr} to probe the black hole stability by positive or negative isobaric heat capacity.  

\begin{figure}[t]
	\begin{center}
		\subfigure[$\
		\alpha<\alpha_c$]{\includegraphics[height=4.65cm]{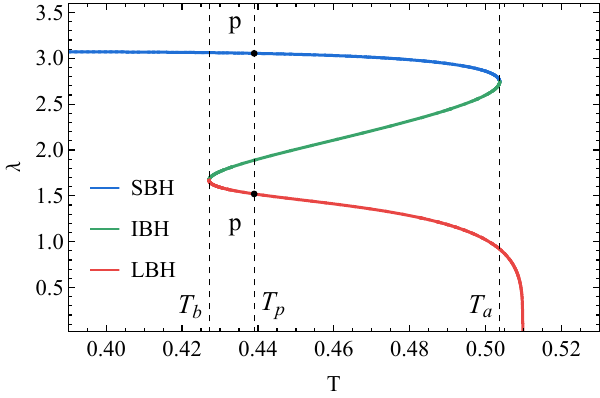}\label{5lt1}}
		\subfigure[$\ \alpha>\alpha_c$]{\includegraphics[height=4.75cm]{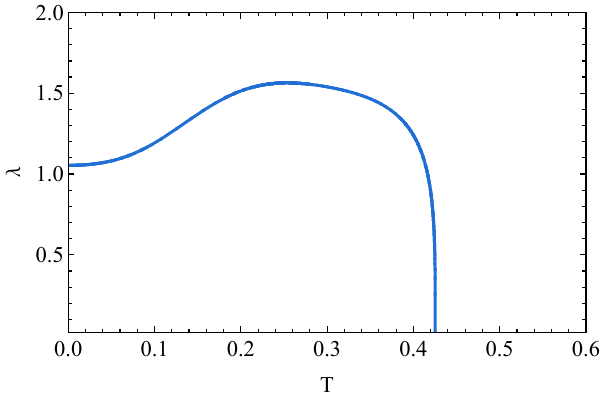}\label{5lt2}}
	\end{center}
	\caption{The $\lambda-T$ diagrams for different $\alpha$ when $q=0.02$ in 5D spacetime. (a) For $\alpha=0.006<\alpha_c$, $\lambda$ forms a multivalued curve against $T$, which corresponds to characteristic swallow tail of the $F-T$ diagram, while red, dark blue and light blue correspond to the Large BH, Intermediate BH and Small BH, respectively. (b) When $\alpha=0.05>\alpha_c$, no phase transition exists, remaining only one phase in the process. }
	\label{L_T_5D}
\end{figure}

The Hawking temperature can be utilized to reveal the chaotic properties of black holes and their phase transitions with Lyapunov exponent, and the $\lambda-T$ diagrams for different $\alpha$ are presented in Fig. \ref{L_T_5D}. Similar to 4D charged Gauss-Bonnet AdS black holes, when the parameter $\alpha=0.006 <\alpha _c$ as shown in Fig. \ref{5lt1}, $\lambda$ versus $T$ is a multivalued function corresponding to the characteristic swallow tail structure in the $F-T$ map in Fig. \ref{5f_t_1}. We could find that as $T$ increases on the Small BH branch before $T_a$, the Lyapunov exponent decreases a little. The Small BH phase transforms into an Intermediate BH at $T_a=0.503702$, while the Intermediate BH is unstable with negative $C_p$ and Small BH is stable for the positive $C_p$ from Fig. \ref{5lc}. As $T$ decreases from $T_a$ to $T_b$, $\lambda$ goes down and the black hole phase transforms from Intermediate BH to Large BH at $T_b=0.42712$, where the Intermediate BH is unstable with negative heat capacity $C_p<0$. For $T$ increasing on the stable Large BH, $\lambda$ will keep decreasing and drop to zero. The Small/Large phase transition will occur when $T=T_p=0.43912$ with the leap of $\lambda$. 
For $\alpha=0.05 >\alpha _c$ presented in Fig. \ref{5lt2}, the Lyapunov exponent will first increase and then continuously drop to zero, which implies no Small/Large phase transition exists and there's only one phase in the progress. 

\subsubsection{6D timelike geodesic with Lyapunov exponents}

\begin{figure}[t]
	\begin{center}
		\centering
		\includegraphics[height=6.5cm]{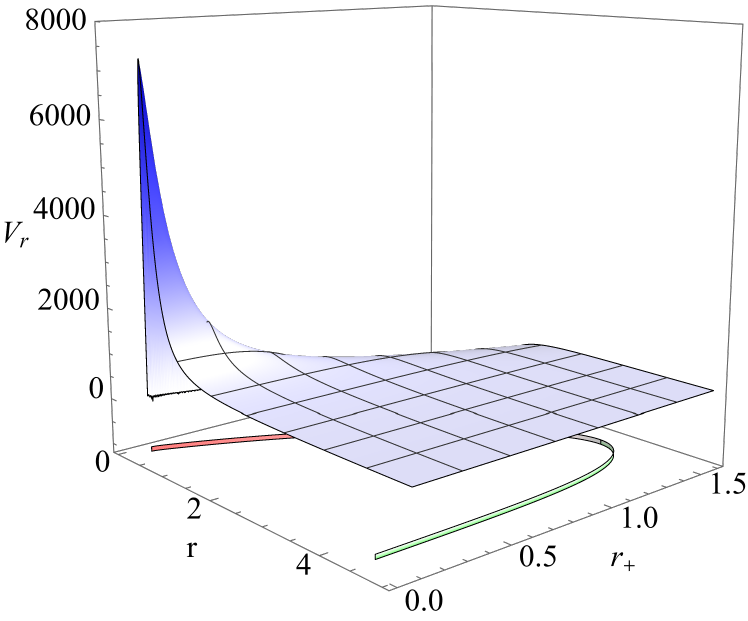}
	\end{center}
	\caption{The diagram of effective potential $V_r$ with the condition $q=0.00205$ in 6D charged Gauss-Bonnet AdS black hole. Analogous to the four and five dimensions, there are red curve for the unstable equilibria and the green curve for the stable equilibria respectively. The stable equilibrium and the unstable equilibrium will meet each other at the intersection of the red and green curves. }
	\label{effective_potential_6D}
\end{figure}

The chaotic features and the thermodynamics of 6D charged Gauss-Bonnet AdS black bole will be investigated with Lyapunov exponent in this subsection. We can derive the formula of effective potential by combining Eq. \eqref{metric_function_dD}, Eq. \eqref{mass_of_dD} and Eq. \eqref{effective_potential} with the finding that, there are $q$, $\alpha$ and $r_+$ influencing the effective potential $V_r$. To observe the trend of $V_r$, we could obtain the diagram of it versus radius of the particle orbit $r$ and the event horizon $r_+$ with fixed $q=0.00205$ and $L=20$ shown in Fig. \ref{effective_potential_6D}. The red and green curves are projected at $V_r=-1000$ in the diagram, which satisfy conditions $V_r=0$ and $V_r'=0$. One can find the red curve represents the stable while the green curve stands for the unstable equilibrium, and the unstable equilibrium is of importance in the derivation of the Lyapunov exponent. Unsurprisingly, we can also find that the unstable equilibrium point will eventually meet the stable equilibrium point at the intersection of the two curves, and after this intersection, there will be no unstable equilibrium and stable equilibrium exists for the massive particles. 

Through the condition that $V_r=0$ and $V_r'=0$, we can get the critical value $r_c$. Combining Eq. \eqref{expression_lambda}, Eq. \eqref{effective_potential} and Eq. \eqref{energy_and_anglur_momentum_4D}, one can obtain the simplified expression of Lyaponuv exponent calculated at the unstable equilibrium as shown in the $V_r$ diagram. It is a function of charge, Gauss-Bonnet constant, and the horizon, which reads as Eq. \eqref{lambda_timelike_4D_2}. 

\begin{figure}[t]
	\begin{center}
		\subfigure[$\ \tilde{\lambda}-r_+$]{\includegraphics[height=6.5cm]{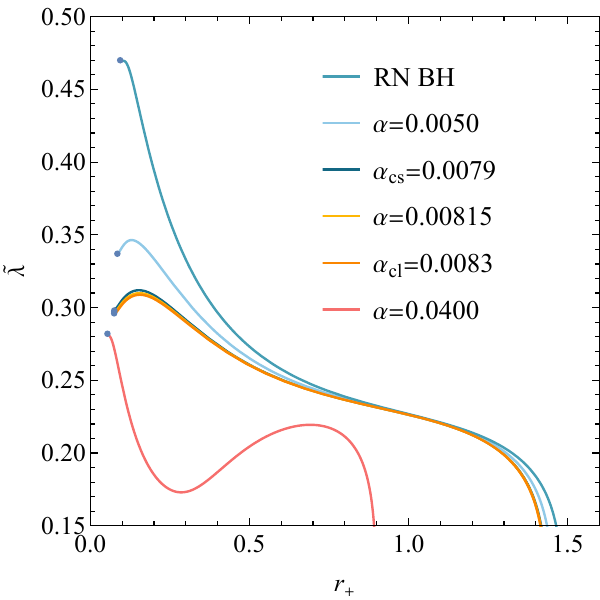}\label{6lr}}
		\subfigure[$\ \tilde{\lambda}(r_+,\alpha)$]{\includegraphics[height=6.5cm]{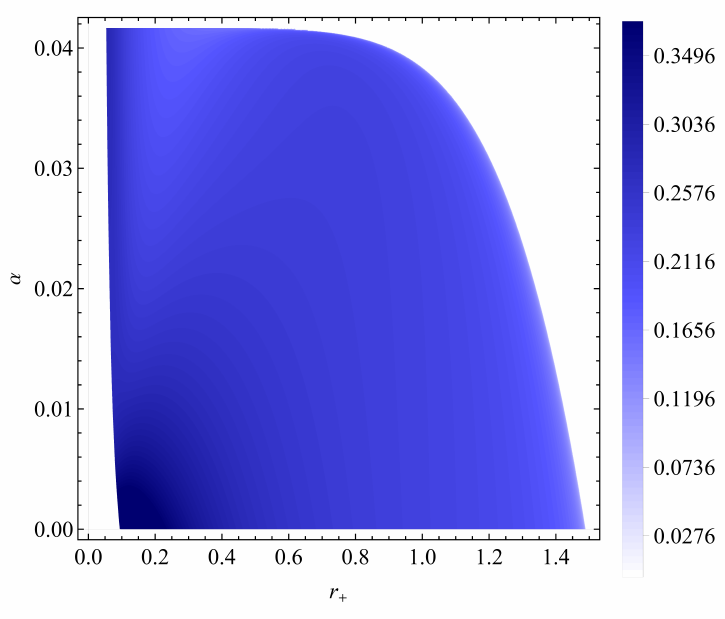}\label{6_lr_c}}
	\end{center}
	\caption{The $\tilde{\lambda}-r_+$ diagram and the contour map for the fixed $q=0.00205$ in 6D spacetime. From the diagrams, one can uncover that there are two branches, the branch with lower $\alpha$ will be a first rising followed by a decreasing function, while the other branch is function that first goes down, then up, then down again to zero. The Lapunov exponent will not be zero overall when $\alpha$ reaches the upper bound $\alpha_M=0.04166$, which is different from the four and five dimension. }
	\label{Lyapunov_r_6D}
\end{figure}

It is difficult for us to get its analytical expression, therefore, we will numerically calculate the Lyapunov exponent with $\alpha$ and $r_+$ in Fig. \ref{Lyapunov_r_6D}. 
One can find that, for different Gauss-Bonnet constant $\alpha$, the $\lambda-r_+$ curves will start from $r_{+0}$ fitting $T(r_{+0}) = 0$, and the starting value $r_{+0}$ depend on $\alpha$ as shown in Fig. \ref{6lr}, which form a different cutoff compared to 5D case in Fig. \ref{Lyapunov_r_5D}. For $\alpha$ is small, the trend of Lyapunov exponent will first increase a little then decrease steadily to zero. For larger $\alpha$ case, $\lambda$ will first decrease and then increase, followed by a continuous decreasing to zero. The equalization of $\lambda$ to zero corresponds to the coincidence of stable equilibrium and unstable equilibrium as shown in Fig. \ref{effective_potential_6D}.  
From Fig. \ref{6_lr_c}, we can infer that the Lyapunov exponents $\lambda$ for the increasing $\alpha$ is declining with fixed $r_+$, indicating that $\alpha$ will reduce chaos in the timelike orbits. The $\lambda$ will not always be zero when $\alpha$ is in the neighborhood of $\alpha_M$, maintaining the same trend with the large $\alpha$ case,, which is different from 4 and 5D case.   

\begin{figure}[t]
		\begin{center}
		\subfigure[$\ \alpha<\alpha_{cs}$]{\includegraphics[height=5.1cm]{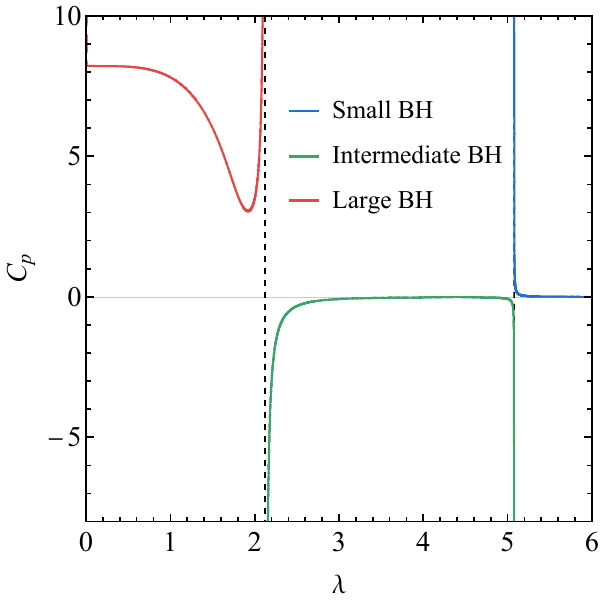}\label{6hc1}}
		\subfigure[$\ \alpha_{cs}<\alpha<\alpha_{cl}$]{\includegraphics[height=5cm]{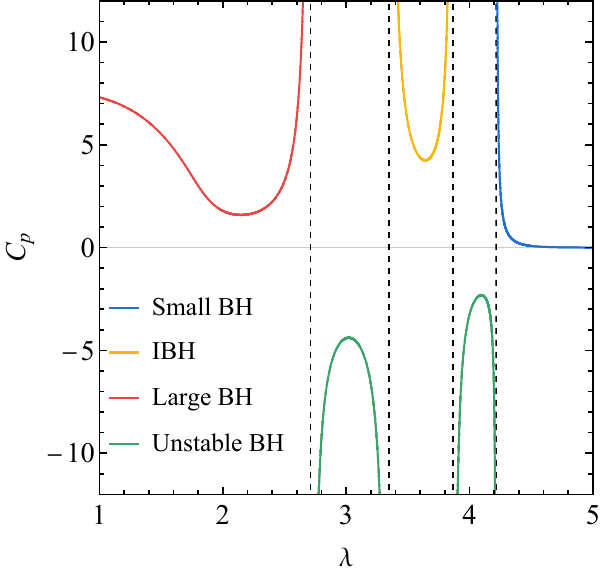}\label{6hc2}}
		\subfigure[$\ \alpha>\alpha_{cl}$]{\includegraphics[height=5cm]{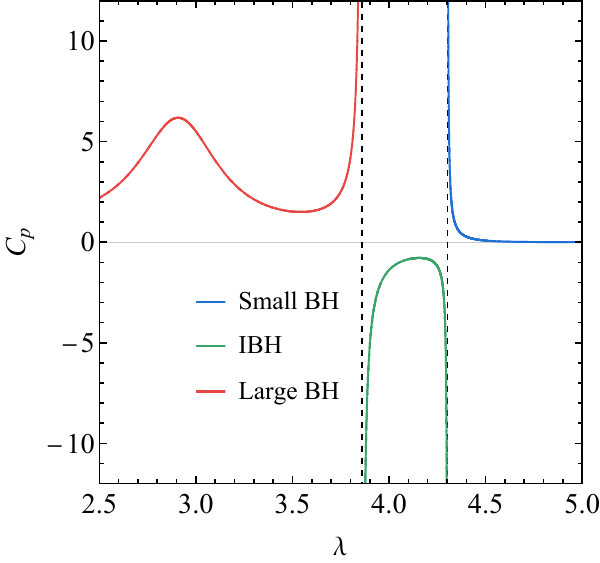}\label{6hc3}}
		\subfigure[$\ \alpha<\alpha_{cs}$]{\includegraphics[height=5cm]{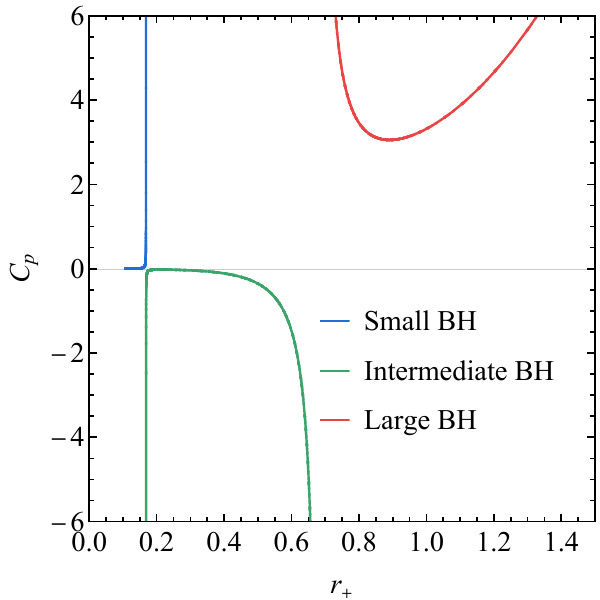}\label{6cr1}}
		\subfigure[$\ \alpha_{cs}<\alpha<\alpha_{cl}$]{\includegraphics[height=5cm]{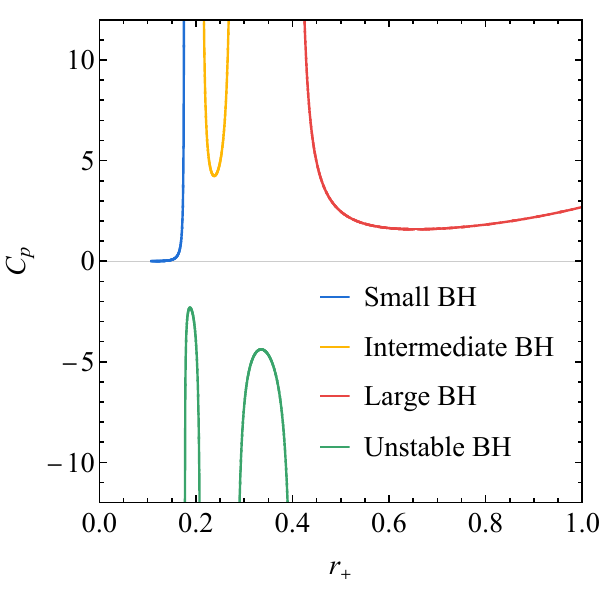}\label{6cr2}}
		\subfigure[$\ \alpha>\alpha_{cl}$]{\includegraphics[height=5cm]{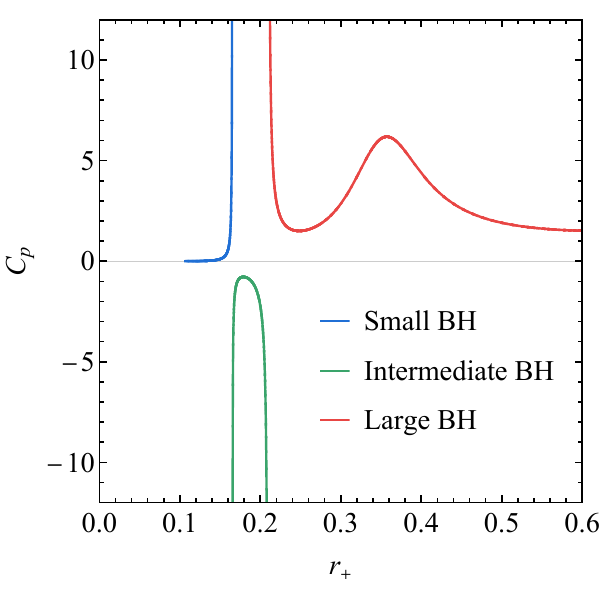}\label{6cr3}}
	\end{center}
	\caption{The $C_p-\lambda$ and $C_p-r_+$ diagrams in 6D spacetime when $q=0.00205$. For $\alpha=0.003<\alpha_{cs}$ and $\alpha=0.084>\alpha_{cl}$, the Small/Large phase transition case, the Large BH and Small BH with $C_p>0$ are stable but the Intermediate BH is unstable with $C_p<0$. However, when $\alpha\in (\alpha_{cs},\alpha_{cl})$ as shown in the middle column, three of the phases (i.e., Large Small and Intermediate BH) are stable, while the green colored phase is unstable with negative $C_p$. }
	\label{6lc}
\end{figure}

To observe the relationship between the Lyapunov exponent and black hole stability, one can obtain diagrams of isobaric heat capacity with the formula Eq. \eqref{ihc4} in Fig. \ref{6lc}. The the red, green and blue curves in Figs. \ref{6hc1}, \ref{6hc3}, \ref{6cr1} and \ref{6cr3} correspond to Large, Intermediate and Small BH, respectively. While the red, yellow and blue curves in Fig. \ref{6hc2} and \ref{6cr2} represent the Large BH, Intermediate BH and Small BH respectively, and the extra green curves correspond to two unstable black holes with negative $C_p$. 

It is shown that the $C_p-\lambda$ diagram corresponds well to the $C_p-r_+$ diagram, where $C_p<0$ corresponds to an unstable black hole and, conversely, a positive heat capacity $C_p>0$ corresponds to a stable black hole phase. For $\alpha\in (0,\alpha_{cs})$ and $\alpha\in (\alpha_{cl}, \alpha_M)$ presented in Figs. \ref{6hc1} and \ref{6hc3} respectively, the Small BH and Large BH are stable, while Intermediate BH are unstable. As $\alpha$ belongs to ($\alpha_{cs},\alpha_{cl}$), there are stable Small, Intermediate and Large BH with $C_p>0$ and specially two unstable phases of black holes, which has negative heat capacity $C_p<0$ shown in Figs. \ref{6hc2} and \ref{6cr2}. %
 
\begin{figure}[t]
	\begin{center}
		\subfigure[$\ \alpha<\alpha_{cs}$]{\includegraphics[height=5cm]{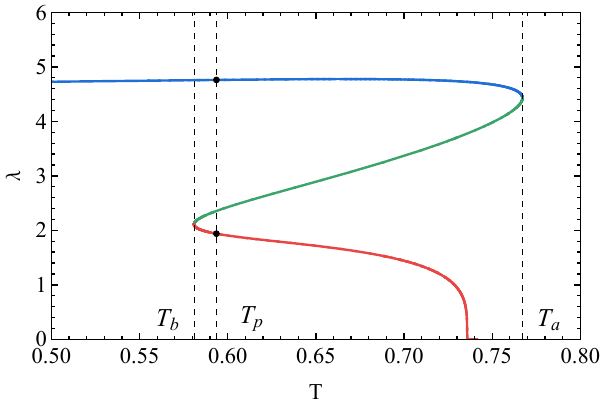}\label{6_lt_1}}
		\subfigure[$\ \alpha<\alpha<\alpha_T$]{\includegraphics[height=5cm]{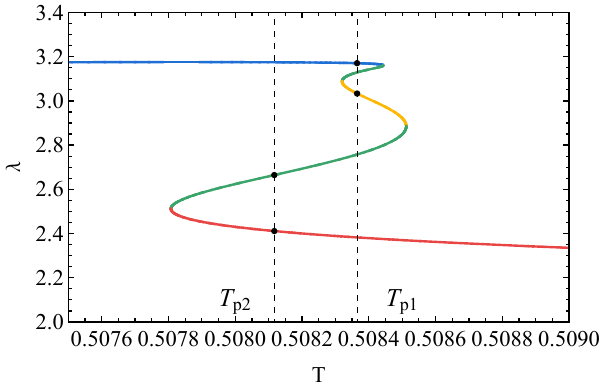}\label{6_lt_2}}
		\subfigure[$\ \alpha_{T}<\alpha<\alpha_{cl}$]{\includegraphics[height=5.1cm]{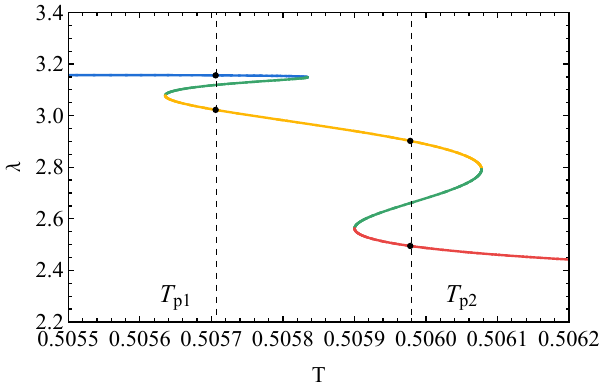}\label{6_lt_4}}
		\subfigure[$\ \alpha>\alpha_{cl}$]{\includegraphics[height=5.1cm]{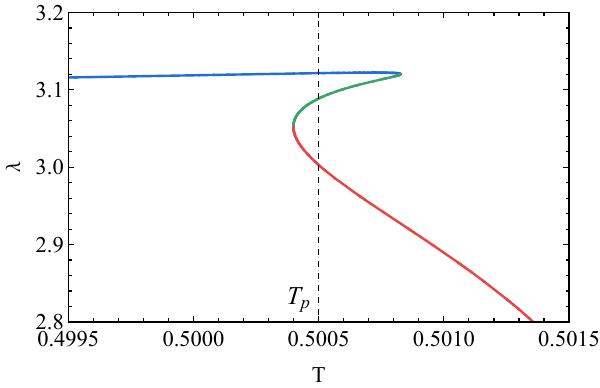}\label{6_lt_5}}
	\end{center}
	\caption{The diagram of Lyapunov exponent and Hawking temperature for different $\alpha$ when $q=0.00205$ in 6D spacetime. For $\alpha=0.003$ case, there are three phases (Small, Intermediate and Large BH) forming a multivalued curve, and the $T_{p}$ is temperature of the Small/Large phase transition point. For $\alpha_{cs}<\alpha<\alpha_{T}$ and $\alpha_{T}<\alpha<\alpha_{cl}$, there are two Small/Large phase transitions and the phase transition temperature $T_{p1}$ and $T_{p2}$ are not coincident. For the case $\alpha=0.0084>\alpha_{cl}$, there's a Small/Large phase transition with the leaping of $\lambda$ at $T_p$.  
	}
	\label{L_T_6D}
\end{figure}

Then we are going to probe the thermodynamic with taking into account both Lyapunov exponents and the Hawking temperature as in the $F-T$ diagrams. By applying the event horizon as the parameter, one can obtain the $\lambda-T$ diagrams for different $\alpha$ in Fig. \ref{L_T_6D}. For $\alpha=0.003 <\alpha _{\text{cs}}$ in Fig. \ref{6_lt_1} 
, we could find that the Lyapunov exponent changes a little with $T$ growing in the interval (0, $T_a$), where it is in the case of stable Small BH with positive $C_p$ as presented in Fig. \ref{6hc1}. As the $\lambda-T$ curve reached the dashed line $T_a=0.76717$, which is an extreme point for the $T-r_+$ diagram, the phase will evolve from Small BH to the unstable Intermediate BH with negative $C_p$. Then the Lyapunov exponent will decrease to $T_b=0.58095$ where the Intermediate BH transforms to stable Large BH, with the increasing of horizon radius $r_+$. During the interval ($T_b$, $T_a$), there exists a Small/Large phase transition occurring at $T_p=0.5936$, accompanied by a discontinuous variation of the Lyapunov exponent. The Lyapunov exponents will finally plunge to zero as it can be predicted by the $\lambda-r_+$ diagram Fig. \ref{effective_potential_6D}. 

For $\alpha_{cs}<\alpha=0.0081<\alpha_{T}$ and $\alpha_T<\alpha=0.0082<\alpha_{cl}$ corresponding to Figs. \ref{6_lt_2} and \ref{6_lt_4}, there are three stable BH with positive $C_p$ shown in Fig. \ref{6hc2}, i.e., the Large, Small and Intermediate BH, and also two unstable black holes marked by green with negative $C_p$. The phase transition between the stable and unstable phases occurs when $T$ takes an extreme value $\frac {\partial T} {\partial r_ +} / \frac {\partial \lambda} {\partial  r_ +} = 0$ in the $\lambda-T$ diagram. More specifically, when $\alpha_{cs}<\alpha<\alpha_T$, there are two Small/Large phase transition, where $T_{p1}$ corresponds to the case with $\lambda$ jumps from Small BH to Intermediate BH and $T_{p2}$ accompanies with $\lambda$ jumps from Small BH to Large BH as shown in Fig. \ref{6_lt_2}. For $\alpha_T<\alpha<\alpha_{cl}$, the Hawking temperature $T_{p1}$ is accompanied with the leaping of $\lambda$ from Small BH to Intermediate BH, while $T_{p2}$ coexists with the discontinuous variation from Intermediate BH to the Large BH, which is different from the $\alpha_{cs}<\alpha<\alpha_T$ case, as presented in Fig. \ref{6_lt_4}. It's worth noting that, for $\alpha<\alpha_{T}$ case, the Hawking temperature of phase transitions fitting $T_{p1}>T_{p2}$, while for $\alpha>\alpha_{T}$ case, the opposite is true. For $\alpha>\alpha_{cl}$ in Fig. \ref{6_lt_5}, the situation is similar to the case $\alpha<\alpha_{cs}$, with the experience of Small/Large phase transition and the variation of $\lambda$ from Small BH to Large BH. 

\begin{figure}
	\centering
	\includegraphics[height=5cm]{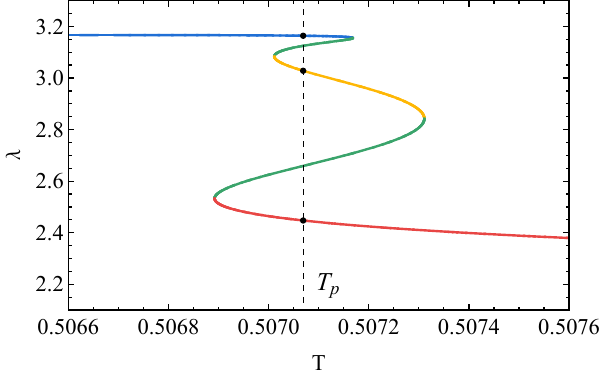}
	\caption{The diagram of Lyapunov exponent versus Hawking temperature for $\alpha=\alpha_T$ when $q=0.00205$ in 6D spacetime. It is a triple point as $T$ approaches $T_p=0.507069$ with the three stable phases coinciding, which is accompanied by a discontinuous leap of $r_+$ and $\lambda$. 
	}
	\label{6_lt_T}
\end{figure}

As $\alpha$ is approaching $\alpha_T$, the two Small/Large phase transitions of the black hole will coincide as demonstrated in Fig. \ref{6_lt_T}. As the Hawking temperature approaches $T_p=0.507069$, there are two Van de Waals-like phase transitions with the coincidence of the two corresponding temperatures $T_{p1}$ and $T_{p2}$. It is so called as the triple point with the coexistence of the Small BH, Intermediate BH and Large BH at the same temperature, as a good correspondence to the two swallow tails in Fig. \ref{6_ft_3}. The Small, Intermediate and Large BH are all stable phases separated by two unstable BHs as shown in Figs. \ref{6hc2} and \ref{6cr2}. Combining Figs. \ref{6_lt_2} and \ref{6_lt_4}, one can find that, with the growing of $\alpha$ in the interval $(\alpha_{cs},\alpha_{cl})$, the relationship between two phase transition temperature will change from $T_{p1}>T_{p2}$ to $T_{p1}<T_{p2}$. There's two discontinuous variations of Lyapunov exponent at $T_p=T_{p1}=T_{p2}$ from the Small BH to Large BH, going through the Intermediate BH. For $\alpha \in (0,\alpha_M)$, the Lyapunov exponent $\lambda$ will always drop to zero when $T$ continue to grow as Fig. \ref{Lyapunov_r_6D} shows. 

\subsection{Lyapunov exponents and thermodynamics for null geodesic}
\label{null_geodesic}

In this section, we will discuss thermodynamics of Gauss-Bonnet AdS black hole with the Lyapunov exponent for the null geodesic. Due to the zero proper time in null geodesic, we set $\gamma$ to be the affine parameter, 
and the Lagrangian can be obtained as 
\begin{align}
	\label{l_null_4D}
	2 \mathcal{L}= f(r)\dot{t}^2 -\frac{\dot{r}^2}{f(r)}+r^2 \dot{\varphi }^2,
\end{align}
in 4D, where dot is the derivative with respect to affine parameter $\gamma$. Similar to the timelike geodesic, the Lagrangian of five dimension null geodesic is 
\begin{align}
	\label{l_null_5D}
	2 \mathcal{L}= f(r)\dot{t}^2-\frac{\dot{r}^2}{f (r)}+r^2 \dot{\varphi }^2+ r^2\dot{\theta _1}^2 +r^2 \dot{\theta _2}^2 cos^2 \varphi \ sin^2\theta _1 ,
\end{align}
and we also have the Lagrangian for 6D gravity expressed as 
\begin{align}
	\label{l_null_6D}
	2 \mathcal{L}= f(r)\dot{t}^2-\frac{\dot{r}^2}{f(r)}+r^2\dot{\varphi }^2+r^2\dot{\theta _2}^2 +r^2\dot{\theta _1}^2+ r^2 \dot{\theta _3}^2 \cos ^2\theta _1 \sin ^2\theta _2  ,
\end{align}
where the coordinates is consistent with the timelike geodesic. Since the Lagrangian is the same for any dimension with the restriction that the null geodesic is in the equatorial plane, we can derive the Hamiltonian of null geodesic with 
\begin{align}
	\label{h_null}
	2 \mathcal{H}=-\frac{\dot{r}^2}{f (r)}-L \dot{\varphi }+E \dot{t}=0,
\end{align}
where $E$ is the energy of photons, and $L$ is the angular momentum. Then, we can get the effective potential for null geodesic according to Eq. \eqref{def_effictive _potential}, 
\begin{align}
	\label{effective_potential_null}
	V_r=f(r) \left[\frac{L}{r^2}-\frac{E}{f (r)}\right].
\end{align}

Satisfying the circular condition $V_r=0$ and $V_r'=0$, 
we can derive the formula of Lyapunov exponent for photons from the definition Eq. \eqref{expression_lambda},
\begin{align}
	\label{lambda_null}
	\lambda =\sqrt{-\frac{ r_c^2 f (r_c)  }{2L^2}V''(r_c)},
\end{align}
where $r_c$ is determined by the conditions $V_r(r_c)=0$ and $V_r'(r_c)=0$. 

\subsubsection{4D null geodesic with Lyapunov exponents}

In this subsection, we will investigate the Lyapunov exponent and thermodynamics of the charged Gauss-Bonnet AdS black holes from the perspective of photon orbits, with the purpose of revealing of the connection between the thermodynamic and the chaotic properties of the black hole. 

\begin{figure}[t]
	\begin{center}
		\subfigure[$\ \tilde{\lambda}-r_+$]{\includegraphics[height=6.5cm]{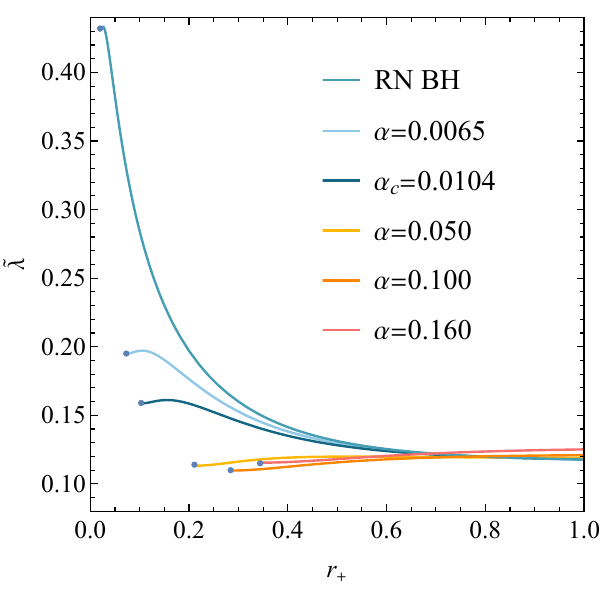}\label{p_4_lr_1}}
		\subfigure[$\ \tilde{\lambda}(r_+,\alpha)  $]{\includegraphics[height=6.5cm]{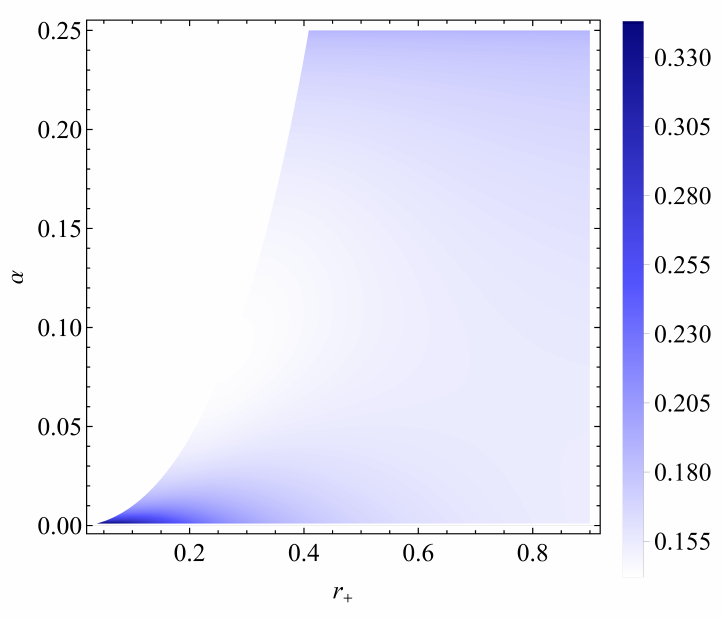}\label{p_4_lr_2}}
	\end{center}
	\caption{The $\log_{100}(\lambda+1)-r_+$ diagram and the contour map in $\alpha-r_+$ space for massless particle with fixed $q=0.02$ in 4D spacetime. The value of $\alpha$ divides the trend of the $\lambda-r_+$ curve into two branches. For $\alpha$ is small, the Lyapunov exponent rises, then falls, and finally tends to a constant. When $\alpha$ is on the other branch, $\lambda$ rises as $r_+$ increases and finally stabilises at a constant. The effect of rising $\alpha$ on the Lyapunov exponents is changing from the pattern that decreasing monotonically, to the pattern that first decrease then increase, finally to the pattern that monotonically increasing. }
	\label{Lyapunov_r_null_4D}
\end{figure}

To began with, we will first investigate the relationship between Lyapunov exponent $\lambda$ and horizon radius $r_+$ with various $\alpha$. 
For the absence of valid analytical solution of Eq. \eqref{lambda_null}, which is computed at the unstable equilibrium, we consider numerical calculation by obtaining the diagram of Lyapunov exponent as a function of horizon $r_+$ presented in Fig. \ref{Lyapunov_r_null_4D} with fixed $q=0.02$ and $L=20$. 

The Lyapunov exponent will first increase a little then gradually decrease with the increase of $r_+$ and finally be equal to a constant, for $\alpha$ is small in Fig. \ref{Lyapunov_r_null_4D}. The reason $\lambda$ remains a constant rather than zero is that, there is always an unstable equilibrium exists regardless of the Gauss-Bonnet constant $\alpha$, which can be derived from the effective potential in Eq. \eqref{effective_potential_null}. Its constant presence makes it different from the timelike geodesic case. For larger $\alpha$, the Lyapunov exponent will monotonically increase and tend to a constant with the rising of horizon $r_+$. During the increase of $r_+$ in Fig. \ref{p_4_lr_2}, the effect of $\alpha$ on the chaotic feature of photon orbits will change from a continuous decrease, to a decrease followed by an increase, and then to a continuous increase, which is different from the massive orbit with a monotonically decreasing effect. 

\begin{figure}[t]
	\begin{center}
		\subfigure[$\ \alpha<\alpha_c$]{\includegraphics[height=4.5cm]{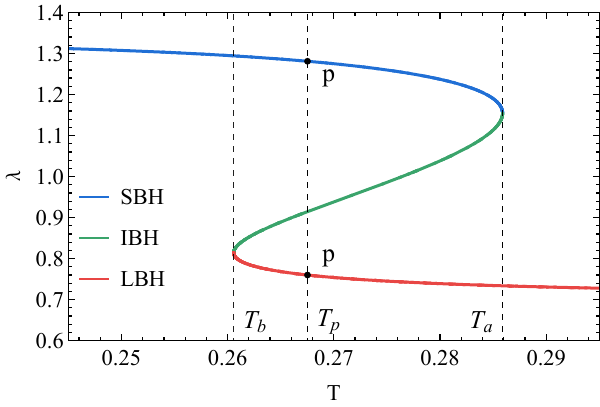}\label{p4lt1}}
		\subfigure[$\ \alpha>\alpha_c$]{\includegraphics[height=4.4cm]{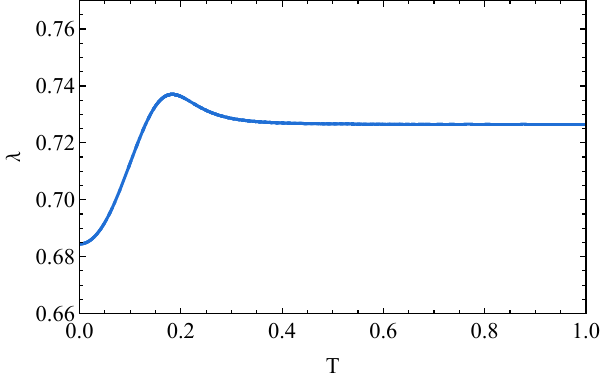}\label{p4lt4}}
	\end{center}
	\caption{The $\lambda-T$ diagram of different $\alpha$ for the fixed $q=0.02$ in 4D spacetime. For $\alpha=0.0065<\alpha_c$, the black hole will go through three stages, i.e., Small BH, Intermediate BH and the Large BH, whose $\lambda-T$ curve forms a multi-valued function indicating a Small/Large phase transition. For $\alpha=0.05>\alpha_c$, the Lyapunov exponent will rises first and falls to a constant with the increasing $T$, which doesn't have phase transition and there's only one phase in the process. }
	\label{Lyapunov_t_null_4D}
\end{figure}

Then we will probe the relationship between Lyapunov exponents $\lambda$ and Hawking temperature $T$ like we did in 4D timelike geodesic. 
By taking the horizon radius $r_+$ as the parameter, one can obtain the $\lambda-T$ diagram in Fig. \ref{Lyapunov_t_null_4D} for different $\alpha$.  

For $\alpha=0.0065<\alpha_c$ that is presented in Fig. \ref{p4lt1}, it is nearly the same as the timelike geodesic of four-dimensional charged Gauss-Bonnet AdS black holes, apart from the behavior when $T$ is large. Note that the dashed lines of Hawking temperature $T_a,T_b \ \text{and} \  T_p$ correspond to the temperature of  $a, b$ and $p$. When in the phase of Small BH, the Lyapunov exponent decreases a little in the interval ($0, T_a$). As the $\lambda-T$ curve reached the dashed line $T_a=0.285884$, which is an extreme point for Hawking temperature, the black holes will evolve from Small BH to the unstable Intermediate BH. For $T$ reaching $T_b=0.260578$, it will transform from unstable Intermediate BH to stable Large BH. There will be a Small/Large phase transition when $T=T_p=0.26752$ with the leaping of $\lambda$. The Lyapunov exponents will finally tend to a non-zero constant as predicted in the $\lambda-r_+$ diagram Fig. \ref{Lyapunov_r_null_4D}, which is different from the behavior of timelike case. For $\alpha=0.05>\alpha_c$, the Lyapunov exponent $\lambda$ will increase first followed by a decrease to a positive constant with the increasing of temperature $T$ as shown in Fig. \ref{p4lt4}, which implies that, there's only one black hole phase and no Small/Large phase transition will exist. 

\subsubsection{5D null geodesic with Lyapunov exponents}

\begin{figure}[t]
	\begin{center}
		\subfigure[$\ \tilde{\lambda}-r_+$]{\includegraphics[height=6.5cm]{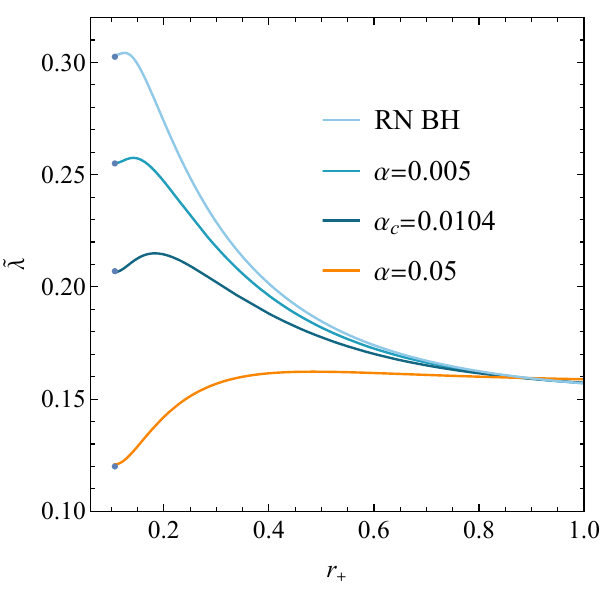}\label{p5lr}}
		\subfigure[$\ \tilde{\lambda}(r_+,\alpha)$]{\includegraphics[height=6.5cm]{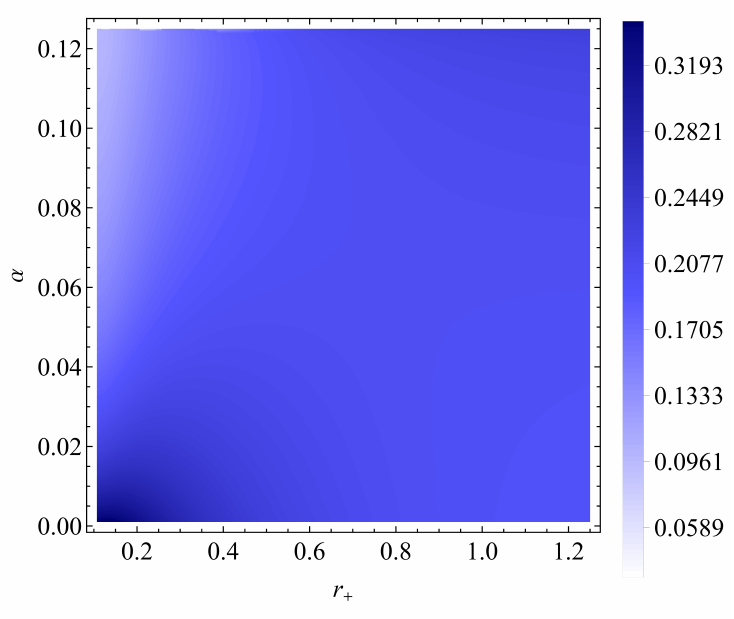}\label{p5clr}}
	\end{center}
	\caption{The $\log_{100}(\lambda+1)-r_+$ diagram and the contour map in $\alpha-r_+$ space with the fixed $q=0.02$ for the massless particles in 5D spacetime. The diagram indicate that when $\alpha$ is small, the Lyapunov exponent will increase first then continuously decrease, finally tending to a constant. When $\alpha$ is large, the Lyapunov exponent will rise and then approach a constant with the growing of horizon. For $r_+$ is fixed to be small, the effect of $\alpha$ makes the value of $\lambda$ smaller, and vice versa, which reflects the different effect of $\alpha$ on the properties of chaos for different horizon. }
	\label{Lyapunov_r_null_5D}
\end{figure}

The relationship between thermodynamic and chaotic properties characterized by the Lyapunov exponents is of great interest for us to investigate in the 5D null geodesic case. With the expression of $\lambda$  derived in Eq.  \eqref{lambda_null}, it is clear to see that the Lyapunov exponent is depending on the black hole charge $q$ and horizon $r_+$ for various Gauss-Bonnet constant $\alpha$. Then numerical diagrams of the Lyapunov exponent $\lambda$ versus horizon $r_+$ are presented in Fig. \ref{Lyapunov_r_null_5D} with the fixed $q=0.02$ and $L=20$. 

Due to the requirement $T(\alpha,r_+) > 0$, the $\lambda-T$ curves will have the same cutoff of $r_+$, as predicted by Eq. \eqref{zpr_+}.  
Similar to four-dimension null geodesic, $\lambda$ will rise a little bit and after an extreme value, it gradually decreases with the increase of $r_+$ and tend to a constant as shown in Fig. \ref{p5lr}. The continuous constant is from the fact that the chaos will always exists during the increase of event horizon, which is different from the timelike geodesic. Notice that when $\alpha$ approaches the upper bound, the trend of $\lambda$ will continuously rise and then tend to a constant as shown in Fig. \ref{p5clr}. With the rising event horizon $r_+$, the trend of the Lyapunov exponent, as a function of the Gauss-Bonnet constant, will changes from decreases to continuing increase. It reflect the influence of $\alpha$ to the chaos of photon orbits and is different from the timelike geodesic, as it is a monotonically decreasing effect.  

\begin{figure}[t]
	\begin{center}
		\subfigure[$\ \alpha<\alpha_c$]{\includegraphics[height=4.5cm]{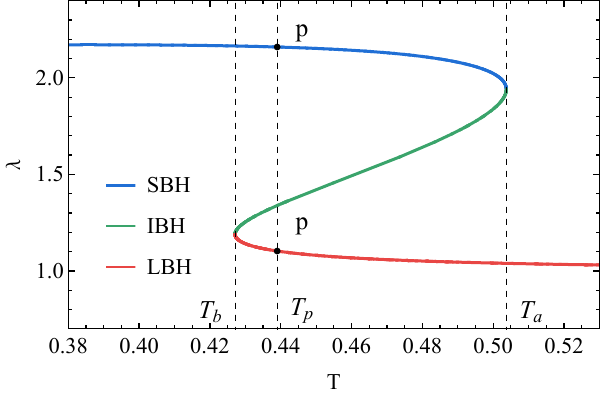}\label{p5lt1}}
		\subfigure[$\ \alpha>\alpha_c$]{\includegraphics[height=4.5cm]{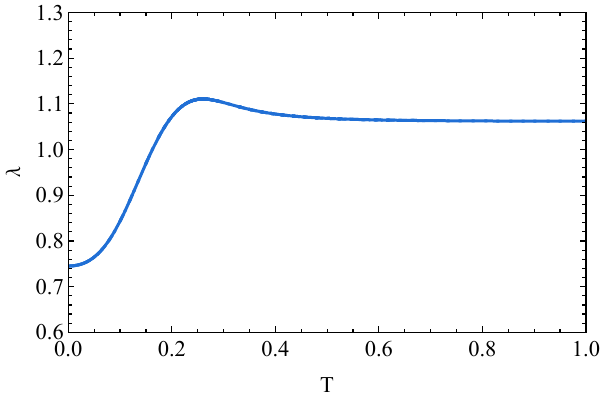}\label{p5lt2}}
	\end{center}
	\caption{The $\lambda-T$ diagram with the fixed $q=0.02$ for the massless particles in 5D spacetime. For $\alpha=0.006<\alpha_c$, the multivalued properties of the $\lambda-T$ curve characterize the Small/Large phase transition process in a 5D charged Gauss-Bonnet AdS black hole, where the temperature $T_p=0.43912$ represents its phase transition temperature. For $\alpha>\alpha_c$, there is a single-valued curve on the graph corresponding to the singular value curve of the $F-T$ diagram, which means no phase transition will occur. }
	\label{Lyapunov_t_null_5D}
\end{figure}

In order to probe the relationship of the black hole chaos and the Hawking temperature, we obtained the $\lambda-T$ diagram in Fig. \ref{Lyapunov_t_null_5D}. As $q=0.02$ and $L=20$ are fixed, we could find the critical value of Gauss-Bonnet constant $\alpha_c$ as calculated in Table \ref{critical_value_5D}. For $\alpha=0.006<\alpha_c$ presented in Fig. \ref{p5lt1}, the dashed lines of Hawking temperature $T_a, T_b\ \text{and } \ T_p$ corresponding to the temperatures of $a, b$ and $p$ in the $F-T$ diagram. We could find that as $r_+$ gradually increases, the Lyapunov exponent slowly drops to the first extreme point ($\lambda_a,T_a$), where the black hole evolves from stable Small BH to the unstable Large BH. The approaching of Hawking temperature to $T_b=0.42712$ is accompanied with the willing of evolution from unstable Intermediate BH to the stable Large BH. Hovering at $T_p=0.43912$, there exists a Small/Large phase transition with discontinuous leaping over $\lambda$ from Large BH to Small BH. For $\alpha=0.05>\alpha_c$ presented in Fig. \ref{p5lt2}, the Lyapunov exponents $\lambda$ will increase followed by continuing decrease to a constant with the increase of $r_+$, indicating that there's only one phase in the process and no Small/Large phase transition exists. The behavior of Lyapunov exponents when the Hawking temperature $T$ increases to the positive infinity is tending to a constant rather than zero in the 5D timelike geodesic.  

\subsubsection{6D null geodesic with Lyapunov exponents}

\begin{figure}[t]
	\begin{center}
		\subfigure[$\ \tilde{\lambda}-r_+$]{\includegraphics[height=6.5cm]{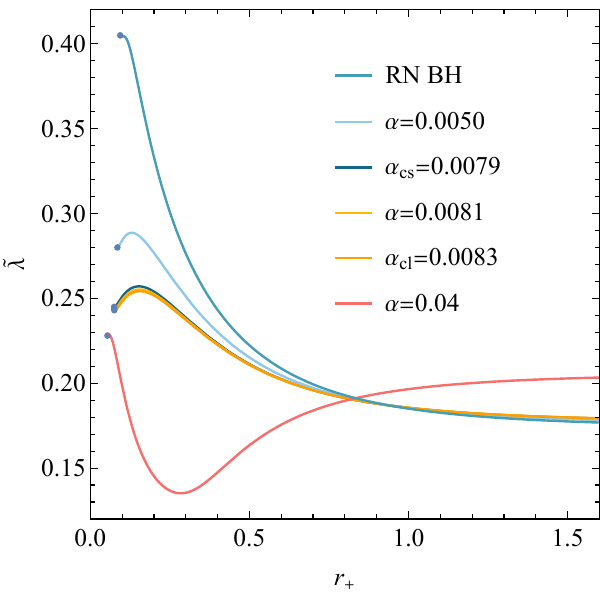}\label{p6lr1}}
		\subfigure[$\ \tilde{\lambda}(r_+,\alpha)$]{\includegraphics[height=6.5cm]{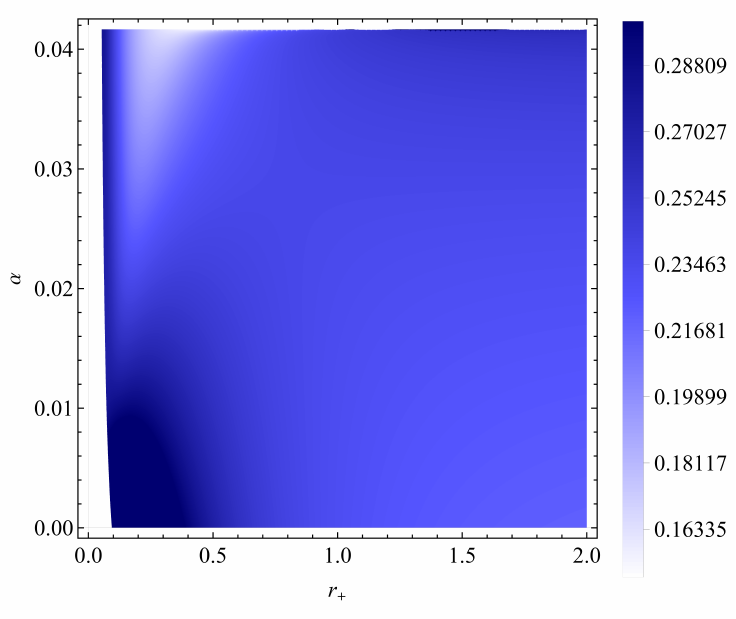}\label{p6lr2}}
	\end{center}
	\caption{The $\log_{100}(\lambda+1)-r_+$ diagram and the contour map in the $\alpha-r_+$ space when $q=0.00205$ for massless particles in 6D. The behavior of Lyapunov exponent can be divided into two branch by the value of $\alpha$. For $\alpha$ is small, we can see that the Lyapunov exponent will first rise and then decrease to a constant. When $\alpha$ is on the other branch, the Lyapunov exponent will first rise then continue to increase and tend to a constant eventually. It's worth noting that the event horizon will also make a difference. As $r_+$ is small, the Lyapunov exponent will decrease with the growth of $\alpha$, and when $r_+$ is large, it is in an opposite situation. }
	\label{Lyapunov_r_null_6D}
\end{figure}

In this subsection, we will investigate the thermodynamic properties of the black hole and the chaos for the null geodesics with Lyapunov exponent. The formula of Lyaponov exponents is obtained in Eq. \eqref{lambda_null}, therefore, the diagrams of $\lambda$ versus horizon $r_+$ for different $\alpha$ are exhibited in Fig. \ref{Lyapunov_r_null_6D}, with fixed $q=0.00205$ and $L=20$. The positive temperature requirement causes the cutoff points on the left in Fig. \ref{p6lr1}, meanwhile forming a cutoff line as presented in Fig. \ref{p6lr2}. As $\alpha$ grows in the interval ($0, \alpha_M$), the trend of $\lambda$ with respect to $r_+$ will change from the pattern that rise and then falls to a constant, to the pattern that falls and then rise and finally tend to a constant. Moreover, the influence of $\alpha$  reduces of the chaos when $r_+$ is small, but conversely, an increase in $\alpha$ makes the chaotic property of black holes increase. 

\begin{figure}[h]
	\begin{center}
		\subfigure[$\ \alpha<\alpha_{cs}$]{\includegraphics[height=5cm]{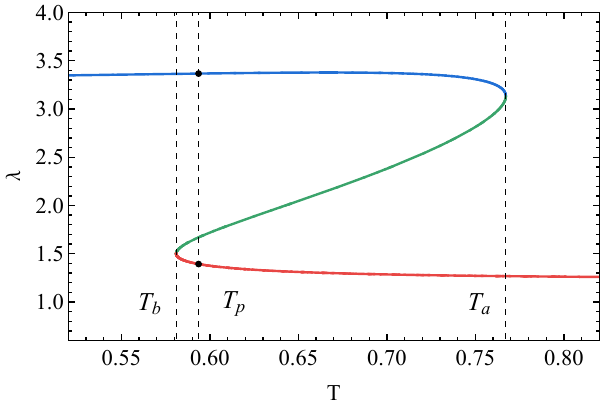}\label{p6lt1}}
		\subfigure[$\ \alpha_{cs}<\alpha<\alpha_{T}$]{\includegraphics[height=5cm]{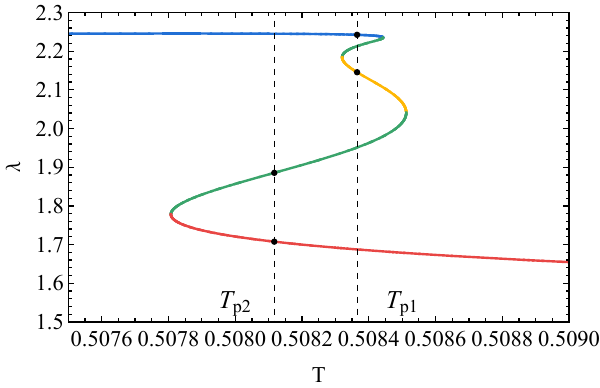}\label{p6lt2}}
		\subfigure[$\ \alpha_T<\alpha<\alpha_{cl}$]{\includegraphics[height=5cm]{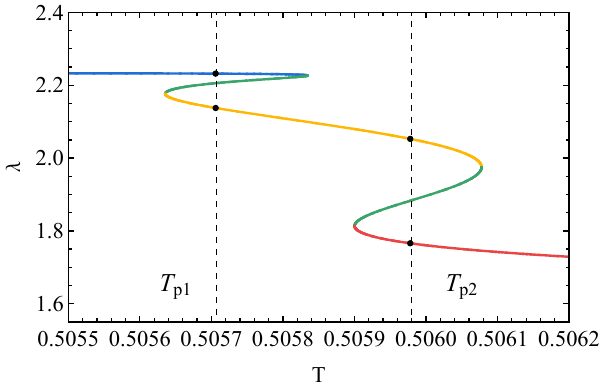}\label{p6lt4}}
		\subfigure[$\ \alpha>\alpha_{cl}$]{\includegraphics[height=5cm]{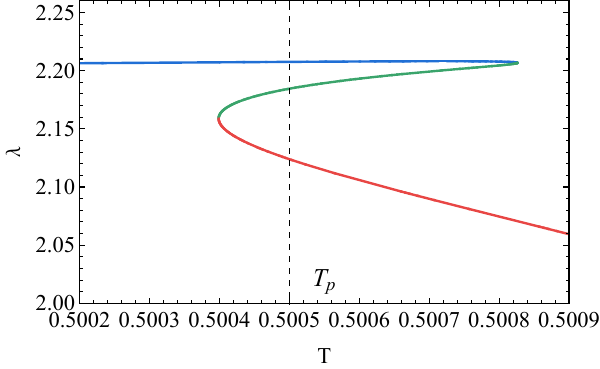}\label{p6lt5}}
	\end{center}
	\caption{The $\lambda-T$ diagram for different $\alpha$ when $q=0.00205$ in 6D spacetime. In the diagrams, the different colors represent different states of black hole as the convention of six dimensional F-T diagram. For $\alpha=0.003<\alpha_{cs}$ and $\alpha=0.0084>\alpha_{cl}$, there are three phases forming a multi-valued curve, which indicate the existence of Small/Large phase transition. For the $\alpha_{cs}<\alpha<\alpha_{cl}$ case, the behaviors of the $\lambda-T$ curve are more complicated, which have two Small/Large phase transitions with the phase transition temperature $T_{p1}$ and $T_{p2}$ marked.    
	}
	\label{Lyapunov_t_null_6D}
\end{figure}

\begin{figure}
	\centering
	\includegraphics[height=5cm]{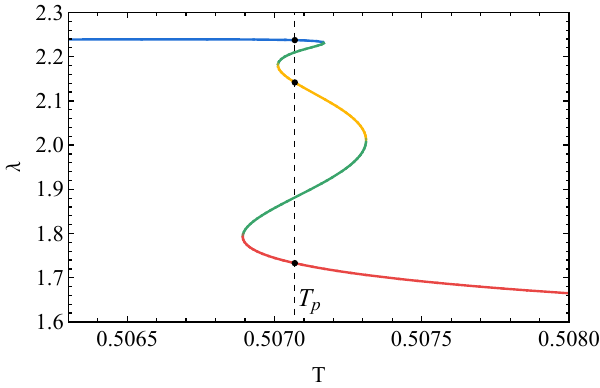}
	\caption{The diagram of Lyapunov exponent and Hawking temperature for $\alpha=\alpha_T$ when $q=0.00205$ in 6D spacetime. There is a triple point as $T$ approaches $T_p$ with the three stable phases and two Small/Large phase transition temperature coinciding, which is accompanied by two discontinuous variations of $r_+$ and $\lambda$. 
	}
	\label{p_6_lt_T}
\end{figure}

Furthermore, we find the relationship between the Lyapunov exponent and the Hawking temperature is valuable to investigate, which connects the chaotic feature and temperature of the black hole. Therefore, we illustrate the $\lambda-T$ diagrams for different $\alpha$ as shown in Fig. \ref{Lyapunov_t_null_6D}. With the fixed $q=0.00205$ and $L=20$, there are two critical values $\alpha_{cs}$ and $\alpha_{cl}$, which can be obtained in Table \ref{critical_value_6D}. There is also a special value $\alpha_T \in (\alpha_{cs}$,$\alpha_{cl}$), corresponding to the triple point.  

As $\alpha=0.003<\alpha_{cs}$ presented in Fig. \ref{p6lt1}, the behavior of $\lambda-T$ curve is multivalued, which is corresponding to the characteristic Small/Large phase transition. 
When $T$ approaches $T_a=0.76717$, it will evolve from a Small BH to the unstable Intermediate BH. If $T=T_b=0.58095$, it will transform from unstable Intermediate BH to stable Large BH. The Small/Large phase transition will occur when $T=T_p=0.5936$, with a discontinuous change of Lyapunov exponents.  
For $\alpha_{cs}<\alpha<\alpha_{cl}$ in Figs. \ref{p6lt2} and \ref{p6lt4}, there are four evolutions among the stable BHs (Small, Intermediate and Large BH) and unstable BHs with continuous variation of horizon and Lyapunov exponents, which will happen as $T$ takes the extrem values $\frac {\partial T} {\partial r_ +} / \frac {\partial \lambda} {\partial  r_ +} = 0$. In the progress, two Small/Large phase transitions are generated, corresponding to when $T$ equals to $T_{p1}$ and $T_{p2}$, which are also accompanied by discontinuous variations of Lyapunov exponent. The discrepancy between when $\alpha<\alpha_T$ and $\alpha>\alpha_T$ can be demonstrated by the quantitative relationship between $T_{p1}$ and $T_{p2}$. When $\alpha<\alpha_T$, the Hawking temperature $T_{p1}>T_{p2}$, while $T_{p1}<T_{p2}$ for $\alpha>\alpha_T$.  
As $\alpha_{cl}<\alpha<\alpha_M$ presented in Fig. \ref{p6lt5}, the behavior is similar to the case when $\alpha<\alpha_{cs}$, with a Small/Large phase transition occurring at $T=T_p$. 

For $\alpha=\alpha_T$ as shown in Fig. \ref{p_6_lt_T}, the behavior of $\lambda-T$ curve is similar to the case $\alpha_{cs}<\alpha<\alpha_{cl}$ in Fig. \ref{6_lt_T}. The property of the triple point is characterized by the coincidence of the two phase transition temperatures $T_{p1}$ and $T_{p2}$, with the two discontinuous variations of $\lambda$ in the $\lambda-T$ diagram, which reflects the conversion of the three stable phases, Small BH, Intermediate BH and Large BH, marked by blue, yellow and red respectively. 

\subsubsection{Lyapunov exponent with observing black hole shadow}
The black hole shadow can be utilized to reveal the thermodynamic properties of black holes. We are expected to investigate the chaotic feature characterized by Lyapunov exponents through the black hole shadow, so that one can probe the possibility of observing the thermodynamic properties of the black hole. In this subsection, we will briefly explore the feasibility of this approach by taking the example of four-dimensional photon orbits, for which our research focuses on the thermodynamic charged Gauss-Bonnet AdS black hole with taking into account the Lyapunov exponents $\lambda$. In particular, the dark area inside the black hole image is called the black hole shadow, One can take $r_{sh}$ as the radius of black hole shadow. Under such spherical symmetry condition from the 4D charged Gauss-Bonnet AdS spacetime, we are prepared to represent the black hole shadow by its radius $r_{sh}$. 

For the effective potential $V_r$ in Eq. \eqref{effective_potential_null} fitting the condition $V_r(r_c)=0, \ V_r'(r_c)=0 \ \text{and} \ V_r''(r_c)<0$, the trajectory it forming a unstable photon sphere. When the radius is smaller than the critical value, photo will fall into the black hole for the observer, and it will escape to infinity for $r>r_c$. 

\begin{figure}[t]
	\begin{center}
		\subfigure[$\ r_{sh}-r_+$]{\includegraphics[height=6.2cm]{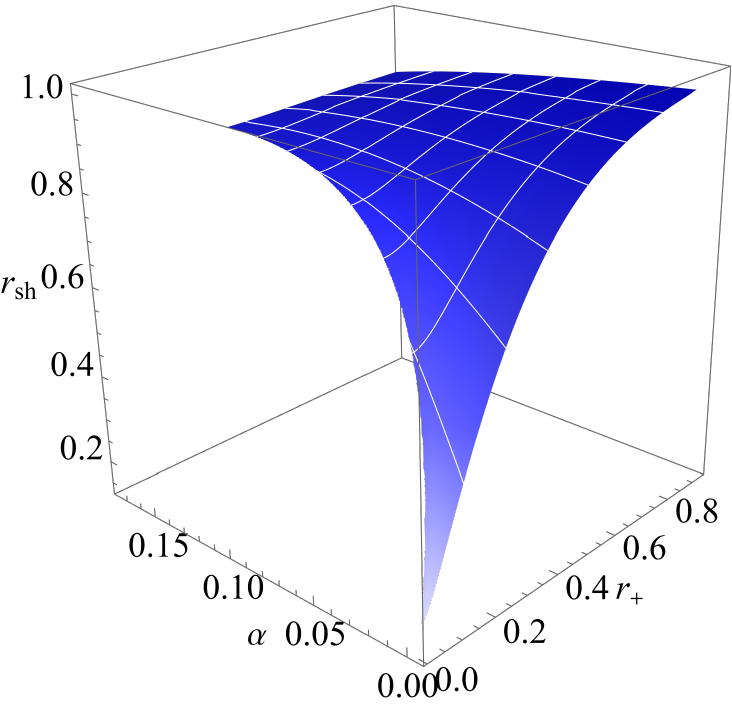}\label{shr4}}
		\subfigure[$\ \lambda-r_{sh} $]{\includegraphics[height=6.2cm]{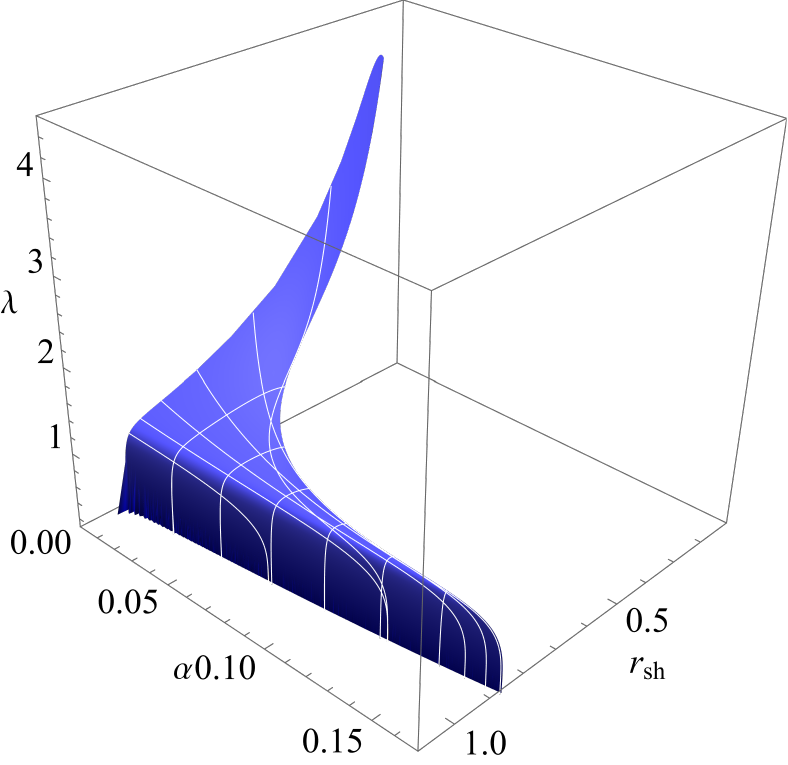}\label{lsh4}}
	\end{center}
	\caption{The diagram of the radius of black hole shadow $r_{sh}$ versus $r_+$ and the diagram of Lyapunov exponent versus radius of the black hole shadow $r_{sh}$ with $q=0.02$ and $T(r_+,\alpha)>0$ in 4D spacetime. (a) The radius $r_{sh}$ will increase with the growing of event horizon $r_+$ for $\alpha \in (0,\alpha_M)$. (b) The Lyapunov exponent is a single-valued function of the black hole shadow radius and it will decrease as the black hole shadow radius increases with the overall trend.}\label{sh}
\end{figure}

As the observer is located at $r_o \gg M$, the radius of the black hole shadow can be given, 
\begin{align}
	\label{radius_shadow}
	r_{sh}=r_o sin\theta,
\end{align} 
where $\theta$ is the inclination angle of the photo at the critical value of the radius, and it can be simplified as,
\begin{align}
	\label{s_radius_shadow}
	r_{sh}=\frac{L}{E},
\end{align}
by substituting the critical condition into Eq. \eqref{effective_potential_null}, one can obtain
\begin{align}
	r_{sh}=\sqrt{\frac{{r_c}^2}{f(r_c)}}.
\end{align}
Due to space constraints, we will use the four dimensional black hole as an example, also because it is more relevant. Since the explicit expression is difficult to derive, we have a $r_{sh}-r_+$ diagram and a diagram representing the variation of the Lyapunov exponent with respect to the shadow of the black hole in Fig. \ref{sh}. 

In Fig. \ref{shr4}, we can see that, with the fixed Gauss-Bonnet constant $\alpha$, the radius of black hole shadow $r_{sh}$ will increase with the rising of event horizon $r_+$. An increase in the Gauss-Bonnet constant will decrease the magnitude of the response of the radius of black hole shadow $r_{sh}$ to the variation in the event horizon $r_+$. The growing $r_+$ will leads to the trend of $r_{sh}$, as a function of $\alpha$, changing from increasing to decreasing. 

We can see a simple and elegant functional relationship between the Lyapunov exponent and the shadow of the black hole presented in Fig. \ref{lsh4}. The Lyapunov exponent will decrease as the radius of the black hole shadow $r_{sh}$ increases. And when the Gauss-Bonnet constant is small, the rate of decrease goes from fast to slow to fast. From Fig. \ref{shr4} that large $\alpha$ causes small magnitude of change of $r_{sh}$, one can find the variations of $\lambda$ and $r_{sh}$ are all confined to small intervals. 

With such a correspondence, a bridge could be built between black hole shadows and black hole thermodynamics through the Lyapunov exponents. We can conceptualize that one could use black hole shadows to obtain chaotic information about the black holes, and the information related to the thermodynamics properties of the black holes. 

\subsection{The critical exponent of Gauss-Bonnet black hole with Lyapunov exponent}
\label{critical_exponent}
The critical exponents determine the qualitative nature of critical phenomenon of a given system, and it is convenient to investigate the phase transition of the charged Gauss-Bonnet AdS black holes with it.
As we have Small/Large phase transition for the proper condition discussed earlier and the Lyapunov exponents as a powerful description, we can use the discontinuous change of Lyapunov exponents $\lambda$ at the first-order phase transition point $p$ to study the critical behavior of black holes \cite{Guo:2022kio,Yang:2023hci}. Without losing generality, we will pay attention to 5D charged Gauss-Bonnet AdS black hole, investigating the critical exponent for Lyapunov exponent with both timelike and null geodesic. 

\begin{figure}[t]
	\begin{center}
		\subfigure[$\ \text{Timelike Geodesic}$]{\includegraphics[height=5cm]{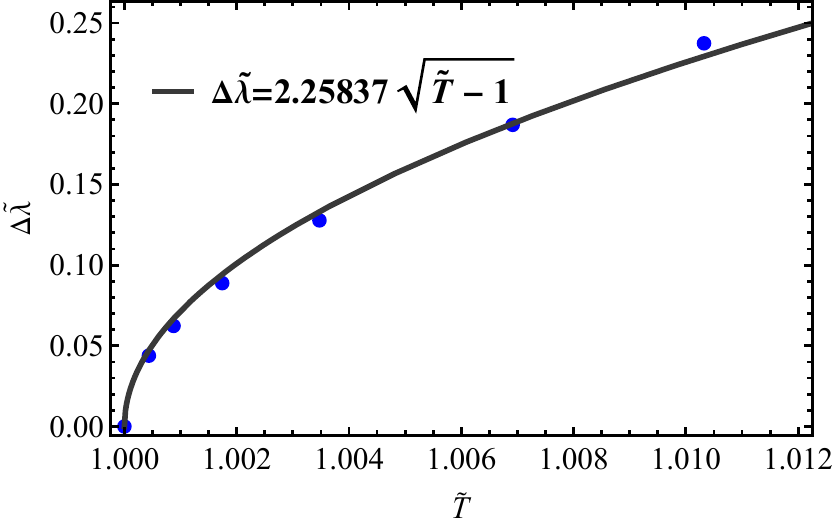}\label{4dl}}
		\subfigure[$\ \text{Null Geodesic}$]{\includegraphics[height=5cm]{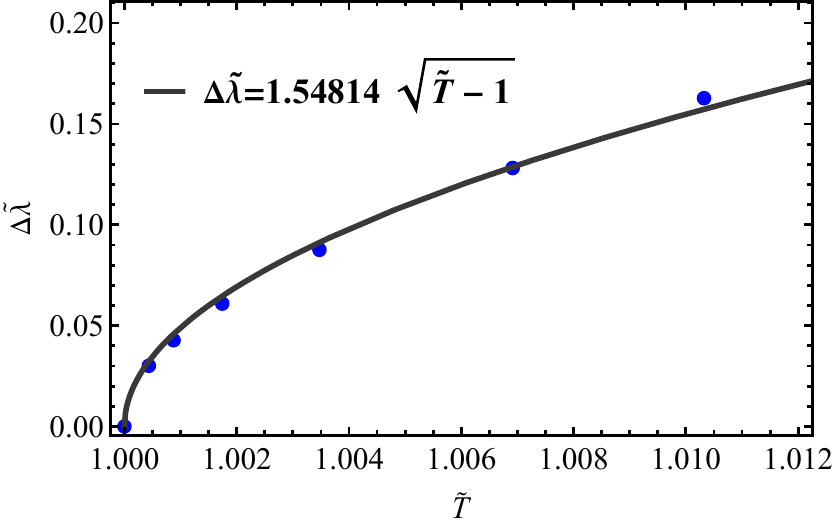}\label{p4dl}}
	\end{center}
	\caption{The diagram of $\Delta \tilde{\lambda}$ and $\tilde{T}$ for 5D timelike and null geodesic when $\alpha=0.006$. The blue dot is the $\Delta \tilde{\lambda}$ of phase transition points found in the neighborhood of critical charge $q_c=0.1491$, and the functions $\Delta \tilde{\lambda}=2.25837\sqrt{\tilde{T}-1}$ and $\Delta \tilde{\lambda}=1.54814\sqrt{\tilde{T}-1}$ are derived by fitting the black points for the timelike and null geodesic cases respectively.  }
	\label{critical_plot}
\end{figure}

We set $\lambda_s$ for the Small BH and $\lambda_l$ for the Large BH, and the difference $\Delta \lambda =\lambda _l-\lambda _s$ can be served as an order parameter. One can focus on the Hawking temperature of phase transition point $T_p$ nearing the critical temperature $T_c$ and $q_c$ in order to calculate the critical Lyapunov exponent $\lambda_c$. For simplicity, we use $\Delta \tilde{\lambda }$ and $\tilde{T}$ to substitute $\Delta \lambda/\lambda _c$ and $T_p/T_c$ respectively. For the intuitive understanding of the critical behavior, we illustrate a diagram of $\Delta \lambda /\lambda _c$ to $T_p/T_c$, as shown in Fig. \ref{critical_plot} with $\alpha=0.0065$. The blue point in the diagram is calculated with different $q$ in the neighborhood of the critical value $q_c$. The black curves is plot by fitting the points, with the approximate form as $\Delta \tilde{\lambda }= 2.25837 \ \sqrt{\tilde{T}-1}$ and $\Delta \tilde{\lambda }= 1.54814 \ \sqrt{\tilde{T}-1}$ corresponding to timelike and null geodesic respectively. 

We have the definition of critical exponent $\delta$ related to the order parameter as
\begin{align}
	\label{criticaldef}
	\Delta \lambda \equiv \lambda _l-\lambda _s\sim \left| T-T_C\right| {}^{\delta }. 
\end{align}
Here we introduce a concise and elegant  approximate approach to calculate the critical exponent \cite{Banerjee:2012zm}. We can make Taylor expand of the Lyapunov exponents at the critical point for as 
\begin{align}
	\label{criticalTe}
	\lambda =\lambda _c+ \left[\frac{\partial \lambda }{\partial r_+}\right]_c dr_+ +\mathcal{O}\left(r_+\right),
\end{align}
where the subscript "c" represents values at critical points. Notice that the horizon radius at phase transition point can be rewrite with critical horizon radius as 
\begin{align}
	\label{delrp}
	r_p=r_c(1+\Delta ),
\end{align}
where $\left|\Delta\right| \ll 1$. 
So $\Delta  \tilde{\lambda }$ is
\begin{align}
	\label{ldiff}
	\Delta  \tilde{\lambda }=\frac{\lambda _l-\lambda _s}{\lambda _c}=\frac{r_c}{\lambda _c}  \left[\frac{\partial \lambda }{\partial r_+}\right]_c\left(\Delta _l-\Delta _s\right), 
\end{align}
where $\Delta _l$ and $\Delta _s$ is the offset of $r_c$ in Eq. \eqref{delrp} on the Large BH branch and the Small BH branch respectively. 
Then we will deal with the Hawking temperature $T$ with Taylor expansion so as to have the order parameter $\Delta \lambda$ the same form as the definition. One can rewrite $ T(r_+)$ as
\begin{align}
	\label{dell}
	T\left(r_+\right)=T_c(1+\epsilon),
\end{align}
where $\left| \epsilon \right|\ll1 $. We Taylor expand the Hawking temperature $T(r_+)$ in a sufficiently small neighborhood of $r_c$.  
Therefore, the relationship between the order parameter $\Delta \lambda$ and Hawking temperature $T$ yielding,
\begin{align}
	\label{dellexp}
	\Delta  \tilde{\lambda }=k \sqrt{\tilde{T}-1},
\end{align}
where,
$$
	k = \frac{\sqrt{ T_c}}{ \lambda _c}   \left[\frac{\partial \lambda }{\partial r_+}\right]_c {\left[  \frac{1}{2} \frac{\partial ^2T}{  \partial r_+^2}\right]_c}^{-1/2}. 
$$
From the expression we can see the critical exponent $\delta$ is 1/2, which is consistent with the order parameter in the van der Waals fluid. From Fig. \ref{critical_plot} and the derivation above, we can infer that the performance of the $\Delta \lambda -T$ diagram near the critical point is in agreement with our theory. Thus, we can investigate the first order phase transition with Lyapunov exponent in $D$-dimensional Gauss-Bonnet AdS black holes, which is characterized by a change in Hawking temperature $T_p$.

\section{Conclusion and discussion\label{concl}}

In this paper, we first investigate the thermodynamics of charged Gauss-Bonnet AdS black hole with different $\alpha$. The thermodynamics of 4D Gauss-Bonnet black holes is analog to the 5D black holes with one critical value $\alpha_c$. The higher dimension Gauss-Bonnet black hole have two critical values $\alpha_{cs}$ and $\alpha_{cl}$ respectively.   
Then we investigate the relationship between chaotic properties and thermodynamic properties with Lyapunov exponents. We find the stability of the black hole can be characterized by the diagram of isobaric heat capacity versus the Lyapunov exponents. The Lyapunov exponent can be applied to reveal the phase transition of the Gauss-Bonnet black holes as its difference can be taken an order parameter.  

The phase transition of 4D and 5D charged Gauss-Bonnet AdS black hole can be divided into two branches by the critical value $\alpha_c$ for $q<q_c$ as shown in Figs. \ref{4par} and \ref{5par}. The Van de Waals-like Small/Large phase transition will occur when $\alpha<\alpha_c$. For $\alpha>\alpha_c$, there's no phase transition. In 6D spacetime, the Small/Large phase transition will always exist when $\alpha \in (0,\alpha_M)$ for $q<0.00211365$. For $\alpha \in (\alpha_{cs}, \alpha_{cl})$ in this case, there are two continuous Small/Large phase transitions, and the triple point will happen in the interval $(\alpha_{cs},\alpha_{cl})$. The phase transition of higher dimension$(D>6)$ spacetime will happen for the case when $q<q_c$ and the case for $\alpha \in (0,\alpha_{cs})\cup(\alpha_{cl},\alpha_M)$ when $q>q_c$ as presented in Table. \ref{critical_value_nD}.  

Then we introduce the Lyapunov exponent from the chaos theory for the investigation of the timelike and null geodesic to indirectly research the thermodynamic properties of the charged Gauss-Bonnet AdS black hole. We obtain the trend of Lyapunov exponents versus the event horizon and the Gauss-Bonnet constant, and we found $\lambda$ can be utilized to probe the stability of the black hole phase and the Small/Large phase transition by the $C_p-\lambda$ and $\lambda-T$ diagrams. The Lyapunov exponents will remain a constant in the null geodesic, while it will be come zero for the timelike geodesic because of the disappearance of unstable equilibrium. Especially, $\lambda$ will always become zero before the upper bound $\alpha_M$ in 4D spacetime, but it will only be always zero in the neighborhood of $\alpha_M$ in 5D. In 6D spacetime, the $\lambda$ presence of non-zero values when $\alpha\rightarrow\alpha_M$. 

In order to well probing the relationship between the Lyapunov exponent and the phase transition, we derive the diagrams of the Lyapunov exponent for timelike and null geodesic versus the Hawking temperature. The diagrams for the Small/Large phase transition case in 4D and 5D spacetime is multivalued, whose Hawking temperature of the phase transition will be accompanied with a discontinuous variation of $\lambda$. The 6D case when $\alpha \in (\alpha_{cs},\alpha_{cl})$ will be more subtle with other two unstable black hole phase, whose two multivalued structures well correspond to the two adjacent swallow tail structures in F-T diagram. The triple point case can be distinguished by the $\lambda$-T diagram presented in Figs. \ref{6_lt_T} and \ref{p_6_lt_T} with the coinciding of the two Small/Large phase transition temperature $T_{p1}$ and $T_{p2}$. We find that the triple point process can be described equivalently with the F-T diagram as shown in Fig. \ref{6_ft_3}. 

Then we find that the difference between the Lyapunov exponents for the Large BH and the Small BH near the critical situation is actually an order parameter, whose critical exponent is $1/2$ equivalent to the van der walls liquid, which has also been investigated in the RN AdS black holes and Born-Infield AdS black holes \cite{Guo:2022kio,Yang:2023hci}. Afterwards, black hole shadows were found to have a single-valued correspondence with the Lyapunov exponent. For a 5D charged Gauss-Bonnet black hole, the Lyapunov exponent decreases with the increasing of black hole shadow. Such a property provides a way to probe the chaotic features of black holes and also builds a bridge to explore thermodynamic properties through the shadow of a black hole. 

In our investigation, we pay attention to the thermodynamic properties of Gauss-Bonnet black holes with varying $\alpha$ and the phase transition characterized by the Lyapunov exponents of timelike and null geodesics. With the fact that the difference of Lyapunov exponents on the two non-adjacent black hole phases (e.g., Small BH and Large BH in the 5D Gauss-Bonnet black holes) can be taken as an order parameter, the diagrams of the Lyapunov exponent versus the Hawking temperature can reveal the phase behavior, as a good correspondence to the F-T diagrams including the triple point case. We can find that the black hole stability can be well reflected by the diagram of isobaric heat capacity versus Lyapunov exponents. By taking $\lambda$ as the bridge of the thermodynamics and the shadow of black hole, we are look forward to have the thermodynamics of black hole to be observable, which can be taken into account in our following studies. 

\begin{acknowledgments}
   The authors are grateful to Haitang Yang and Shaojie Yang for useful discussions. This work is supported by NSFC Grant Nos. 12175212 ,12275183 and 12275184. 
   
\end{acknowledgments}



\begin{thebibliography}{99} 
   \bibitem{Hawking:1971tu}
   S.~W.~Hawking,
   Gravitational radiation from colliding black holes,
   Phys. Rev. Lett. \textbf{26} (1971), 1344-1346. 
   
   \bibitem{Bekenstein:1973ur}
   J.~D.~Bekenstein,
   Black holes and entropy,
   Phys. Rev. D \textbf{7} (1973), 2333-2346. 
   
   \bibitem{Bekenstein:1974ax}
   J.~D.~Bekenstein,
   Generalized second law of thermodynamics in black hole physics,
   Phys. Rev. D \textbf{9} (1974), 3292-3300. 
   
   \bibitem{Hawking:1974rv}
   S.~W.~Hawking,
   Black hole explosions,
   Nature \textbf{248} (1974), 30-31.
   
   \bibitem{Bardeen:1973gs}
   J.~M.~Bardeen, B.~Carter and S.~W.~Hawking,
   The Four laws of black hole mechanics,
   Commun. Math. Phys. \textbf{31} (1973), 161-170.
   
   \bibitem{Maldacena:1997re}
   J.~M.~Maldacena,
   The Large N limit of superconformal field theories and supergravity,
   Adv. Theor. Math. Phys. \textbf{2} (1998), 231-252.
   
   \bibitem{Cai:1998vy}
   R.~G.~Cai and K.~S.~Soh,
   Topological black holes in the dimensionally continued gravity,
   Phys. Rev. D \textbf{59} (1999), 044013.
   
   \bibitem{Hawking:1982dh}
   S.~W.~Hawking and D.~N.~Page,
   Thermodynamics of Black Holes in anti-De Sitter Space,
   Commun. Math. Phys. \textbf{87} (1983), 577.
   
   \bibitem{Chamblin:1999tk}
   A.~Chamblin, R.~Emparan, C.~V.~Johnson and R.~C.~Myers,
   Charged AdS black holes and catastrophic holography,
   Phys. Rev. D \textbf{60} (1999), 064018.
   
   \bibitem{Chamblin:1999hg}
   A.~Chamblin, R.~Emparan, C.~V.~Johnson and R.~C.~Myers,
   Holography, thermodynamics and fluctuations of charged AdS black holes,
   Phys. Rev. D \textbf{60} (1999), 104026.
   
   \bibitem{He:2010zb}
   X.~He, B.~Wang, R.~G.~Cai and C.~Y.~Lin,
   Signature of the black hole phase transition in quasinormal modes,
   Phys. Lett. B \textbf{688} (2010), 230-236.
   
   \bibitem{Guo:2021enm}
   G.~Guo, P.~Wang, H.~Wu and H.~Yang,
   Quasinormal modes of black holes with multiple photon spheres,
   JHEP \textbf{06} (2022), 060.
   
   \bibitem{Yao:2020ftk}
   F.~Yao and J.~Tao,
   Extended phase space thermodynamics for dyonic black holes with a power Maxwell field,
   Phys. Rev. D \textbf{105} (2022) no.12, 124018.
   
   \bibitem{He:2016fiz}
   S.~He, L.~F.~Li and X.~X.~Zeng,
   Holographic Van der Waals-like phase transition in the Gauss\textendash{}Bonnet gravity,
   Nucl. Phys. B \textbf{915} (2017), 243-261.
   
   \bibitem{Huang:2021iyf}
   Y.~Huang, H.~Jing, J.~Tao and F.~Yao,
   Phase structures and transitions of quintessence surrounding RN black holes in a grand canonical ensemble,
   Chin. Phys. C \textbf{45} (2021) no.7, 075101.
   
   \bibitem{Bai:2022hti}
   N.~Bai, A.~He and J.~Tao,
   Microstructure of charged AdS black hole with minimal length effects*,
   Chin. Phys. C \textbf{46} (2022) no.12, 125105.
   
   \bibitem{Wang:2019urm}
   P.~Wang, H.~Yang and S.~Ying,
   Thermodynamics and phase transition of a Gauss-Bonnet black hole in a cavity,
   Phys. Rev. D \textbf{101} (2020) no.6, 064045.
   
   \bibitem{Wang:2018xdz}
   P.~Wang, H.~Wu and H.~Yang,
   Thermodynamics and Phase Transitions of Nonlinear Electrodynamics Black Holes in an Extended Phase Space,
   JCAP \textbf{04} (2019), 052.
   
   \bibitem{Li:2019dai}
   H.~Li, Y.~Chen and S.~J.~Zhang,
   Photon orbits and phase transitions in Born-Infeld-dilaton black holes,
   Nucl. Phys. B \textbf{954} (2020), 114975.
   
   \bibitem{Wang:2019kxp}
   P.~Wang, H.~Wu and H.~Yang,
   Thermodynamics and Phase Transition of a Nonlinear Electrodynamics Black Hole in a Cavity,
   JHEP \textbf{07} (2019), 002.
   
   \bibitem{Bai:2023woh}
   N.~C.~Bai, L.~Li and J.~Tao,
   Superfluid \ensuremath{\lambda} transition in charged AdS black holes,
   Sci. China Phys. Mech. Astron. \textbf{66} (2023) no.12, 120411.
    
   \bibitem{Boulware:1985wk}
   D.~G.~Boulware and S.~Deser,
   String Generated Gravity Models,
   Phys. Rev. Lett. \textbf{55} (1985), 2656.
   
   \bibitem{Zwiebach:1985uq}
   B.~Zwiebach,
   Curvature Squared Terms and String Theories,
   Phys. Lett. B \textbf{156} (1985), 315-317.
   
   \bibitem{Lovelock:1971yv}
   D.~Lovelock,
   The Einstein tensor and its generalizations,
   J. Math. Phys. \textbf{12} (1971), 498-501.
   
   \bibitem{Lovelock:1972vz}
   D.~Lovelock,
   The four-dimensionality of space and the einstein tensor,
   J. Math. Phys. \textbf{13} (1972), 874-876.
   
   \bibitem{Wang:2020eln}
   P.~Wang, H.~Wu, H.~Yang and S.~Ying,
   Derive Lovelock Gravity from String Theory in Cosmological Background,
   JHEP \textbf{05} (2021), 218.
   
   \bibitem{Bai:2022klw}
   N.~C.~Bai, L.~Li and J.~Tao,
   Topology of black hole thermodynamics in Lovelock gravity,
   Phys. Rev. D \textbf{107} (2023) no.6, 064015.
   
   \bibitem{Liu:2022aqt}
   C.~Liu and J.~Wang,
   Topological natures of the Gauss-Bonnet black hole in AdS space,
   Phys. Rev. D \textbf{107} (2023) no.6, 064023.
  
   \bibitem{Cai:2001dz}
   R.~G.~Cai,
   Gauss-Bonnet black holes in AdS spaces,
   Phys. Rev. D \textbf{65} (2002), 084014.
   
   \bibitem{Oikonomou:2020sij}
   V.~K.~Oikonomou and F.~P.~Fronimos,
   Reviving non-minimal Horndeski-like theories after GW170817: kinetic coupling corrected Einstein\textendash{}Gauss\textendash{}Bonnet inflation,
   Class. Quant. Grav. \textbf{38} (2021) no.3, 035013.
   
   \bibitem{Oikonomou:2021kql}
   V.~K.~Oikonomou,
   A refined Einstein\textendash{}Gauss\textendash{}Bonnet inflationary theoretical framework,
   Class. Quant. Grav. \textbf{38} (2021) no.19, 195025.
   
   \bibitem{Odintsov:2020sqy}
   S.~D.~Odintsov, V.~K.~Oikonomou and F.~P.~Fronimos,
   Rectifying Einstein-Gauss-Bonnet Inflation in View of GW170817,
   Nucl. Phys. B \textbf{958} (2020), 115135.
   
   \bibitem{Banados:1993ur}
   M.~Banados, C.~Teitelboim and J.~Zanelli,
   Dimensionally continued black holes,
   Phys. Rev. D \textbf{49} (1994), 975-986.
   
   \bibitem{Wiltshire:1985us}
   D.~L.~Wiltshire,
   Spherically Symmetric Solutions of Einstein-maxwell Theory With a {Gauss-Bonnet} Term,
   Phys. Lett. B \textbf{169} (1986), 36-40.
   
   \bibitem{Hendi:2017lgb}
   S.~H.~Hendi, B.~E.~Panah and S.~Panahiyan,
   Black Hole Solutions in Gauss-Bonnet-Massive Gravity in the Presence of Power-Maxwell Field,
   Fortsch. Phys. \textbf{66} (2018) no.3, 1800005.
   
   \bibitem{Cvetic:2001bk}
   M.~Cvetic, S.~Nojiri and S.~D.~Odintsov,
   Black hole thermodynamics and negative entropy in de Sitter and anti-de Sitter Einstein-Gauss-Bonnet gravity,
   Nucl. Phys. B \textbf{628} (2002), 295-330.
   
   \bibitem{Cai:2013qga}
   R.~G.~Cai, L.~M.~Cao, L.~Li and R.~Q.~Yang,
   P-V criticality in the extended phase space of Gauss-Bonnet black holes in AdS space,
   JHEP \textbf{09} (2013), 005.
   
   \bibitem{Wei:2014hba}
   S.~W.~Wei and Y.~X.~Liu,
   Triple points and phase diagrams in the extended phase space of charged Gauss-Bonnet black holes in AdS space,
   Phys. Rev. D \textbf{90} (2014) no.4, 044057.
   
   \bibitem{Hendi:2016yof}
   S.~H.~Hendi, G.~Q.~Li, J.~X.~Mo, S.~Panahiyan and B.~Eslam Panah,
   New perspective for black hole thermodynamics in Gauss\textendash{}Bonnet\textendash{}Born\textendash{}Infeld massive gravity,
   Eur. Phys. J. C \textbf{76} (2016) no.10, 571.
   
   \bibitem{Haroon:2020vpr}
   S.~Haroon, R.~A.~Hennigar, R.~B.~Mann and F.~Simovic,
   Thermodynamics of Gauss-Bonnet-de Sitter Black Holes,
   Phys. Rev. D \textbf{101} (2020), 084051.
   
   \bibitem{Qu:2022nrt}
   Y.~Qu, J.~Tao and H.~Yang,
   Thermodynamics and phase transition in central charge criticality of charged Gauss-Bonnet AdS black holes,
   Nucl. Phys. B \textbf{992} (2023), 116234.
   
   \bibitem{Wei:2012ui}
   S.~W.~Wei and Y.~X.~Liu,
   Critical phenomena and thermodynamic geometry of charged Gauss-Bonnet AdS black holes,
   Phys. Rev. D \textbf{87} (2013) no.4, 044014.
   
   \bibitem{Glavan:2019inb}
   D.~Glavan and C.~Lin,
   Einstein-Gauss-Bonnet Gravity in Four-Dimensional Spacetime,
   Phys. Rev. Lett. \textbf{124} (2020) no.8, 081301.
   
   \bibitem{Fernandes:2020rpa}
   P.~G.~S.~Fernandes,
   Charged black holes in AdS spaces in 4D Einstein Gauss-Bonnet gravity,
   Phys. Lett. B \textbf{805} (2020), 135468.
   
   \bibitem{Yang:2020jno}
   K.~Yang, B.~M.~Gu, S.~W.~Wei and Y.~X.~Liu,
   Born\textendash{}Infeld black holes in 4D Einstein\textendash{}Gauss\textendash{}Bonnet gravity,
   Eur. Phys. J. C \textbf{80} (2020) no.7, 662.
   
   \bibitem{Kumar:2020owy}
   R.~Kumar and S.~G.~Ghosh,
   Rotating black holes in $4D$ Einstein-Gauss-Bonnet gravity and its shadow,
   JCAP \textbf{07} (2020), 053.
   
   \bibitem{Konoplya:2020qqh}
   R.~A.~Konoplya and A.~Zhidenko,
   Black holes in the four-dimensional Einstein-Lovelock gravity,
   Phys. Rev. D \textbf{101} (2020) no.8, 084038.
   
   \bibitem{Wang:2020pmb}
   Y.~Y.~Wang, B.~Y.~Su and N.~Li,
   Hawking\textendash{}Page phase transitions in four-dimensional Einstein\textendash{}Gauss\textendash{}Bonnet gravity,
   Phys. Dark Univ. \textbf{31} (2021), 100769.
   
   \bibitem{Marks:2021fpe}
   G.~A.~Marks, F.~Simovic and R.~B.~Mann,
   Phase transitions in 4D Gauss\textendash{}Bonnet\textendash{}de Sitter black holes,
   Phys. Rev. D \textbf{104} (2021) no.10, 104056.
   
   \bibitem{Suzuki:1996gm}
   S.~Suzuki and K.~i.~Maeda,
   Chaos in Schwarzschild space-time: The motion of a spinning particle,
   Phys. Rev. D \textbf{55} (1997), 4848-4859.
   
   \bibitem{Lu:2018mpr}
   F.~Lu, J.~Tao and P.~Wang,
   Minimal Length Effects on Chaotic Motion of Particles around Black Hole Horizon,
   JCAP \textbf{12} (2018), 036.
   
   \bibitem{Hartl:2002ig}
   M.~D.~Hartl,
   Dynamics of spinning test particles in Kerr space-time,
   Phys. Rev. D \textbf{67} (2003), 024005.
   
   \bibitem{Bombelli:1991eg}
   L.~Bombelli and E.~Calzetta,
   Chaos around a black hole,
   Class. Quant. Grav. \textbf{9} (1992), 2573-2599.
   
   \bibitem{Wang:2016wcj}
   M.~Wang, S.~Chen and J.~Jing,
   Chaos in the motion of a test scalar particle coupling to the Einstein tensor in Schwarzschild\textendash{}Melvin black hole spacetime,
   Eur. Phys. J. C \textbf{77} (2017) no.4, 208.
   
   \bibitem{Chen:2016tmr}
   S.~Chen, M.~Wang and J.~Jing,
   Chaotic motion of particles in the accelerating and rotating black holes spacetime,
   JHEP \textbf{09} (2016), 082.
   
   \bibitem{Yang:2023hci}
   S.~Yang, J.~Tao, B.~Mu and A.~He,
   Lyapunov exponents and phase transitions of Born-Infeld AdS black holes,
   JCAP \textbf{07} (2023), 045.
   
   \bibitem{EventHorizonTelescope:2019dse}
   K.~Akiyama \textit{et al.} [Event Horizon Telescope],
   First M87 Event Horizon Telescope Results. I. The Shadow of the Supermassive Black Hole,
   Astrophys. J. Lett. \textbf{875} (2019), L1.
   
   \bibitem{EventHorizonTelescope:2019uob}
   K.~Akiyama \textit{et al.} [Event Horizon Telescope],
   First M87 Event Horizon Telescope Results. II. Array and Instrumentation,
   Astrophys. J. Lett. \textbf{875} (2019) no.1, L2.
   
   \bibitem{EventHorizonTelescope:2019jan}
   K.~Akiyama \textit{et al.} [Event Horizon Telescope],
   First M87 Event Horizon Telescope Results. III. Data Processing and Calibration,
   Astrophys. J. Lett. \textbf{875} (2019) no.1, L3.
   
   \bibitem{EventHorizonTelescope:2019ths}
   K.~Akiyama \textit{et al.} [Event Horizon Telescope],
   First M87 Event Horizon Telescope Results. IV. Imaging the Central Supermassive Black Hole,
   Astrophys. J. Lett. \textbf{875} (2019) no.1, L4.
   
   \bibitem{EventHorizonTelescope:2019pgp}
   K.~Akiyama \textit{et al.} [Event Horizon Telescope],
   First M87 Event Horizon Telescope Results. V. Physical Origin of the Asymmetric Ring,
   Astrophys. J. Lett. \textbf{875} (2019) no.1, L5.
   
   \bibitem{EventHorizonTelescope:2019ggy}
   K.~Akiyama \textit{et al.} [Event Horizon Telescope],
   First M87 Event Horizon Telescope Results. VI. The Shadow and Mass of the Central Black Hole,
   Astrophys. J. Lett. \textbf{875} (2019) no.1, L6.
   
   \bibitem{EventHorizonTelescope:2022wkp}
   K.~Akiyama \textit{et al.} [Event Horizon Telescope],
   First Sagittarius A* Event Horizon Telescope Results. I. The Shadow of the Supermassive Black Hole in the Center of the Milky Way,
   Astrophys. J. Lett. \textbf{930} (2022) no.2, L12.
   
   \bibitem{EventHorizonTelescope:2022apq}
   K.~Akiyama \textit{et al.} [Event Horizon Telescope],
   First Sagittarius A* Event Horizon Telescope Results. II. EHT and Multiwavelength Observations, Data Processing, and Calibration,
   Astrophys. J. Lett. \textbf{930} (2022) no.2, L13.
   
   \bibitem{EventHorizonTelescope:2022wok}
   K.~Akiyama \textit{et al.} [Event Horizon Telescope],
   First Sagittarius A* Event Horizon Telescope Results. III. Imaging of the Galactic Center Supermassive Black Hole,
   Astrophys. J. Lett. \textbf{930} (2022) no.2, L14.
   
   \bibitem{EventHorizonTelescope:2022exc}
   K.~Akiyama \textit{et al.} [Event Horizon Telescope],
   First Sagittarius A* Event Horizon Telescope Results. IV. Variability, Morphology, and Black Hole Mass,
   Astrophys. J. Lett. \textbf{930} (2022) no.2, L15.
   
   \bibitem{EventHorizonTelescope:2022urf}
   K.~Akiyama \textit{et al.} [Event Horizon Telescope],
   First Sagittarius A* Event Horizon Telescope Results. V. Testing Astrophysical Models of the Galactic Center Black Hole,
   Astrophys. J. Lett. \textbf{930} (2022) no.2, L16.
   
   \bibitem{EventHorizonTelescope:2022xqj}
   K.~Akiyama \textit{et al.} [Event Horizon Telescope],
   First Sagittarius A* Event Horizon Telescope Results. VI. Testing the Black Hole Metric,
   Astrophys. J. Lett. \textbf{930} (2022) no.2, L17.
   
   \bibitem{Synge:1966okc}
   J.~L.~Synge,
   The Escape of Photons from Gravitationally Intense Stars,
   Mon. Not. Roy. Astron. Soc. \textbf{131} (1966) no.3, 463-466.
   
   \bibitem{deVries:1999tiy}
   A.~de Vries,
   The apparent shape of a rotating charged black hole, closed photon orbits and the bifurcation set $A_4$,
   Class. Quant. Grav. \textbf{17} (1999) no.1, 123-144.
   
   \bibitem{Bardeen:1972fi}
   J.~M.~Bardeen, W.~H.~Press and S.~A.~Teukolsky,
   Rotating black holes: Locally nonrotating frames, energy extraction, and scalar synchrotron radiation,
   Astrophys. J. \textbf{178} (1972), 347.
   
   \bibitem{Grenzebach:2014fha}
   A.~Grenzebach, V.~Perlick and C.~L\"ammerzahl,
   Photon Regions and Shadows of Kerr-Newman-NUT Black Holes with a Cosmological Constant,
   Phys. Rev. D \textbf{89} (2014) no.12, 124004.
   
   \bibitem{Guo:2018kis}
   M.~Guo, N.~A.~Obers and H.~Yan,
   Observational signatures of near-extremal Kerr-like black holes in a modified gravity theory at the Event Horizon Telescope,
   Phys. Rev. D \textbf{98} (2018) no.8, 084063.
   
   \bibitem{Hennigar:2018hza}
   R.~A.~Hennigar, M.~B.~J.~Poshteh and R.~B.~Mann,
   Shadows, Signals, and Stability in Einsteinian Cubic Gravity,
   Phys. Rev. D \textbf{97} (2018) no.6, 064041.
   
   \bibitem{Amir:2017slq}
   M.~Amir, B.~P.~Singh and S.~G.~Ghosh,
   Shadows of rotating five-dimensional charged EMCS black holes,
   Eur. Phys. J. C \textbf{78} (2018) no.5, 399.
   
   \bibitem{Jusufi:2020cpn}
   K.~Jusufi, M.~Jamil and T.~Zhu,
   Shadows of Sgr A$^{*}$ black hole surrounded by superfluid dark matter halo,
   Eur. Phys. J. C \textbf{80} (2020) no.5, 354.
   
   \bibitem{Bousder:2021aek}
   M.~Bousder, K.~El Bourakadi and M.~Bennai,
   Charged 4D Einstein-Gauss-Bonnet black hole: Vacuum solutions, Cauchy horizon, thermodynamics,
   Phys. Dark Univ. \textbf{32} (2021), 100839.
   
   \bibitem{Hawking:1975vcx}
   S.~W.~Hawking,
   Particle Creation by Black Holes,
   Commun. Math. Phys. \textbf{43} (1975), 199-220
   [erratum: Commun. Math. Phys. \textbf{46} (1976), 206].
   
   \bibitem{Bekenstein:1972tm}
   J.~D.~Bekenstein,
   Black holes and the second law,
   Lett. Nuovo Cim. \textbf{4} (1972), 737-740.
   
   \bibitem{Guo:2022kio}
   X.~Guo, Y.~Lu, B.~Mu and P.~Wang,
   Probing phase structure of black holes with Lyapunov exponents,
   JHEP \textbf{08} (2022), 153.
   
   \bibitem{Cardoso:2008bp}
   V.~Cardoso, A.~S.~Miranda, E.~Berti, H.~Witek and V.~T.~Zanchin,
   Geodesic stability, Lyapunov exponents and quasinormal modes,
   Phys. Rev. D \textbf{79} (2009) no.6, 064016.
   
   \bibitem{Wei:2023fkn}
   S.~W.~Wei and Y.~X.~Liu,
   Aschenbach effect and circular orbits in static and spherically symmetric black hole backgrounds.
   
   \bibitem{Banerjee:2012zm}
   R.~Banerjee and D.~Roychowdhury,
   Critical behavior of Born Infeld AdS black holes in higher dimensions,
   Phys. Rev. D \textbf{85} (2012), 104043.

\end{thebibliography}
\end{document}